\documentclass[fleqn,usenatbib]{mnras}
\usepackage{newtxtext,newtxmath}
\usepackage[T1]{fontenc}

\DeclareRobustCommand{\VAN}[3]{#2}
\let\VANthebibliography\thebibliography
\def\thebibliography{\DeclareRobustCommand{\VAN}[3]{##3}\VANthebibliography}

\usepackage{graphicx}	% Including figure files
\usepackage{amsmath}	% Advanced maths commands

\title[$\beta$ in galaxies at $1.5 < z < 4.2$]{The dust emissivity index $\beta$ in infrared-bright galaxies at $1.5 < z < 4.2$}

\author[G. J. Bendo et al.]{
G. J. Bendo$^{1}$\thanks{E-mail: george.bendo@manchester.ac.uk}, 
T. J. L. C. Bakx$^{2}$,
H. S. B. Algera$^{3,4,5}$,
A. Amvrosiadis$^{6}$,
S. Berta$^{7}$,
L. Bonavera$^{8,9}$,\newauthor
P. Cox$^{10}$,
G. De Zotti$^{11}$,
S. Eales$^{12}$,
J. Gonz\'alez-Nuevo$^{8,9}$,
M. Hagimoto$^{13}$,
D. Ismail$^{14}$,
D. A. Riechers$^{15}$,\newauthor
S. Serjeant$^{16}$,
M. W. L. Smith$^{12}$,
P. Temi$^{17}$,
T. Tsukui$^{18,19}$,
S. A. Urquhart$^{16}$,
C. Vlahakis$^{20}$
\\
$^1$ UK ALMA Regional Centre Node, Jodrell Bank Centre for Astrophysics, Department of Physics and Astronomy, The University of Manchester,\\ Oxford Road, Manchester M13 9PL, United Kingdom\\
$^{2}$ Department of Space, Earth, \& Environment, Chalmers University of Technology, Chalmersplatsen 4 412 96 Gothenburg, Sweden \\
$^{3}$Institute of Astronomy and Astrophysics, Academia Sinica, 11F of Astronomy-Mathematics Building, No.1, Sec. 4, Roosevelt Rd, Taipei 106216, Taiwan, R.O.C. \\
$^{4}$ Hiroshima Astrophysical Science Center, Hiroshima University, 1-3-1 Kagamiyama, Higashi-Hiroshima, Hiroshima 739-8526, Japan \\
$^{5}$ National Astronomical Observatory of Japan, 2-21-1, Osawa, Mitaka, Tokyo, Japan \\
$^{6}$  Institute for Computational Cosmology, Department of Physics, Durham University, South Road, Durham DH1 3LE, United Kingdom\\
$^{7}$ Institut de Radioastronomie Millim\'etrique (IRAM), 300 Rue de la Piscine, 38400 Saint-Martin-d’H\`{e}res, France\\
$^{8}$ Departamento de Fisica, Universidad de Oviedo, C. Federico Garcia Lorca 18, 33007 Oviedo, Spain\\
$^{9}$ Instituto Universitario de Ciencias y Tecnolog\'ias Espaciales de Asturias (ICTEA), C. Independencia 13, 33004 Oviedo, Spain\\
$^{10}$ Institut d’Astrophysique de Paris, Sorbonne Universit\'{e},UPMC Universit\'{e} Paris 6 and CNRS, UMR 7095, 98 bis boulevard Arago, F-75014 Paris, France\\
$^{11}$ INAF, Osservatorio Astronomico di Padova, Vicolo Osservatorio 5, I-35122 Padova, Italy\\
$^{12}$ School of Physics and Astronomy, Cardiff University, The Parade, Cardiff, CF24 3AA, UK\\
$^{13}$ Department of Physics, Graduate School of Science, Nagoya University, Nagoya 464-8602, Japan\\
$^{14}$ Observatoire Astronomique de Strasbourg, Université de Strasbourg, CNRS, UMR 7550, 67000 Strasbourg, France\\
$^{15}$ Institut f\"ur Astrophysik, Universit\"at zu K\"oln, Z\"ulpicher Stra{\ss}e 77, D-50937 K\"oln, Germany\\
$^{16}$ School of Physical Sciences, The Open University, Milton Keynes, MK7 6AA, UK\\
$^{17}$ NASA - Ames Research Center, MS 245-6, Moffett Field, CA 94035, USA\\
$^{18}$ Research School of Astronomy and Astrophysics, Australian National University, Cotter Road, Weston Creek, ACT 2611, Australia\\
$^{19}$ ARC Centre of Excellence for All Sky Astrophysics in 3 Dimensions (ASTRO 3D), Australia\\
$^{20}$ National Radio Astronomy Observatory, 520 Edgemont Road, Charlottesville, VA 22903, USA\\
}

\pubyear{2025}

\begin{document}
\label{firstpage}
\pagerange{\pageref{firstpage}--\pageref{lastpage}}
\maketitle

\begin{abstract}
We have measured the dust emissivity index $\beta$ for 21 infrared-bright sources (including several gravitationally lensed galaxies) at $1.5 < z < 4.2$ using Atacama Large Millimeter/submillimeter Array (ALMA) 101-199 GHz data sampling the Rayleigh-Jeans side of the SED.  These data are largely insensitive to temperature variations and therefore should provide robust measurements of $\beta$.  We obtain a mean $\beta$ of 2.2 with a standard deviation of 0.6 that is at the high end of the range of values that had previously been measured in many galactic and extragalactic sources.  We find no systematic variation in $\beta$ versus redshift.  We also demonstrate with a subset of our sources that these higher $\beta$ values have significant implications for modelling dust emission and in particular for calculating dust masses or the wavelength at which dust becomes optically thick.
\end{abstract}


\section{Introduction}
\label{s_intro}

Extragalactic surveys with multiple far-infrared and submillimetre single-dish telescopes, including the {\it Herschel} Space Observatory \citep{pilbratt2010}; the James Clerk Maxwell Telescope, and the South Pole Telescope \citep[SPT;][]{carlstrom2011}, detected the far-infrared dust emission from a large population of infrared-bright sources at high redshifts ($z \gtrsim 2$).  A signficant fraction of sources with 500~$\mu$m flux densities above 100~mJy are expected to be gravitationally-lensed galaxies \citep[e.g.,][]{negrello2017}, although others may potentially be galaxies that are simply intrinsically bright or protoclusters \citep[e.g.,][]{miller2018, oteo2018}.  The gravitationally lensed systems are of particular interest, as they magnify the light from objects at higher redshifts that would otherwise be difficult to detect \citep[e.g.,][]{swinbank2010,dye2015,dye2022}, as the lensed light can be used to derive the mass profiles and dark matter content of the lensing galaxies \citep[e.g.][]{treu2010}, and as statistical information about the lenses can be used to place constraints on cosmological parameters \citep{grillo2008,eales2015,gonzaleznuevo2017,gonzaleznuevo2021,bonavera2020,bonavera2021,cueli2021,cueli2024}.

Unfortunately, these sources were unresolved in the {\it Herschel} beam, and the detections were primarily limited to photometric measurements.  Follow-up observations with ground-based interferometers such as the Atacama Large Millimeter/submillimeter Array (ALMA) and NOrthern Extended Millimetre Array (NOEMA) have been used to resolve the emission, to identify sources affected by confusion effects, and to measure spectral line emission that can be used to confirm the redshift of these objects.  While multiple interferometric observations have been performed on individual or small sets of gravitational lens candidates from {\it Herschel}, the most notable surveys at this time include the $z$-GAL project with NOEMA \citep{berta2023, cox2023, ismail2023}, the ongoing ALMA follow-up observations of dusty star-forming galaxies identified by the SPT \citep{viera2013, weiss2013, spilker2016, reuter2020}, the ALMA Spectroscopic Survey of the Brightest Submillimeter Galaxies in the SCUBA-2-COSMOS Field \citep{liao2024}, and the Bright Extragalactic ALMA Redshift Survey (BEARS), which is the focus of our analysis.

BEARS performed spectral scan observations with ALMA of 85 fields containing gravitational lens candidates identified in the South Galactic Pole field observed by the {\it Herschel} Astrophysical Terahertz Large Area Survey \citep[H-ATLAS;][]{eales2010}.  These ALMA observations covered frequency ranges from 89.6 to 112.8~GHz in ALMA Band 3 (with some exceptions\footnote{The data for nine fields used in the analysis were acquired with the ALMA 7-m Array and covered frequency ranges of 86.6 to 115.7~GHz.  The HerBS-49 field was observed over the same frequency range but with notable gaps at 97.0-98.6 and 108.9-112.2~GHz.}) and from 139.0 to 162.2~GHz in ALMA Band 4.  The primary goal was to detect and measure spectral line emission from redshifted CO and other spectral lines to determine the spectroscopic redshifts to the objects in these fields, and this analysis was published by \citet{urquhart2022}.  Additionally, \citet{hagimoto2023} published additional analyses of the spectral line emission, including the application of photodissociation models, examinations of multiple scaling relations, and calculations of gas-to-dust ratios, and \citet{bendo2023} used the continuum data to study the multiplicities in the observed fields and the spectral energy distributions (SEDs) of the galaxies.

A total of 142 millimetre sources with redshifts ranging from 1.5 to 4.2 were detected in the BEARS fields.  Most of the sources with measured redshifts have infrared luminosities not including magnification corrections of $>10^{13}$~L$_\odot$, which would be comparable to what is seen in similar surveys of gravitational lens candidates \citep[e.g.][]{ismail2023}.  When magnification corrections are applied, the infrared luminosities of the sources with measured redshifts mostly lie in between $10^{12}$ and $10^{13}$~L$_\odot$, which would make the objects ultraluminous infrared galaxies.  These sources tend to be slightly brighter than the main sequence galaxies at these redshifts that have been selected in ultraviolet, optical, or near-infrared bands \citep[e.g.,][]{schreiber2018, bouwens2020}, although the luminosity ranges overlap.  However, the magnification-corrected infrared luminosities of our sample are comparable to those from galaxies selected just by their submillimetre flux densities \citep[e.g.,][]{dudzeviciute2021}.

While this first set of observations of the BEARS fields was critical for identifying the redshifts of many of the sources in the fields, the full-width at half-maxima (FWHMs) of the beams in the images were $\geq$2~arcsec.  Most individual sources were not resolved at this resolution, and in particular, it was not possible to resolve any Einstein rings in these data, which would confirm that these sources are gravitational lenses.  Additionally, the observations did not cover higher frequency lines, including higher order CO lines, that could be used to extend the analysis of the gas within these galaxies.

\citet{bakx2024}, in a project named A Novel high-$z$ submillimetre Galaxy Efficient Line Survey (ANGELS) performed follow-up ALMA observations in ALMA Bands 3, 4, 5, 6, 7, and 8 (ranging from 98.4 to 419.6~GHz or 714~$\mu$m to 3.05 mm) of a subset of 16 BEARS fields that had superior angular resolutions (with FWHM of 0.10-0.30~arcsec) compared to the original BEARS data.  The continuum measurements from ANGELS have provided ample amounts of new data that have been quite useful in further defining the SEDs, particularly the Rayleigh-Jeans slopes of the SEDs, of the objects in the BEARS field.  These data provide an excellent opportunity to measure the dust emissivity index $\beta$ in the power law that describes dust emissivity as a function of frequency.  Measuring this quantity is critically important for calculating accurate dust masses, but, in addition, its value is also related to the properties of the dust grains, particularly the composition of the grains.  Variations in $\beta$ would be indicative of variations in interstellar dust grain properties.

Theoretical models of dust emission typically set $\beta \cong 2$ \citep[e.g,][]{draine2003}, but the quantity has been poorly constrained observationally, mainly because submillimetre and millimetre data are needed to constrain $\beta$, and observing in these bands has been technically challenging because of the limited atmospheric transmissivity at $<$3~mm.  The mean $\beta$ values measured for the Milky Way and other nearby galaxies have ranged from 1.4 to 2.3\citep{dupac2003, yang2007, boselli2012, paradis2010, juvela2011, planckxxiii2011, smith2012, galametz2014, planck2014, planck2015, tabatabaei2014, lamperti2019}, although the analysis presented by \citet{desert2008} suggested that $\beta$ could be as high as 4 in some locations in the Milky Way.  Observations of $z>1$ galaxies have yielded $\beta$ values with a similar range \citep{beelen2006, magnelli2012, dacunha2021, cooper2022, ismail2023, tsukui2023, witstok2023, liao2024, tripodi2024, ward2024, algera2024}.  Additionally, multiple analyses have found that $\beta$ varies with temperature within the Milky Way \citep[e.g.,][]{dupac2003, desert2008, paradis2010, juvela2011, planckxxiii2011}, among or within nearby galaxies \citep[e.g.,][]{yang2007, gordon2014, tabatabaei2014, lamperti2019, whitworth2019}, and among more distant galaxies \citep[e.g.][]{dacunha2021, ismail2023, mckay2023}, although variations in $\beta$ with redshift have not been seen \citep{bendo2023, ismail2023, witstok2023, liao2024, tripodi2024, ward2024}.  

One of the limitations with most of the above studies is that the values of $\beta$ have been derived using modified blackbodies or other functions fit to the peak and Rayleigh-Jeans sides of the dust SEDs.  Degeneracies in these fits as well as oversimplifications of the dust SED that do not properly account for warmer dust could lead to false relations between $\beta$ and temperature \citep[e.g.,][]{shetty2009a, shetty2009b, juvela2012, kelly2012, juvela2013} or otherwise inaccurate $\beta$ values \citep[e.g.,][]{kirkpatrick2014, hunt2015, bendo2023}.

With the ALMA Band 3-5 data (corresponding to 98.4-206.6~GHz or 1.45-3.04~mm) from ANGELS, we can make measurements of the slope of the Rayleigh-Jeans side of the dust SED that will primarily depend on $\beta$ and that will be largely insensitive of temperature variations or redshift variations (over the range $0 \leq z \leq 5$, which encompasses the measured redshifts of the BEARS sample from \citealt{urquhart2022}).  These measurements will therefore avoid the degeneracies with temperature that affect the $\beta$ measurements published in most other papers and therefore could potentially be more robust.  Additionally, we can use these data to search for variations in $\beta$ related to the wavelength where the SED peaks ($\lambda_\mathrm{peak}$), which can be used as an empirical measurement related to dust temperature, and variations in $\beta$ related to redshift.  Our full analysis of the $\beta$ for these galaxies is presented in this paper.

\section{Data}

\subsection{Observations and data processing}

The BEARS sample selection and observations are described by \citet{urquhart2022}, and the ANGELS observations and data reduction are described by \citet{bakx2024}.  To summarize, ANGELS selected a set of 16 objects from BEARS as a pilot study for high angular resolution follow-up observations in six different ALMA Bands.  These observations were intended to demonstrate the feasibility of using relatively short observations to make multiple line and continuum measurements of a broad sample of objects at $z \geq 2$.  To improve the efficiency of the observations, ANGELS mostly selected fields with objects with identified redshifts where various spectral lines of interest (including multiple CO lines, [C{\sc I}] at 492~GHz, [O{\sc I}] at 2060 GHz, [N{\sc II}] at 1461~GHz, and multiple H$_2$O lines) would fall within the same set of spectral tunings used across ALMA Bands 3, 4, 5, 6, 7, and 8.   Note that the targeted objects were at different redshifts, so for each object, different spectral lines may fall within the spectral tunings, but the expectation was that some type of spectral line emission would be detected for each object.  Also note that redshifts had not been measured for several objects in these fields, most notably HerBS-87, HerBS-104B, and HerBS-170, when the proposal for these observations was submitted.

The observations were executed in program 2021.1.01628.S (PI: T. Bakx) in ALMA Cycle 8 using the 12~m Array in extended configurations that could achieve the requested angular resolutions ranging from 0.10 to 0.40~arcsec (although the final beams may differ from what was requested).  Within each band, all targets as well as a bandpass/flux calibrator (J2258-2758) and a phase calibrator (J2359-3133) were observed within a single Scheduling Block that was performed using just one execution.  Because this program was demonstrating the feasibility of snapshot-like observations for spectral line and continuum measurements, the requested sensitivity levels were set relatively high (which means that the rms noise in the final images is also relatively high), and each of the science targets themselves were observed for only 90-150~s (depending on the specific target and waveband).  The details of the observations in each band are listed in Table~\ref{t_obsspw}.

\begin{table*}
\centering
\begin{minipage}{170mm}
\caption{Observing information}
\label{t_obsspw}
\begin{tabular}{@{}lcccccccccc@{}}
\hline
ALMA &
  Frequencies &
  $uv$ &
  Beam &
  Maximum &
  Primary &
  Pixel &
  \multicolumn{2}{c}{Image} &
  Typical &
  Calibration \\
Band &
  (GHz) &
  coverage &
  FWHM &
  recoverable &
  beam &
  scale &
  \multicolumn{2}{c}{size} &
  rms &
  uncertainty$^a$ \\
&
  &
  (m) &
  (arcsec) &
  scale &
  diameter &
  (arcsec) &
  (pixels) &
  (arcsec) &
  noise\\
&
  &
  &
  &
  (arcsec) &
  (arcsec) &
  &
  &
  &
  (mJy/beam) &
  \\
\hline
3 &
  98.36 - 102.09 &
  31 - 3058 &
  0.52 $\times$ 0.32 &
  4.9 &
  82.0 &
  0.05 &
  1800 $\times$ 1800 &
  90  $\times$ 90 &
  0.05 &
  5\% \\
&
  110.37 - 114.10 &
  &
  &
  &
  &
  &
  &
  \\  
4 &
  132.44 - 136.16 &
  14 - 3065 &
  0.39 $\times$ 0.35&
  4.2 &
  62.0 &
  0.05 &
  1500 $\times$ 1500 &
  75 $\times$ 75 &
  0.05 &
  5\% \\ 
&
  144.44 - 148.17 &
  &
  &
  &
  &
  &
  &
  \\  
5 &
  190.86 - 194.59 &
  14 - 2613 &
  0.31 $\times$ 0.26 &
  3.0 &
  44.0 &
  0.04 &
  1350 $\times$ 1350 &
  54 $\times$ 54 &
  0.05 &
  5\% \\ 
&
  202.86 - 206.59 &
  &
  &
  &
  &
  &
  &
  \\
6 &
  222.94 - 226.67 &
  13 - 3620 &
  0.21 $\times$ 0.19 &
  2.7 &
  37.5 &
  0.03 &
  1500 $\times$ 1500 &
  45 $\times$ 45 &
  0.07 &
  10\% \\ 
&
  237.44 - 241.17 &
  &
  &
  &
  &
  &
  &
  \\
7 &
  279.22 - 282.95 &
  15 - 3628 &
  0.19 $\times$ 0.15 &
  2.2 &
  30.5 &
  0.02 &
  1600 $\times$ 1600 &
  32 $\times$ 32 &
  0.09 &
  10\% \\ 
&
  291.22 - 294.95 &
  &
  &
  &
  &
  &
  &
  \\
8 &
  403.91 - 407.64 &
  14 - 2515 &
  0.17 $\times$ 0.16 &
  1.5 &
  21 &
  0.02 &
  1600 $\times$ 1600 &
  32 $\times$ 32 &
  0.18 &
  20\% \\ 
&
  415.92 - 419.64 &
  &
  &
  &
  &
  &
  &
  \\
\hline
\end{tabular}
$^a$ The calibration uncertainties come from the ALMA Technical Handbook \citep[][; \url{https://almascience.eso.org/documents-and-tools/cycle11/alma-technical-handbook}]{cortes2024}.
\end{minipage}
\end{table*}

The data were processed using the {\sc common astronomy software applications} ({\sc casa}) package version 6.2.1 \citep{mcmullin2007, casateam2022}.  First, the pipeline calibration was restored using the scripts downloaded from the ALMA Science Archive.  Next, preliminary image cubes were created using {\sc tclean} to identify spectral lines that were detectable at the 5$\sigma$ level in at least one individual frequency slice of the image cubes.  All channels in the visibility data not covering detectable line emission were used to create final continuum images.

The continuum images were created using {\sc tclean} using multiple settings that were optimized for imaging and detecting faint continuum emission; although some of the targeted gravitational lens candidates are detected at a high signal-to-noise level, other sources are detected in the individual bands at just above the $5\sigma$ level.  Natural weighting and the Hogbom deconvolver \citep{hogbom1974} were used to optimize for signal detection.  Additionally, a {\it uv} taper of 0.05~arcsec was applied to remove high spatial frequency noise from the data.  Images with primary beam corrections, which adjust for lowered sensitivity towards the edges of the fields of view, were used for photometry, while images without primary beam corrections were used for display purposes.  The pixel scales of the images were set so that the FWHM of the beam is sampled by at least 5 pixels.  The fields of view of the images (including the primary beam diameters, which related to the size of the regions where emission could effectively be detected), the beam sizes, the pixel scales, the maximum recoverable scales, the achieved rms noise levels, and the calibration uncertainties are listed in Table~\ref{t_obsspw}.  Final Band 5 images of all of the detected sources are shown in Figure~\ref{f_map} in Appendix~\ref{a_map}.

\subsection{Photometry}

Because the objects were frequently resolved and often had complex structures including multiple lobes or Einstein rings, we simplified the photometry by measuring flux densities within circular apertures that included all of the associated emission from each source.  An exception was made for HerBS-21A, which is a gravitationally lensed object that was resolved into a point-like image and an extended arc separated by $\sim$2.8~arcsec.  In this specific situation, we used a circular aperture for the point-like source, but for the arc, we used an ellipse with an axis ratio of $\sim$1.8/1 and a position angle of 65$^\circ$ from north to east, which closely matched the dimensions of the structure.  We then added the flux densities from the two apertures together for our final photometry measurement.

For objects with a central peaked source, the centres of the measurement apertures correspond to the peak of the Band 6 emission (or the Band 5 emission if the source is either near or outside the edge of the imaged area in the Band 6 data).  We chose Bands 5 and 6 for position measurements because they lie closest to the midpoint of our frequency measurements, and we gave preference to measurements in Band 6 because it usually has a higher signal-to-noise ratio.  The diameters of the measurement apertures for these sources were adjusted using a curve of growth analysis to ensure that we were measuring all of the emission from the targets while minimizing the amount of excess background noise included within the apertures.  For other sources, including Einstein rings and multi-lobed objects, we first identified the best place to centre a measurement aperture by shifting a small aperture (equivalent in size to source emission $>$5$\sigma$ above the background) within each Band 6 image around the source until we found the location where the signal was maximized.  We then applied the same curve of growth analysis as used for the sources with the centrally peaked emission to determine the best aperture diameter for measuring the flux densities.  Note that this method for determining optimal measurement apertures is applied to each band independently, but the resulting apertures are usually very close to the same diameter in all bands.

For all sources in all images, background noise levels were determined by calculating the standard deviations from measurements of the total flux density within randomly placed apertures in the versions of the images without the primary beam correction.  These apertures had the same diameters as what we used to measure the source emission.  We then scaled these standard deviations by the values of the primary beam at the location of each source. This method should allow us to include correlated noise in our uncertainties.

We made measurements of all sources previously identified by \citet{bendo2023} with the exception of five sources.  HerBS-41B, HerBS-42C, HerBS-106B, and both sources in the HerBS-159 field.  HerBS-41B is not detected in the new Band 3 and 4 data, it lies too close to the edge of the field of view in the Band 5 data for a reliable photometry measurement, and it is outside the field of view in the Band 6-8 images.   HerBS-42C is a unique case where the peak emission is detected at above 5 times the rms noise levels in Bands 5-8 but where it seems to have extended emission on the basis that measurements in $\sim$0.8-1.2~arcsec diameter apertures yield a power law consistent with the \citet{bendo2023} flux densities, but this extended emission is too low to measure at the $\geq5\sigma$ level, which is why we excluded it from the analysis.  HerBS-106B and both sources in the HerBS-159 fields are simply never detected at above 5 times the rms noise levels of the images.  We also checked all fields for additional sources that were not identified by \citet{bendo2023} but that were detected at above the $5\sigma$ level in two or more bands (thus excluding any random noise spikes in any image), but we saw no such sources.

Table~\ref{t_coord} lists the coordinates of the sources, which were used as the central apertures for the photometry measurements, and the redshifts from either \citet{urquhart2022} or \citet{bakx2024}.  Table~\ref{t_fd} lists the photometry measurements from this paper as well as the additional Band 3 (101~GHz) and Band 4 (151~GHz) photometry measurements from \citet{bendo2023}.  The uncertainties in the flux densities are based solely on the background noise measurements; the calibration uncertainties are listed in Table~\ref{t_obsspw}.  All upper limits are equivalent to $5\sigma$.  The SEDs for the sources as sorted by field are shown in Figure~\ref{f_sed} in Appendix~\ref{a_sed}.

\begin{table}
\caption{Coordinates and redshifts for the ANGELS sample}
\label{t_coord}
\begin{center}
\begin{tabular}{@{}lccc@{}}
\hline
Object &
  \multicolumn{2}{c}{Coordinates (ICRS)} &
  Redshift$^a$ \\
&
  R.A. &
  Dec. &
  \\
\hline
HerBS-21A &
  23:44:18.07$^b$ &
  -30:39:37.6$^b$ &
  3.323$^c$ \\
HerBS-21B &
  23:44:18.262 &
  -30:39:34.83 &
  3.323$^d$ \\
HerBS-22A &
  00:26:25.002 &
  -34:17:38.10 &
  3.050 \\
HerBS-22B &
  00:26:25.555 &
  -34:17:23.29 &
  \\
HerBS-25 &
  23:58:27.507 &
  -32:32:44.96 &
  2.912 \\
HerBS-36 &
  23:56:23.087 &
  -35:41:19.66 &
  3.095 \\
HerBS-41A &
  00:01:24.796 &
  -35:42:11.07 &
  4.098 \\
HerBS-42A &
  00:00:7.459 &
  -33:41:3.05 &
  3.307 \\
HerBS-42B &
  00:00:7.435 &
  -33:40:55.88 &
  3.314 \\
HerBS-81A &
  00:20:54.201 &
  -31:27:57.35 &
  3.160 \\
HerBS-81B &
  00:20:54.737 &
  -31:27:50.96 &
  2.588 \\
HerBS-86 &
  23:53:24.568 &
  -33:11:11.78 &
  2.564 \\
HerBS-87 &
  00:25:33.676 &
  -33:38:26.17 &
  2.059 \\
HerBS-93 &
  23:47:50.437 &
  -35:29:30.13 &
  2.400 \\
HerBS-104A &
  00:18:39.438 &
  -35:41:48.19 &
  \\
HerBS-104B &
  00:18:38.847 &
  -35:41:33.08 &
  1.536 \\
HerBS-106A &
  00:18:2.463 &
  -31:35:5.16 &
  2.369 \\
HerBS-155A &
  00:03:30.643 &
  -32:11:35.00 &
  3.077 \\
HerBS-155B &
  00:03:30.073 &
  -32:11:39.36 &
  \\
HerBS-170 &
  00:04:55.447 &
  -33:08:12.90 &
  4.182 \\
HerBS-184 &
  23:49:55.667 &
  -33:08:34.37 &
  2.507 \\
\end{tabular}
\end{center}
$^a$ The redshifts are from \citet{urquhart2022} or \citet{bakx2024}.\\
$^b$ Since this source consists of two images separated by $\sim$2.8~arcsec, two separate apertures were used to measure the flux densities of the separate images, and then the flux densities were added together.  The coordinates for this source point to the approximate midpoint between the two images.\\
$^c$ One redshift was reported by \citet{urquhart2022} for both sources in the HerBS-21 field because the emission from both sources fell within one beam in one of the image cubes used for the redshift identification.    However, spectral line emission (at very similar frequencies) has been measured from both sources in images with beams small enough to separate the emission from the sources, so we treat both sources as having the same redshift.\\
\end{table}

\begin{table*}
\centering
\begin{minipage}{167mm}
\caption{Photometry for the ANGELS sample}
\label{t_fd}
\begin{tabular}{@{}lcccccccc@{}}
\hline
Object &
  \multicolumn{6}{c}{Flux Density (this paper; mJy)$^a$} &
  \multicolumn{2}{c}{Flux Density (\citealt{bendo2023}; mJy)$^a$} \\
&
  106 GHz &
  140 GHz &
  199 GHz &
  232 GHz &
  287 GHz &
  412 GHz &
  101 GHz &
  151 GHz \\
\hline
HerBS-21A &
  $<2.1 $  &
  $ 2.1 \pm 0.3 $  &
  $ 8.6 \pm 0.5 $  &
  $ 15 \pm 2 $ &
  $ 26 \pm 2 $ & 
  $ 68 \pm 4 $ &
  $^b$ &
  $3.01 \pm 0.02$ \\
HerBS-21B &
  $< 0.40 $  &
  $< 0.63 $  &
  $ 1.72 \pm 0.15 $  &
  $ 4.4 \pm 0.8 $  &
  $ 6.6 \pm 0.9 $  &
  $ 17.1 \pm 1.8 $ &
  $^b$ &
  $0.93 \pm 0.02$ \\
HerBS-22A &
  $< 1.4 $  &
  $ 2.6 \pm 0.3 $  &
  $ 9.2 \pm 0.5 $  &
  $ 13.6 \pm 1.2 $  &
  $ 29 \pm 3 $  &
  $ 80 \pm 3 $  &
  $0.66 \pm 0.02$ &
  $3.10 \pm 0.02$ \\
HerBS-22B &
  $< 0.22 $  &
  $< 0.28 $  &
  $ 0.93 \pm 0.18 $  &
  $< 3.3 $  &
  &
  &
  $<0.20$ &
  $0.35 \pm 0.04$ \\
HerBS-25 &
  $< 2.0 $  &
  $ 2.4 \pm 0.3 $  &
  $ 10.3 \pm 0.5 $  &
  $ 17.2 \pm 1.8 $  &
  $ 36 \pm 2 $  &
  $ 89 \pm 3 $ &
  $0.91 \pm 0.07$ &
  $3.46 \pm 0.03$ \\
HerBS-36 &
  $ 1.38 \pm 0.13 $  &
  $ 3.52 \pm 0.19 $  &
  $ 11.8 \pm 0.3 $  &
  $ 19.0 \pm 0.9 $  &
  $ 34.9 \pm 1.3 $  &
  $ 76 \pm 2 $ &
  $1.16 \pm 0.02$ &
  $4.81 \pm 0.03$ \\
HerBS-41A &
  $ 1.0 \pm 0.2 $  &
  $ 3.0 \pm 0.3 $  &
  $ 8.8 \pm 0.3 $  &
  $ 13.8 \pm 0.7 $  &
  $ 23.8 \pm 1.4 $  &
  $ 50 \pm 3 $  &
  $0.71 \pm 0.02$ &
  $3.79 \pm 0.03$ \\
HerBS-42A &
  $< 0.93 $  &
  $ 1.20 \pm 0.20 $  &
  $ 4.2 \pm 0.3 $  &
  $ 6.9 \pm 0.6 $  &
  $ 14.9 \pm 1.1 $  &
  $ 42 \pm 2 $  &
  $^b$ &
  $1.84 \pm 0.03$ \\
HerBS-42B &
  $< 0.39 $  &
  $ 0.38 \pm 0.07 $  &
  $ 1.6 \pm 0.2 $  &
  $ 2.6 \pm 0.3 $  &
  $ 4.8 \pm 0.7 $  &
  $ 10.1 \pm 1.3 $  &
  $^b$ &
  $0.54 \pm 0.02$ \\
HerBS-81A &
  $< 1.0 $  &
  $< 1.1 $  &
  $ 2.3 \pm 0.3 $  &
  $ 3.2 \pm 0.6 $  &
  $ 8.6 \pm 1.3 $  &
  $ 19 \pm 3 $ &
  $0.20 \pm 0.02$ &
  $0.76 \pm 0.03$ \\
HerBS-81B &
  $< 0.67 $  &
  $ 0.41 \pm 0.08 $  &
  $ 2.11 \pm 0.19 $  &
  $ 4.4 \pm 0.4 $  &
  $ 6.0 \pm 0.6 $  &
  $ 17.2 \pm 1.0 $ &
  $<0.20$ &
  $0.68 \pm 0.02$ \\
HerBS-86 &
  $< 0.85 $  &
  $ 0.89 \pm 0.16 $  &
  $ 3.9 \pm 0.4 $  &
  $ 6.1 \pm 0.8 $  &
  $ 13.0 \pm 1.1 $  &
  $ 35 \pm 2 $ &
  $0.26 \pm 0.02$ &
  $1.53 \pm 0.02$ \\
HerBS-87 &
  $< 0.82 $  &
  $< 0.88 $  &
  $ 2.9 \pm 0.3 $  &
  $ 5.2 \pm 0.6 $  &
  $ 11.1 \pm 1.2 $  &
  $ 33.1 \pm 1.5 $  &
  $<0.20$ &
  $1.24 \pm 0.02$ \\
HerBS-93 &
  $< 1.1 $  &
  $< 1.1 $  &
  $ 3.4 \pm 0.3 $  &
  $ 6.1 \pm 0.8 $  &
  $ 10.9 \pm 1.3 $  &
  $ 32 \pm 2 $  &
  $0.16 \pm 0.01$ &
  $1.37 \pm 0.02$ \\
HerBS-104A &
  $< 0.71 $  &
  $< 0.83 $  &
  $ 2.2 \pm 0.4 $  &
  $< 6.7 $  &
  &
  &
  $<0.20$ &
  $0.64 \pm 0.03$ \\
HerBS-104B &
  $< 0.83 $  &
  $< 0.84 $  &
  $ 1.4 \pm 0.2 $  &
  $ 2.5 \pm 0.5 $  &
  $ 6.3 \pm 1.0 $  &
  $ 23 \pm 2 $  &
  $<0.20$ &
  $0.52 \pm 0.02$ \\
HerBS-106A &
  $< 1.0 $  &
  $ 1.3 \pm 0.2 $  &
  $ 4.5 \pm 0.3 $  &
  $ 8.5 \pm 1.1 $  &
  $ 14.8 \pm 1.5 $  &
  $ 41 \pm 3 $  &
  $0.29 \pm 0.02$ &
  $1.45 \pm 0.03$ \\
HerBS-155A &
  $< 1.5 $  &
  $< 1.5 $  &
  $ 5.6 \pm 0.4 $  &
  $ 9.6 \pm 0.8 $  &
  $ 18.3 \pm 1.7 $  &
  $ 50 \pm 2 $  &
  $0.29 \pm 0.01$ &
  $2.16 \pm 0.04$ \\
HerBS-155B &
  $< 0.32 $  &
  $< 0.37 $  &
  $ 0.67 \pm 0.13 $  &
  $ 1.5 \pm 0.2 $  &
  $ 3.0 \pm 0.5 $  &
  $ 5.2 \pm 1.0 $  &
  $<0.20$ &
  $0.54 \pm 0.03$ \\
HerBS-170 &
  $< 1.6 $  &
  $ 2.4 \pm 0.4 $  &
  $ 7.4 \pm 0.4 $  &
  $ 12.6 \pm 1.4 $  &
  $ 24 \pm 3 $  &
  $ 39 \pm 4 $  &
  $0.81 \pm 0.02$ & 
  $3.50 \pm 0.03$ \\
HerBS-184 &
  $< 1.1 $  &
  $ 1.00 \pm 0.18 $  &
  $ 4.4 \pm 0.4 $  &
  $ 6.5 \pm 0.7 $  &
  $ 15.0 \pm 1.1 $  &
  $ 45 \pm 2 $  &
  $0.46 \pm 0.02$ &
  $1.47 \pm 0.02$ \\
\hline
\end{tabular}
$^a$ Non-detections are reported as $5\sigma$ upper limits.\\
$^b$ In the \citet{bendo2023} data, the emission from these sources at 101~GHz was blended with emission from other sources, so we did not use the photometry data in our analysis.
\end{minipage}
\end{table*}

\section{Emissivity indices derived from ratios of flux densities}
\label{s_beta_ratio}

\subsection{Methodology of deriving $\beta$ from flux density ratios}
\label{s_beta_ratio_method}

\begin{figure}
\begin{center}
\includegraphics[width=8cm]{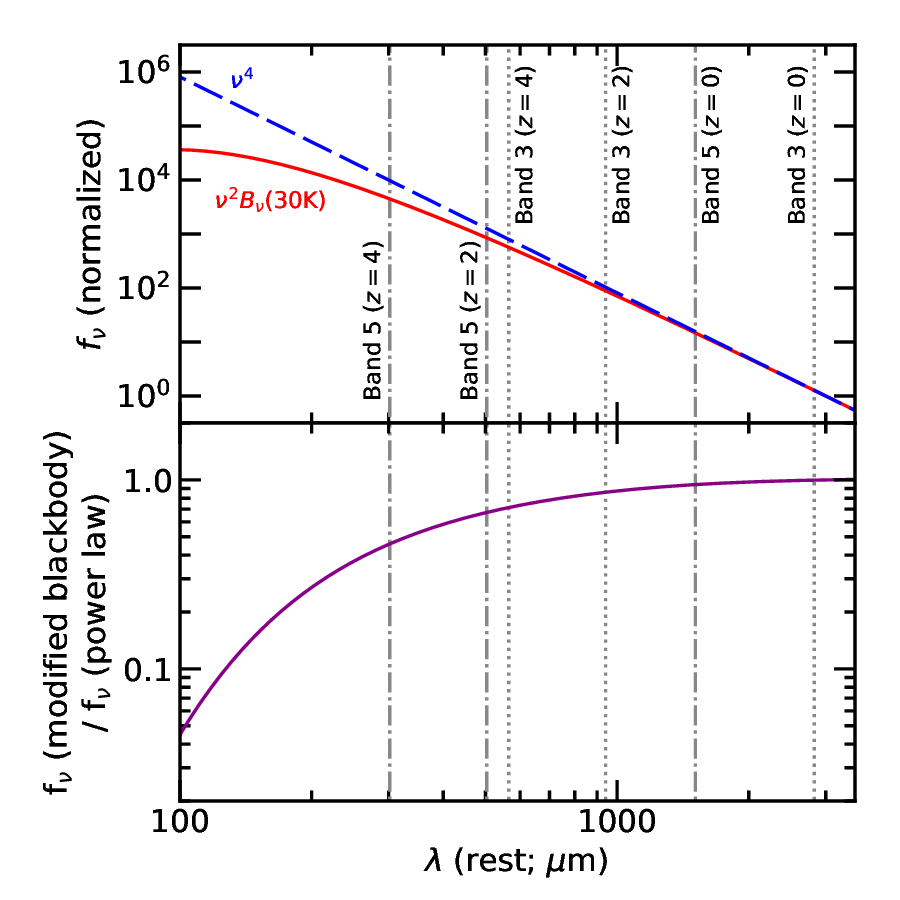}
\end{center}
\caption{Example figure illustrating the deviation of modified blackbodies from pure power laws in the frequency range of our observations.  The top panel shows an example plot of a modified blackbody with a temperature of 30~K and an emissivity index $\beta$ (solid red line) of 2 and a power law that varies as $\nu^{(2+\beta)}$ (dashed blue line).  Both functions have been normalized to 1 at 100 GHz (3 mm).  The vertical dotted grey lines show the frequencies sampled by our Band 3 (106~GHz) data at $z$ of 0, 2, and 4, while the vertical dash-dot grey lines show the frequencies sampled by our Band 5 (199~GHz) data at those same $z$.  The bottom panel shows the ratio of the modified blackbody to the power law.}
\label{f_bbdemo}
\end{figure}

Because the ALMA data sample the dust emission on the Rayleigh-Jeans side of the dust SED, it may seem like simply fitting a power law to the data would be sufficient for deriving $\beta$.  However, for dust emission with temperatures in the range 15-40~K and $\beta$ in the range 1-3 (which, based on the results from \citealt{bendo2023}, may be broader than but similar to the emission that we expect from these objects), the modified blackbody curves may deviate significantly from pure power laws in the frequency ranges that we are observing.  This is shown in Figure~\ref{f_bbdemo}, which plots a power law that scales as $\nu^{\beta+2}$ and a modified blackbody with a temperature of 30~K and $\beta$ of 2.  Both functions have been normalized to 1 at 100 GHz (3 mm).  At $z=0$, we could expect the slope between the ALMA Band 3 and ALMA Band 5 data for a 15-40~K modified blackbody to deviate $\lesssim$10\% from the expected slope from a $\nu^{\beta+2}$ power law.  At $z=2$, however, this deviation increases to $\sim$15-40\%, and at $z=4$, this deviation is $\sim$35-75\%. See \citet{dacunha2021} for an alternate version of this analysis.

Given this, it is necessary to use the complete modified blackbody equation to compute the expected slopes of the ALMA data for deriving $\beta$.  The higher frequency data from ALMA Bands 6-8 cover observed frequencies of 232-412 GHz (728-1293~$\mu$m), which at $z=3$ would correspond to rest frequencies of 928-1648~GHz (182-323~$\mu$m).  Hence, these data are relatively close to the peak of the dust SED, so the $\beta$ derived from the slope of these data would be sensitive to the effects of temperature variations.  Additionally, the calibration uncertainties in these bands are relatively high, which would affect the uncertainties in $\beta$ values derived from the data.  However, the slopes of the lower frequency ALMA data (Bands 3-5), which correspond to observed frequencies of 106-199~GHz (1508-2830~$\mu$m) are much less dependent on temperature variations, even at $z=3$ where the rest frequencies correspond to 424-796~GHz (377-708~$\mu$m). The lower frequency data also have much lower calibration uncertainties, although the signal-to-noise ratios of some measurements are still relatively low.   

\begin{figure*}
\begin{center}
\includegraphics[width=8cm]{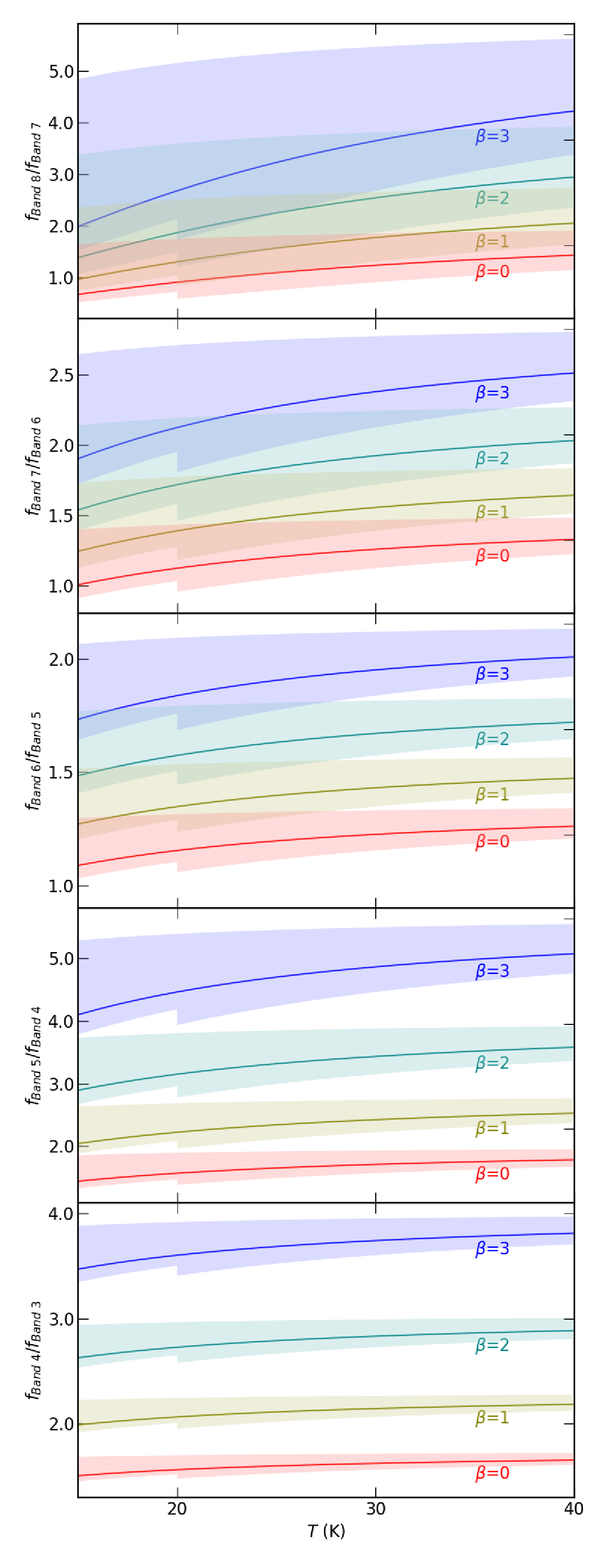}
~
\includegraphics[width=8cm]{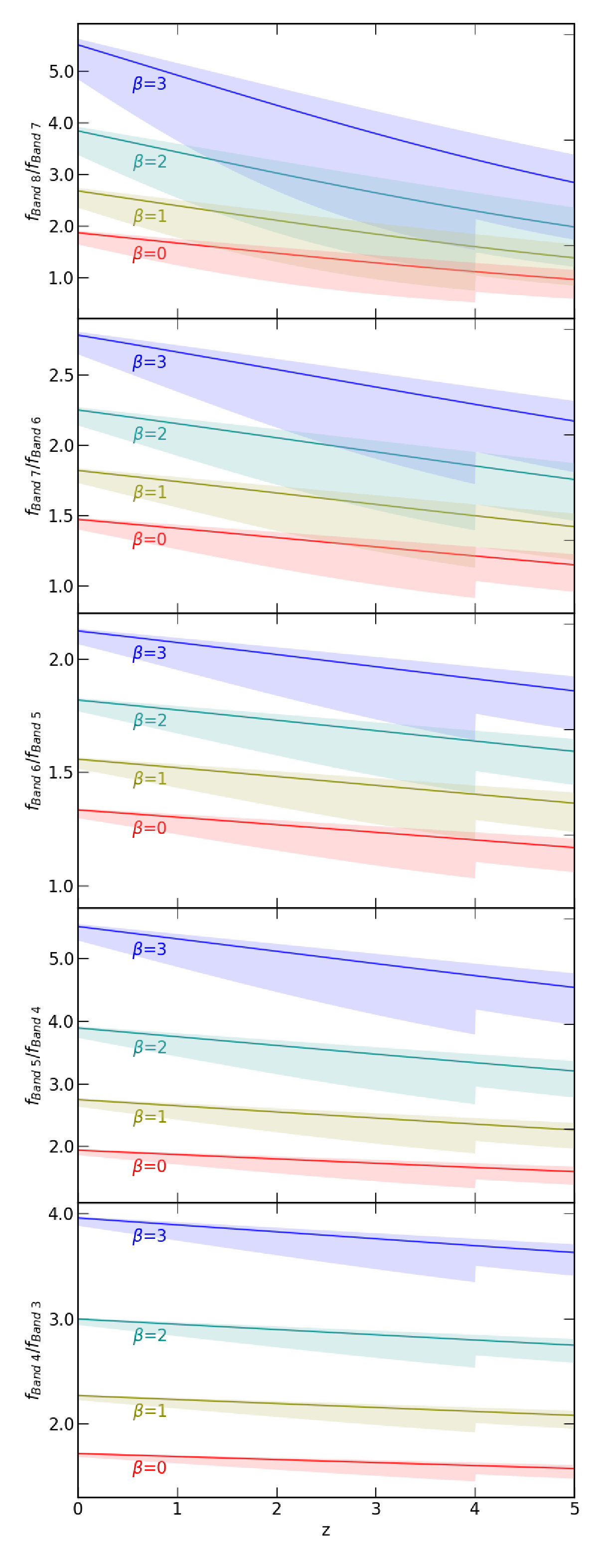}
\end{center}
\caption{The ALMA band ratios for $\beta$ values of 0, 1, 2, and 3.  The plot on the left shows how the ratios vary with temperature.  The shaded areas show how the ratio changes between $z=0$ and $z=5$, and the solid lines show the ratios for $z=3$, which is close to the mean $z$ for the sample in this paper.  The plot on the right shows how the ratios vary with redshift.  The shaded areas show how the ratios vary between temperatures of 15 and 40~K (but this is truncated to a range of 20 to 40~K for $z>4$, leading to discontinuities in the shaded regions).  The solid line shows the ratios corresponding to temperatures of 32~K, which is the midrange of the values derived for the BEARS sample by \citet{bendo2023}.}
\label{f_ratiodemo}
\end{figure*}

The variations in several sets of ALMA flux density ratios (which correspond to the slopes of the data) are shown in Figure~\ref{f_ratiodemo} for various blackbodies with varing ranges of temperatures, $\beta$, and redshift.  The temperature range was set to 15-40~K, which we noted above is similar to the range of temperatures found by \citet{bendo2023}, but this range was truncated to 20-40~K for $z\geq4$ to avoid issues with dust temperatures falling below the CMB temperature.  The redshift vary from 0 to 5, which encompasses the range of values for the objects in our sample with known redshifts.  In the bottom panels, which correspond to the Band 4/Band 3 or Band 5/Band 4 flux density ratios, the ranges of the ratios corresponding to $\beta$ of 0, 1, 2, and 3 do not overlap in spite of the broad variations in temperature and redshift.  In the upper panels, which correspond to flux density ratios involving the higher frequency bands, the ranges begin to overlap.  Even when the redshift of a source is known (as shown by the lines corresponding to $z=3$ in the left set of panels), a given ratio based on high frequency bands may correspond to either a modified blackbody with a large $\beta$ and low temperature or a low $\beta$ and high temperature, whereas the ratios based on the lower frequency data hardy vary with temperature and are more strongly dependent on $\beta$.

We therefore derived $\beta$ using the flux density ratios between two pairs of ALMA bands: the Band 4/Band 3 ratio and the Band 5/Band 4 ratio.  We have one set of Band 3 and 4 measurements from the data in the paper and another set from \citet{bendo2023}.  We could have potentially chosen one set of Band 3 data and one set of Band 4 data for calculating the flux density ratios, but this would mean that an outlying measurement in one of the datasets would give an inaccurate result when calculating $\beta$.  Instead, we decided to calculate separate $\beta$ values for all possible combinations involving any Band 3 and Band 4 data where the flux densities were measured at above the $5\sigma$ level, thus giving us potentially four separate $\beta$ values based on the Band 4/Band 3 ratios and two separate $\beta$ values based on the Band 5/Band 4 ratios.  We could then use the weighted mean of all of the $\beta$ derived from the different Band 4/Band 3 ratios to get one value that we label as $\beta_{\mathrm{4/3~ratio}}$ and use the weighted mean of both of the $\beta$ from the two Band 5/Band 4 ratios to get one value that we label as $\beta_{\mathrm{5/4~ratio}}$.  We could then use the weighted mean of $\beta_{\mathrm{4/3~ratio}}$ and $\beta_{\mathrm{5/4~ratio}}$ to calculate a weighted mean of the two values that we have labelled $\beta_{\mathrm{mean~ratio}}$.  This general approach is effectively similar to fitting a power law to the data and converting that to a $\beta$ value.  However, reporting the $\beta$ from the separate ratios as well as their weighted mean allows us to check the consistency of the measurements and look for bends in the SED that may indicative of emission from sources other than dust.

To calculate $\beta$ from these ratios, we created a series of modified blackbodies from 15-40~K (but with the range truncated to 20-40~K for modified blackbodies calculated at $z\geq4$) and with $\beta$ ranging from 0 to 5.  For objects with known redshifts, we generated these modified blackbodies at the redshift of that object.  For objects with unknown redshifts, we generated these modified blackbodies for $z$ ranging from 0 to 5.  All modified blackbodies incorporated adjustments for the effects of the CMB as described by \citet{dacunha2013}.  Next, we calculated the Band 4/Band 3 and Band 5/Band 4 flux density ratios for each modified blackbody and determined the mean, minimum, and maximum values of each ratio for all blackbodies with a given $\beta$.  This gave us $\beta$ as functions of the Band 4/Band 3 and Band 5/Band 4 flux density ratios that we could use to derive the $\beta_{\mathrm{4/3~ratio}}$ and $\beta_{\mathrm{5/4~ratio}}$ values.

The uncertainties in the $\beta_{\mathrm{4/3~ratio}}$ and $\beta_{\mathrm{5/4~ratio}}$ values have two sources.  One source is from random noise related to both the calibration and measurement uncertainties.  This was calculated using a Monte Carlo approach where, in each iteration, random noise is added to all Band 3, 4, and 5 data, the ratios and weighted mean $\beta$ values are derived for each iteration, and then the standard deviation in the weighted mean $\beta$ values among the iterations is used as the uncertainties.  (This approach accounts for correlated noise among the sets of ratios.)  Typically, this gave uncertainties of 0.2-0.3 for $\beta_{\mathrm{4/3~ratio}}$, 0.3-0.5 for $\beta_{\mathrm{5/4~ratio}}$ when the redshift is known, and 0.8-0.9 for $\beta_{\mathrm{5/4~ratio}}$ when the redshift is unknown.  The other source of uncertainties is based on range in $\beta_{\mathrm{4/3~ratio}}$ and $\beta_{\mathrm{5/4~ratio}}$ values that may correspond to any given Band 4/Band 3 and Band 5/Band 4 ratios.  This is related to the underlying assumptions used regarding the potential range of temperatures that could apply to the dust seen in these bands.  These uncertainties are typically $\sim0.1$ for $\beta_{\mathrm{4/3~ratio}}$ and 0.1-0.3 for $\beta_{\mathrm{5/4~ratio}}$.

\subsection{Results from deriving $\beta$ from flux density ratios}

\begin{table}
\caption{$\beta$ values derived from the Band 4/Band 3 and Band 5/Band 4 ratios}
\label{t_rat}
\begin{center}
\begin{tabular}{@{}lccc@{}}
\hline
Object &
  $\beta_{\mathrm{4/3~ratio}}$ &
  $\beta_{\mathrm{5/4~ratio}}$ &
  $\beta_{\mathrm{mean~ratio}}^a$ \\
\hline
HerBS-21A &
  &
  2.5 $\pm$ 0.4 &
  2.5 $\pm$ 0.4 \\
HerBS-21B &
  &
  0.9 $\pm$ 0.5 &
  0.9 $\pm$ 0.5 \\
HerBS-22A &
  2.2 $\pm$ 0.2 &
  2.4 $\pm$ 0.4 &
  2.3 $\pm$ 0.1 \\
HerBS-22B &
  &
  2.0 $\pm$ 0.9 &
  2.0 $\pm$ 0.9 \\
HerBS-25 &
  1.5 $\pm$ 0.3 &
  2.6 $\pm$ 0.4 &
  2.0 $\pm$ 0.2 \\
HerBS-36 &
  1.8 $\pm$ 0.2 &
  2.0 $\pm$ 0.3 &
  1.9 $\pm$ 0.1 \\
HerBS-41A &
  2.5 $\pm$ 0.2 &
  1.7 $\pm$ 0.3 &
  2.3 $\pm$ 0.1 \\
HerBS-42A &
  &
  1.8 $\pm$ 0.4 &
  1.8 $\pm$ 0.4 \\
HerBS-42B &
  &
  2.6 $\pm$ 0.6 &
  2.6 $\pm$ 0.6 \\
HerBS-81A &
  1.6 $\pm$ 0.3 &
  2.6 $\pm$ 0.6 &
  1.9 $\pm$ 0.3 \\
HerBS-81B &
  &
  2.8 $\pm$ 0.5 &
  2.8 $\pm$ 0.5 \\
HerBS-86 &
  2.6 $\pm$ 0.3 &
  2.2 $\pm$ 0.5 &
  2.5 $\pm$ 0.2 \\
HerBS-87 &
  &
  1.5 $\pm$ 0.5 &
  1.5 $\pm$ 0.5 \\
HerBS-93 &
  3.6 $\pm$ 0.3 &
  1.8 $\pm$ 0.5 &
  3.2 $\pm$ 0.2 \\
HerBS-104A &
  &
  3.0 $\pm$ 0.8 &
  3.0 $\pm$ 0.8 \\
HerBS-104B &
  &
  2.0 $\pm$ 0.6 &
  2.0 $\pm$ 0.6 \\
HerBS-106A &
  2.4 $\pm$ 0.3 &
  2.4 $\pm$ 0.4 &
  2.4 $\pm$ 0.2 \\
HerBS-155A &
  3.3 $\pm$ 0.2 &
  2.1 $\pm$ 0.4 &
  3.0 $\pm$ 0.1 \\
HerBS-155B &
  &
  $^b$ &
  $^b$ \\
HerBS-170 &
  2.0 $\pm$ 0.2 &
  1.5 $\pm$ 0.3 &
  1.9 $\pm$ 0.1 \\
HerBS-184 &
  1.1 $\pm$ 0.2 &
  2.6 $\pm$ 0.5 &
  1.4 $\pm$ 0.2 \\
\hline
\end{tabular}
\end{center}
$^a$ $\beta_{\mathrm{mean~ratio}}$ is the weighted mean of $\beta_{\mathrm{4/3~ratio}}$ and $\beta_{\mathrm{5/4~ratio}}$.  When the ALMA Band 3 emission was not detected at the $\geq5\sigma$ level, no $\beta_{\mathrm{4/3~ratio}}$ value is reported, and $\beta_{\mathrm{mean~ratio}}$ is set to $\beta_{\mathrm{5/4~ratio}}$.\\
$^b$ The slope of these data was consistent with $\beta<0$ for the range of temperatures and redshifts used in this analysis.
\end{table}

\begin{figure}
\begin{center}
\includegraphics[width=8cm]{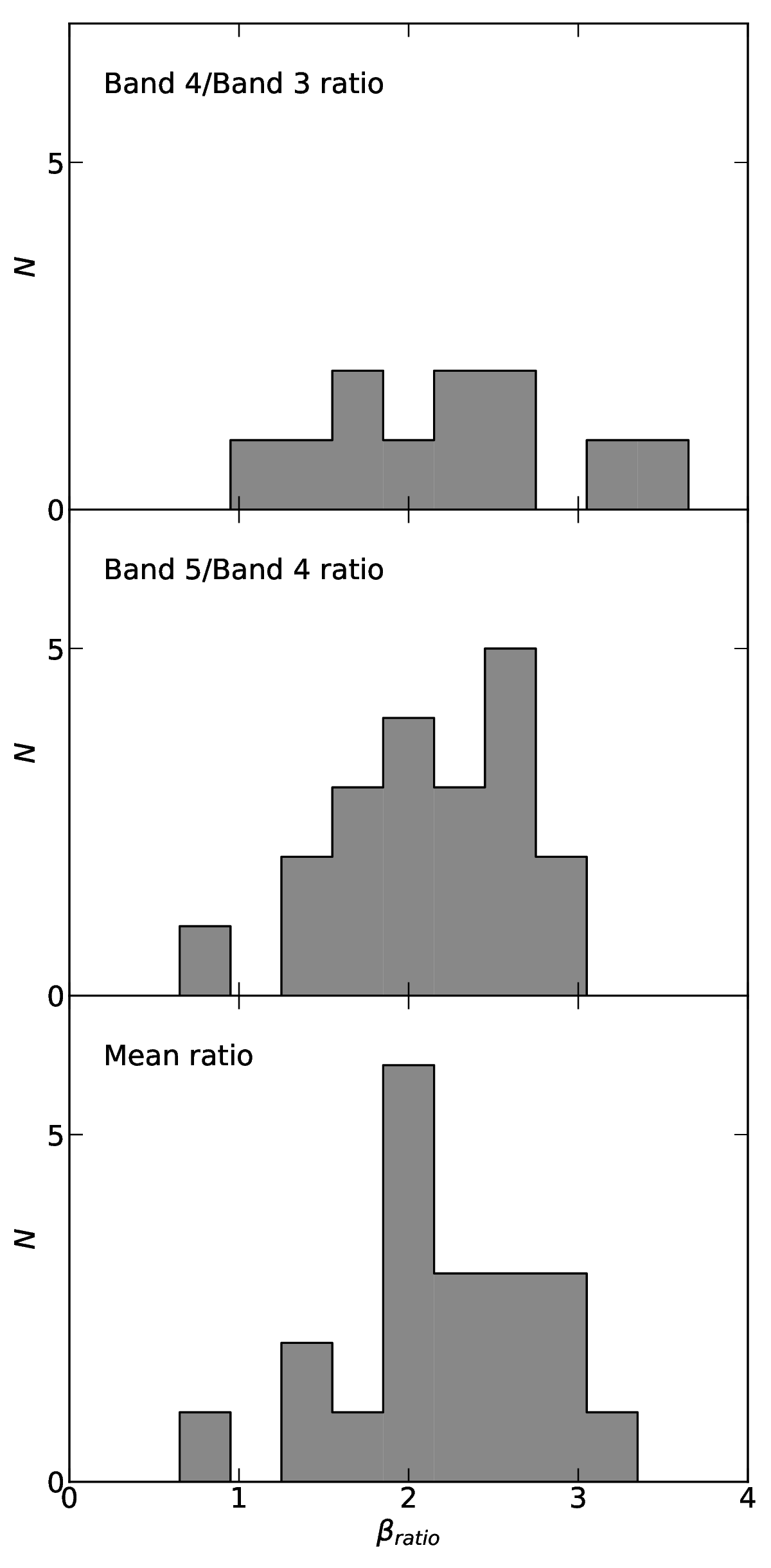}
\end{center}
\caption{Histogram of the $\beta$ values derived from the ALMA Band 4/Band 3 ratios (top panel) and from the ALMA Band 5/Band 4 ratios (middle panel) as well as the weighted mean of those values (bottom panel).  $\beta_{\mathrm{Band~4/Band~3}}$ values were not derived for objects that were not detected at the $\geq5\sigma$ level, in which case $\beta_{\mathrm{mean~ratio}}$ is set to $\beta_{\mathrm{Band~5/Band~4}}$.  The values for HerBS-104A fall outside the maximum range of $\beta$ values plotted here.  All of these $\beta$ values are derived in Section~\ref{s_beta_ratio} and are listed in Table~\ref{t_rat}.}
\label{f_betarathist}
\end{figure}

The $\beta_{\mathrm{4/3~ratio}}$, $\beta_{\mathrm{5/4~ratio}}$, and $\beta_{\mathrm{mean~ratio}}$ values along with the final uncertainties are listed in Table~\ref{t_rat}.  Histograms of these three different $\beta$ values are shown in Figure~\ref{f_betarathist}.  The derived values range from 0.9 to 3.0.  The mean $\beta_{\mathrm{4/3~ratio}}$, $\beta_{\mathrm{5/4~ratio}}$, and $\beta_{\mathrm{mean~ratio}}$ values are all 2.2.  The standard deviation in the $\beta_{\mathrm{4/3~ratio}}$ values is 0.7, which is slightly higher than the standard deviation of 0.5 for the $\beta_{\mathrm{5/4~ratio}}$ values.  The $\beta_{\mathrm{mean~ratio}}$ values have a standard deviation of 0.6.

The very high $\beta_{\mathrm{4/3~ratio}}$ and $\beta_{\mathrm{5/4~ratio}}$ values may be unreliable.  HerBS-93 and HerBS-155A both have $\beta_{\mathrm{Band~4/Band~3}}$ values greater than 3 but $\beta_{\mathrm{Band~5/Band~4}}$ values of $\sim$2.  This may indicate that the SED becomes steeper at lower frequencies or that the Band 3 measurements are problematic (and \citet{bendo2023} also indicated that the steep slope between Bands 3 and 4 for HerBS-93 was suspect), although we have not identified any specific problems with the Band 3 photometry for these objects.  As for HerBS-104A, the high $\beta_{\mathrm{5/4~ratio}}$ is based on two data points, one of which has a relatively low signal-to-nose, and the uncertainties in the derived $\beta_{\mathrm{5/4~ratio}}$ value reflects that.

The low $\beta_{\mathrm{5/4~ratio}}$ of $0.9\pm0.5$ for HerBS-21B appears to be more trustworthy.  Values of $\beta$ less than 1 are physically implausible for optically thin dust emission, as the values violate the Kramers-Kronig Relation \citep{li2004, galliano2022}, but such low values could be consistent with emission from sources other than dust.  However, the uncertainty in the $\beta_{\mathrm{5/4~ratio}}$ value is sufficiently large that it is possible that the emission is consistent with $\beta>1$ and that the emission originates from optically thin dust that has a rather unusually low $\beta$ compared to at least the other galaxies in this sample.  Also note that the Band 5/Band 4 ratio for HerBS-155B was not consistent with $\beta\geq0$, implying that the Band 4 emission may originate from sources other than dust, but the relatively high uncertainty in the Band 5 data means that this result may be unreliable.

When we were able to calculate $\beta_{\mathrm{4/3~ratio}}$ values, some of the differences between these values and the $\beta_{\mathrm{5/4~ratio}}$ values seemed attributable to noise.  For four sources (HerBS-22A, HerBS-36, HerBS-86, and HerBS-106A), the $\beta_{\mathrm{4/3~ratio}}$ and $\beta_{\mathrm{5/4~ratio}}$ values differed by less than $1\sigma$, while for two more sources (HerBS-81A and HerBS-170), the differences were less than $2\sigma$.   HerBS-93 is the only case where the difference between $\beta_{\mathrm{4/3~ratio}}$ and $\beta_{\mathrm{5/4~ratio}}$ is $>3\sigma$, although it and HerBS-155A (where the $\beta$ values differ by $\sim2.7\sigma$) were already discussed above.  As for the other three sources (HerBS-25, HerBS-41A, and HerBS-184), the difference  between $\beta_{\mathrm{4/3~ratio}}$ and $\beta_{\mathrm{5/4~ratio}}$ is not significant enough to draw any firm conclusions about either the quality of the data and $\beta$ values or any possible physical processes that could cause the values to differ.

If $\beta_{\mathrm{4/3~ratio}}$ was significantly lower than $\beta_{\mathrm{5/4~ratio}}$, this could be consistent with the presence of emission from sources other than dust.  This could be the case for HerBS-25, HerBS-81A, and HerBS-184.  Synchrotron emission would be the most likely emission source in the lower frequency bands.  This is discussed more in Section~\ref{s_discuss}.  However, it is more difficult to explain the physical phenomena that could cause $\beta_{\mathrm{4/3~ratio}}$ to be significantly higher than $\beta_{\mathrm{5/4~ratio}}$.  Aside from potential issues with the measurements, this would not be explainable by other sources of emission; it could only be explained by changes in the dust emissivity at lower frequencies.

\section{Discussion}
\label{s_discuss}

To summarize, the emissivity index that we derived from the weighed mean of the ALMA Band 5/Band 4 and Band 4/Band 3 flux density ratios ($\beta_{\mathrm{mean~ratio}}$) has a mean of 2.2 and a standard deviation of 0.6.  This is higher than but still statistically consistent with the $\beta$ of $\sim$2.0 originally derived by \citet{bendo2023} from the ALMA Band 4/Band 3 ratios for a different but overlapping subset of the BEARS sample.  The additional data and improved frequency coverage provided by ANGELS and our improved techniques for deriving $\beta$ should make our new results more reliable.

Our $\beta_{\mathrm{mean~ratio}}$ values are slightly higher than but consistent with the value of 2 typically used in dust models.  Additionally, the mean of our $\beta_{\mathrm{mean~ratio}}$ values is at the high end of but consistent with the mean $\beta$ values measured in the Milky Way, in nearby galaxies, and in galaxies at redshifts similar to those for our sample (as listed by the references in Section~\ref{s_intro}).  However, most of the $\beta$ values from these other studies should be treated very cautiously, mainly because most of these $\beta$ values are derived using fits to both the peak and Rayleigh-Jeans side of the SED where both the temperature and $\beta$ are treated as free parameters.  

First of all, as also stated in Section~\ref{s_intro}, such SED fits are prone to inherent degeneracies between these two quantities, and noise in the data can create apparent relations between the quantities.  Second, integrating the emission from dust with different temperatures along the line of sight, including integrating the emission across and through the disc of a galaxy, could lead to the peaks of dust SEDs appearing overly broad, which could give the appearance that the bulk of the dust within galaxies is hotter than it actually is, that $\beta$ is lower than it actually is, and also that $\beta$ could vary among galaxies.  Moreover, if a set of galaxies with widely varying redshifts are observed at the same (observed-frame) wavelengths, and if data covering the peak of the dust emission are fit with modified blackbodies where both the temperatures and $\beta$ are treated as free parameters, more emission from warmer dust will be blended with emission from colder dust for the higher redshift galaxies in the dataset.  This will not only make $\beta$ appear lower than it actually is but also create a false correlation between $\beta$ and redshift, as explained by \citet{bendo2023}.  Some studies have used tests with simple model SEDs to try to illustrate that their SED fits should be robust against the $\beta$-temperature degeneracies caused by noise, but it is unclear if such testing can also account for the line-of-sight integration issues, especially if the models, which often assume that the dust can be described by just one thermal component or one thermal component added to a power law, are unrealistic descriptions of the range of dust temperatures in the targets.  Our methods for measuring $\beta$ should be much more robust, mainly because our methods depend entirely on measurements from the Rayleigh-Jeans side of the dust SED.

The slope of the dust SED in our ALMA Band 3-5 data, which would correspond to rest wavelengths of $\sim$500-1000~$\mu$m for a source at $z=2$ and $\sim$270-550~$\mu$m for a source at $z=4.5$, could be altered by the inclusion of very strong free-free emission, very strong synchrotron emission, very large masses of dust at temperatures of $<$10~K, or other forms of exotic and poorly-quantified sources of submillimetre emission such as spinning dust \citep[e.g.,][]{draine1998}.  If such emission is present in our target sources, this would imply that our $\beta$ values are only lower limits and that the dust emissivity function is actually much steeper than expected.  However, all of the sources in our sample were covered by both the Very Large Array Sky Survey \citep{gordon2021}, which has a completeness limit of 3 mJy, or the Rapid ASKAP Continuum Survey \citep{duchesne2024}, which is 95\% complete for sources brighter than 2mJy, and no radio sources were detected in either survey within $\sim7$~arcsec of any source in our sample.

While most standard models use theoretical dust grain models where $\beta$ is fixed to $\sim$2 \citep[e.g.][]{draine2003}, laboratory experiments on astrophysical dust analogues have revealed that the dust emissivity functions are often more complicated than simple power laws \citep[e.g.][]{boudet2005, coupeland2011, demyk2017a, demyk2017b}.  This could arise in part because the grains are not spherical or otherwise smooth but actually amorphous, although the chemical composition of the grains also plays a role.  Some of these laboratory measurements have also shown that temperature may affect dust emissivity, with the spectral slope sometimes being anticorrelated with temperature.

If the values of $\beta$ are indeed greater than 2 in at least some objects at $z>1$, this has multiple ramifications for the analysis of the dust in these objects.  The greatest impact would be on calculating dust masses, which are highly dependent on the emissivity function.  Since most standard dust models set $\beta$ to $\sim$2 to calculate emissivity functions, these functions cannot be used to calculate dust masses if the actual $\beta$ differs from this value.  Multiple authors circumvent this issue by rescaling the emissivity function using a new power law, but that requires fixing the amplitude of the emissivity function at a specific wavelength  \citep[e.g.,][]{beelen2006, yang2007, desert2008, magnelli2012, smith2012, planck2015, lamperti2019, dacunha2021, tsukui2023, liao2024, ward2024}.  The emissivity at that selected wavelength is often taken from a model that uses $\beta\cong2$, such as \citet{draine2003}, which seems inconsistent, contradictory, and potentially unreliable \citep{bianchi2013}.  Additionally, the wavelength at which the emissivity function is fixed to the function from one of these models is often arbitrarily selected or poorly justified.  These choices ultimately bias the dust masses calculated.  Unfortunately, it is not straightforward to determine exactly how the dust masses would be affected if $\beta$ is variable or greater than 2, mainly because of the issues with trying to determine exactly how to rescale the emissivity function.    What is really needed to calculate dust masses accurately are measurements of the amplitude of the dust emissivity in the infrared and submillimetre that are model-independent, such as the measurements from \citet{james2002}, but even then, such emissivity measurements would need to be validated across multiple galaxies and multiple environments.

Another potentially complex implication of $\beta$ varying among galaxies is that it could affect the derivation of dust temperatures and the application dust emission models.  This compounds the issues with the dust mass calculations, which are temperature-dependent, but additionally, it also complicates the overall description of how energy is absorbed and re-radiated by dust, as dust grains with differing $\beta$ values will potentially reach different equilibrium temperatures.  If these variations in $\beta$ in the infrared are related to variations in absorption in the ultraviolet and optical, then this would also indicate variations in dust extinction in other galaxies and therefore affect the derivations of the general properties of those galaxies' stellar populations.

\subsection{Variations in $\beta$ with redshift}

Variations in dust emissivity as a function of redshift could be indicative of changes in dust properties over time.  However, even though $\beta$ potentially varies among galaxies, most observational results show no trend in $\beta$ versus redshift \citep{bendo2023, witstok2023, liao2024, tripodi2024, ward2024}.  While \citet{ismail2023} reported that $\beta$ may decrease with redshift, they indicated that their relation was statistically insignificant.  Unfortunately, a major limitation with many of these prior studies (except the work by \citealt{bendo2023}) is that they mostly depend on $\beta$ values derived from SED fits to both the peak and Rayleigh-Jeans side of the dust emission, and, as we discussed earlier, such SED fits are potentially subject to degeneracies between $\beta$ and temperature and with issues related to mixing emission from dust at different temperatures.  Consequently, the SED for the higher-redshift objects could appear broader and the derived $\beta$ from SED fitting could be lower, thus creating an artificial evolution with redshift.

\begin{figure*}
\begin{center}
\includegraphics[width=17cm]{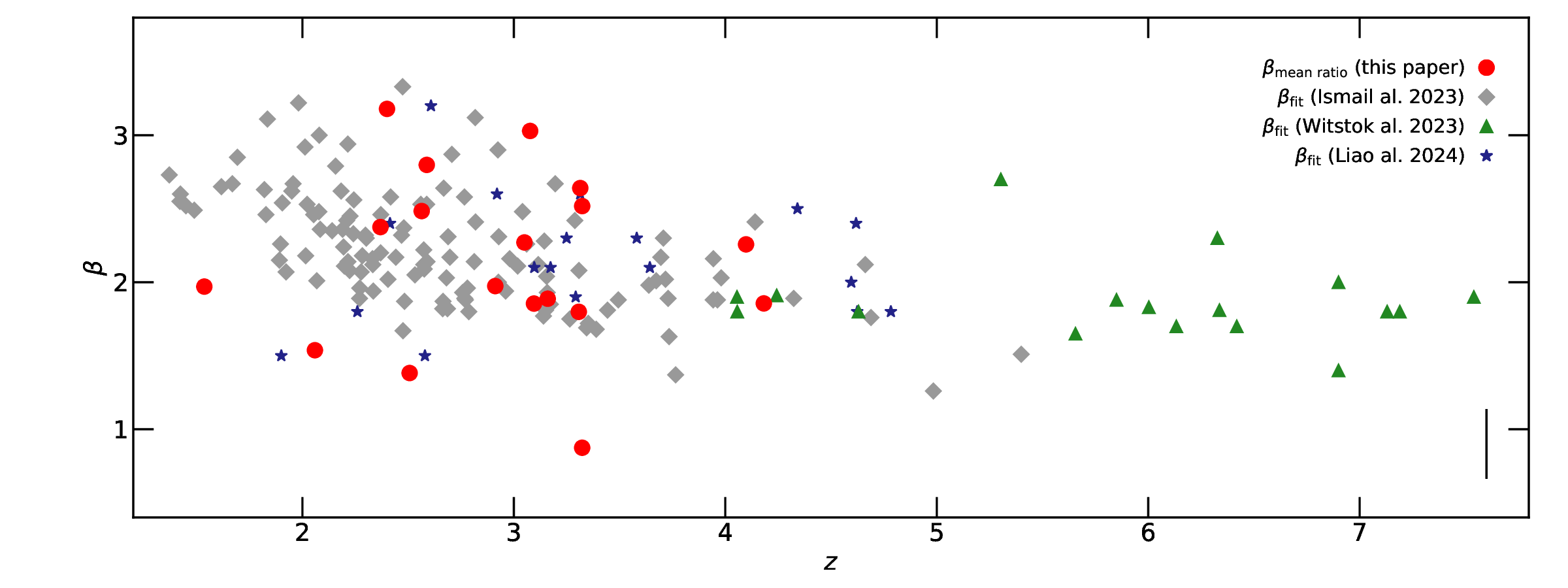}
\end{center}
\caption{The relation between $\beta$ (as measured from ALMA flux density ratios) and redshift for four different samples of data.  The red circles represent measurements for the sample in this paper, the grey diamonds represent data from \citet{ismail2023}, the green triangles represent data from \citet{witstok2023}, and the blue stars represent data from \citet{liao2024}.  A representative error bar for the $\beta$ values are shown in the bottom right corner of the plot; the uncertainties in redshift are negligible.  Only objects with spectroscopic redshifts were used in this plot.  The $\beta_\mathrm{mean~ratio}$ values were used for our sample.  The $\beta$ values from \citet{ismail2023} and \citet{liao2024} are all based on SED fits using an optically thin modified blackbody, while the values from \citet{witstok2023} correspond to dust that becomes optically thick at shorter wavelengths.}
\label{f_beta_z}
\end{figure*}

To bypass these problems, we can use the $\beta_\mathrm{mean~ratio}$ values from Section~\ref{s_beta_ratio} to examine the relation between $\beta$ and redshift.  These data are plotted in Figure~\ref{f_beta_z}.  For comparison, we also included the $\beta$ values and redshifts measured by \citet{ismail2023}, \citet{witstok2023}, and \citet{liao2024} for other samples of galaxies\footnote{Although \citet{tripodi2024} and \citet{ward2024} also present relations between redshift and $\beta$, the \citet{tripodi2024} sample largely overlaps with the \citet{witstok2023} sample, and the \citet{ward2024} sample largely overlaps with ours.}.  Although these $\beta$ values are derived using SED fits and are affected by degeneracies between the derived $\beta$ and temperature values, they still allow us to place our measurements into the context of what has already been published.  The $\beta$ values from either our combined sample or either subsample do not appear to vary with redshift.  The weighted Pearson correlation coefficient is -0.13 for our sample, which is  low enough to indicate that less than 25\% of the variance in $\beta$ is related to redshift.  

The range of our $\beta$ values seem similar to the values from \citet{witstok2023} and \citet{liao2024}, both of which also indicated that they found no relation between $\beta$ and redshift.  The \citet{ismail2023} $\beta$ values for the optically thin blackbody, however, exhibit a weak relation with redshift; the corresponding weighted Pearson correlation coefficient is -0.57, indicating that $\sim$30\% of the variation in $\beta$ is related to redshift.  \citet{ismail2023} fitted both their {\it Herschel} and NOEMA data with a single modified blackbody to derive $\beta$, which could have produced a bias in the $\beta$ values for the reasons discussed by \citet{bendo2023} that we discussed earlier in Section~\ref{s_discuss}.  Ideally, the $\beta$ values for the objects in the \citet{ismail2023}, \citet{witstok2023}, and \citet{liao2024} samples should be recalculated using the same method, and that method should not depend on SED fitting if possible, as the original values may include biases or offsets related to the derivation technique that was applied.  (This is why we have not fit a single relation to all of the data in Figure~\ref{f_beta_z} or attempted to calculate any statistics for that relation.)  However, deriving the $\beta$ values for the other samples is beyond the scope of our paper.

Our ultimate conclusion is that we do not observe any notable evolution in $\beta$ with redshift, at least among gravitational lenses and infrared-bright galaxies at $z$ between 1.5 and 4.5, and this is largely consistent with what has been found in other studies, including ones that extend their observations to redshifts of $\sim$7.5.  This in turn implies that the dust grain properties have not evolved over this time period or if the dust grain properties do evolve, it does not affect their emissivity.

\subsection{SED fitting using $\beta_\mathrm{mean~ratio}$}
\label{s_discuss_sedapp}

\subsubsection{SED fitting methodology}

As discussed in Section~\ref{s_intro}, variations in $\beta$ can also affect the temperatures derived from SED fitting, and this will also affect the derived masses.  We explored this in more detail by comparing SED fits using $\beta_\mathrm{mean~ratio}$ versus SED fits where $\beta$ is treated as a free parameter the fits (which we will label $\beta_\mathrm{free}$).  We will work solely with the sources from the 7 fields in our sample that contain only one ALMA source with a spectroscopic redshift.  The ALMA Band 3-8 data for these sources can be combined with the {\it Herschel} 250-500~$\mu$m data from \citet{valiante2016} to create well-sampled SEDs spanning both sides of the peak in the dust emission, which will be very useful for constraining the functions fit to the data.

In a general situation, dust emission can be related to dust temperature and mass by the equation
\begin{equation}
f_\nu = \frac{(1+z)\mu A}{D^2} (1-e^{-\kappa_\nu M_\mathrm{dust} / A}) (B_\nu(T_\mathrm{dust})-B_\nu(T_\mathrm{CMB})).
\label{e_dustbasic}
\end{equation}
In this equation, $\mu$ represents the magnification factor, $A$ represents the area of the emitting source, $D$ is the luminosity distance to the source, $\kappa_\nu$ is the emissivity as a function of $\nu$, $M_\mathrm{dust}$ is the dust mass, and $B_\nu(T_\mathrm{dust})$ is the blackbody function for a temperature $T_\mathrm{dust}$.  This generalized equation may be applied to dust that becomes optically thick at infrared wavelengths.  

The emissivity may be written as
\begin{equation}
\kappa_\nu = \kappa_0 \left( \frac{\lambda}{\lambda_0} \right)^{-\beta}    
\end{equation}
where the function is fixed to a constant value $\kappa_0$ at a specific wavelength $\lambda_0$.  As discussed earlier, people generally adopt a $\kappa_0$ and $\lambda_0$ from a model where $\beta\cong2$ even though the derived $\beta$ may differ from 2, which would mean that the $\kappa_0$ from that model is unreliable.  For this exercise, we will set $\kappa_0$ to 6.37~cm$^2$~g$^{-1}$ at 200~$\mu$m as specified by \citet{draine2003}, but we will also discuss the implications of fixing this to a value at a shorter or longer wavelength.

For optically thin dust where the exponential term is very small, Equation~\ref{e_dustbasic} can be simplified as
\begin{equation}
f_\nu = \frac{(1+z)\mu \kappa_0 M_\mathrm{dust}}{D^2} \left( \frac{\lambda}{\lambda_0} \right)^{-\beta} (B_\nu(T_\mathrm{dust})-B_\nu(T_\mathrm{CMB})).
\label{e_dustthin}
\end{equation}
For the optically thick case, we can define the wavelength $\lambda_\mathrm{thick}$ as where the dust becomes optically thick.  This gives $A$ as
\begin{equation}
A=\kappa_0  \left( \frac{\lambda_0}{\lambda_\mathrm{thick}} \right)^{\beta} M_\mathrm{dust}.
\end{equation}
Substituting this into Equation~\ref{e_dustbasic} gives
\begin{multline}
f_\nu = \frac{(1+z)\mu \kappa_0 M_\mathrm{dust}}{D^2} \left( \frac{\lambda_0}{\lambda_\mathrm{thick}} \right)^{\beta} (1-e^{(\lambda_\mathrm{thick}/\lambda )^\beta} ) \\ \times(B_\nu(T_\mathrm{dust})-B_\nu(T_\mathrm{CMB})).
\label{e_dustthick}
\end{multline}
We performed four fits using Equations~\ref{e_dustthin} and \ref{e_dustthick}.  Note that using a single modified blackbody to represent the dust emission is an oversimplification, as each object may contain dust at a range of temperatures.  However, such SED fits are commonly applied to SEDs to attempt to characterize the dust, as can be seen in the references in Section~\ref{s_intro}; our goal is to understand how decisions in SED fitting, particularly regarding $\beta$, affect the results.

In the first fit, we fixed $\beta$ to $\beta_\mathrm{mean~ratio}$ and then sought to fit the data at observed wavelengths $>1500$~$\mu$m (which are equivalent to the ALMA Band 3-5 data and which were selected because those data were used to derive $\beta_\mathrm{mean~ratio}$) with optically thin modified blackbody with the highest $T_\mathrm{dust}$ that were consistent with the peaks of the SEDs.  These modified blackbodies were not allowed to exceed the data at rest wavelengths $\leq300$~$\mu$m (which covered the peak of the SED and included the {\it Herschel} data points, the ALMA Band 8 data point, and, for the higher redshift sources, the ALMA Band 7 data point) by more than $1\sigma$.  We did this by iterating through a series of $T_\mathrm{dust}$ values starting with 15~K and increasing by 1~K in each iteration.  We rescaled each of these modified blackbodies using the ALMA Band 3-5 data and compared the flux densities from the extrapolated modified blackbody to the observed flux densities to ensure that the model did not exceed the observed data by $1\sigma$.\footnote{Alternately, it may have been possible to use a standard nonlinear least squares fitter to fit the SEDs with optically thin modified blackbodies while setting all data points except for the Band 3-5 data to upper limits.  However, it is not clear whether the algorithm would have converged on the highest temeperature possible (i.e., whether it would have fit a curve through the upper limits or significanty below them).}  The reported $T_\mathrm{dust}$ and $M_\mathrm{dust}$ correspond to not only the warmest possible dust component with an emissivity index of $\beta_\mathrm{mean~ratio}$ but also to a lower limit on the dust mass present in these objects if the dust is optically thin.  Note that the dust masses could potentially be much higher if the Rayleigh-Jeans side of the dust SED originates from dust with a range of temperatures, including temperatures colder than what we derive using this single optically thin modified blackbody.

The other three SED fits are performed to all of the available {\it Herschel} and ALMA data in a more standard way.  In the second fit, we fit all of the data with a single optically thin modified blackbody where $\beta$ is a free parameter, which is a common practice in many studies of dust in other galaxies.  For the third fit, we used the optically thick modified blackbody equation, treated $\lambda_\mathrm{thick}$ as a free parameter (except for two objects), and fixed $\beta$ to $\beta_\mathrm{mean~ratio}$.  In the fourth fit, we  we used the optically thick modified blackbody equation, treated $\beta$ as a free parameter, and also treated $\lambda_\mathrm{thick}$ as free parameter  (except for two objects).  With HerBS-87 and HerBS-184, where we encountered other fitting issues discussed below, $\lambda_\mathrm{thick}$ was effectively unconstrained in the SED fits, so we fixed the value to 220~$\mu$m, which is the approximate average of the value from the other five objects using either a fixed or variable $\beta$.

\subsubsection{SED fitting results}

\begin{figure*}
\begin{center}
\includegraphics[width=5.75cm]{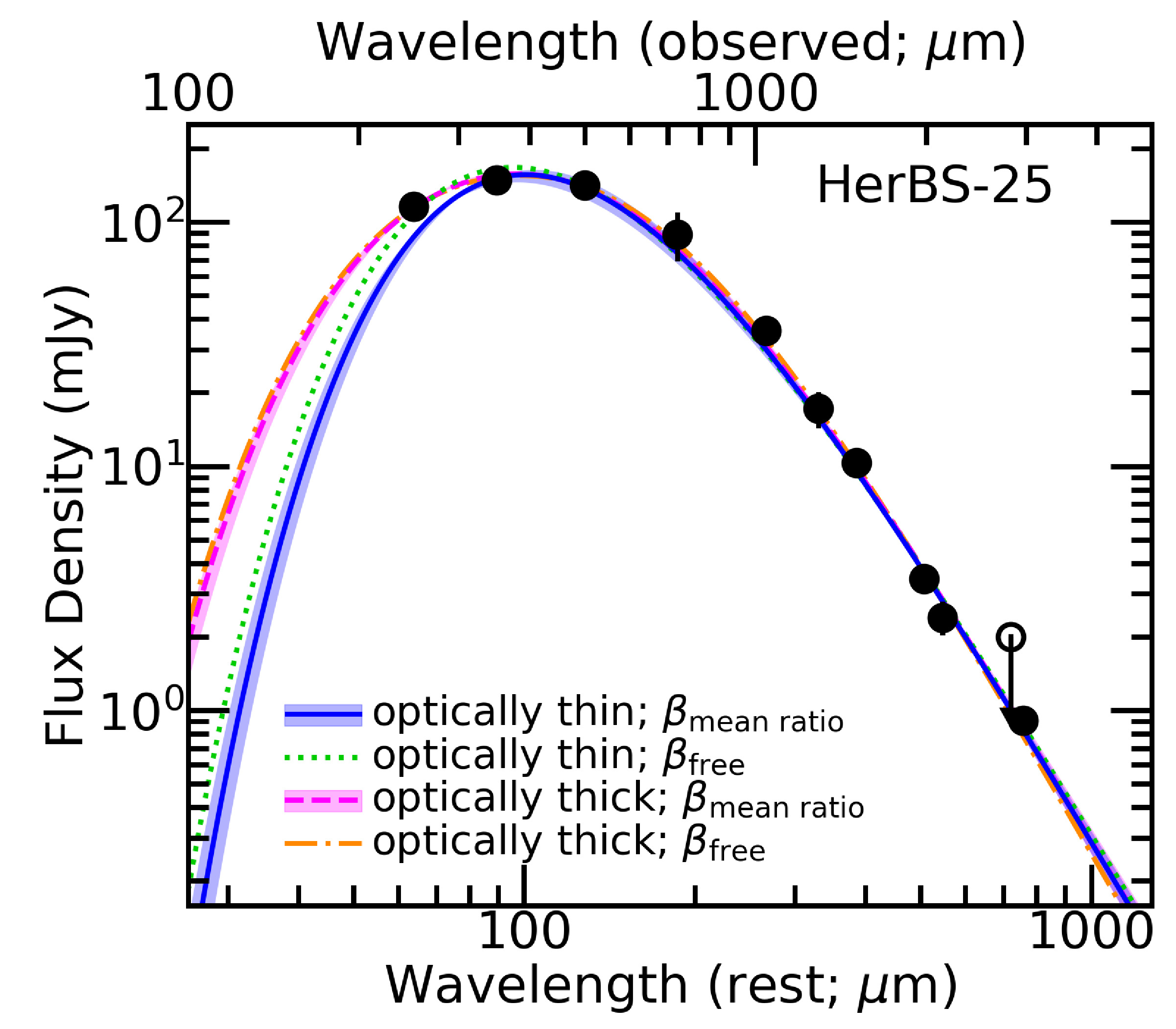} ~
\includegraphics[width=5.75cm]{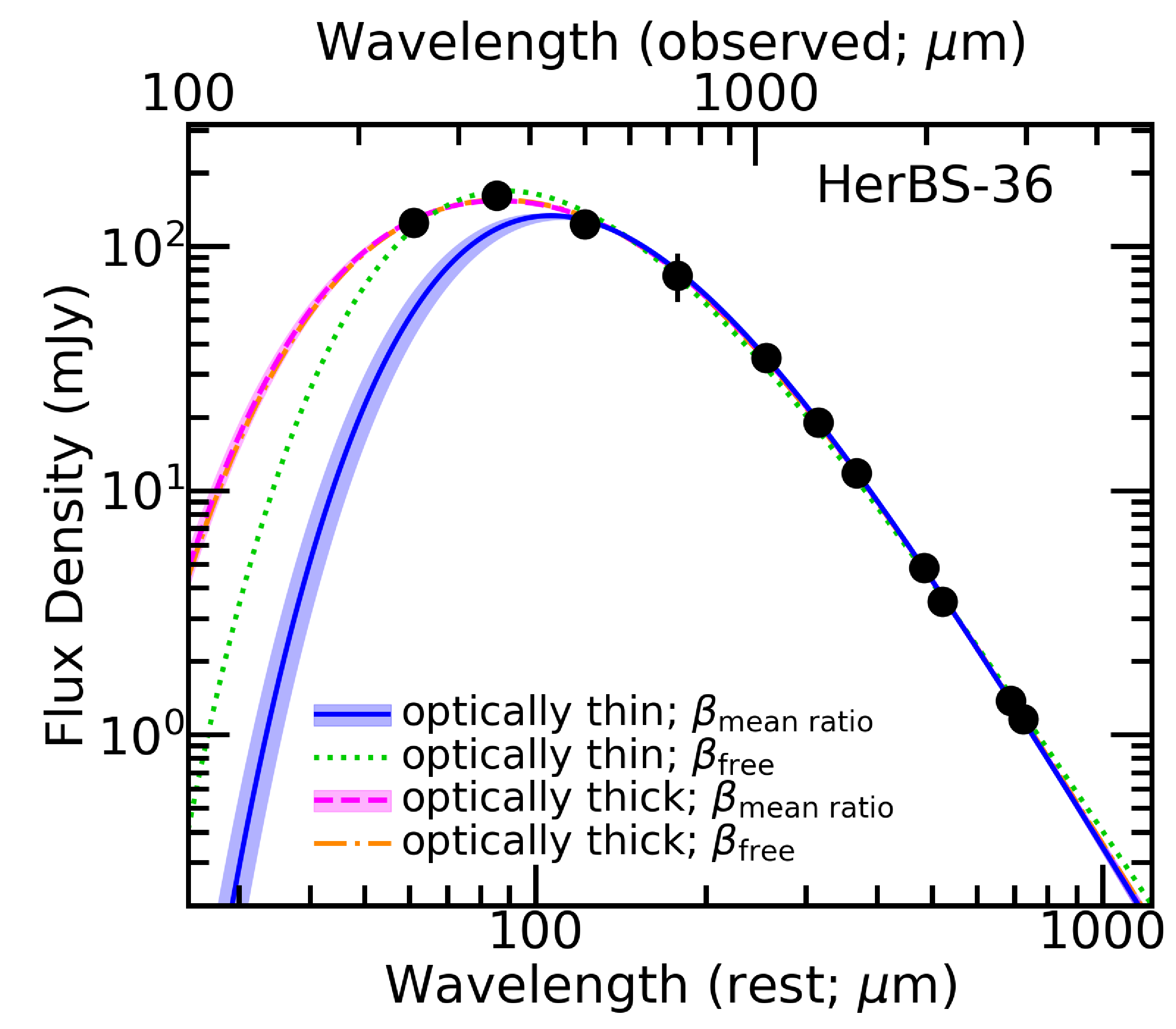} ~
\includegraphics[width=5.75cm]{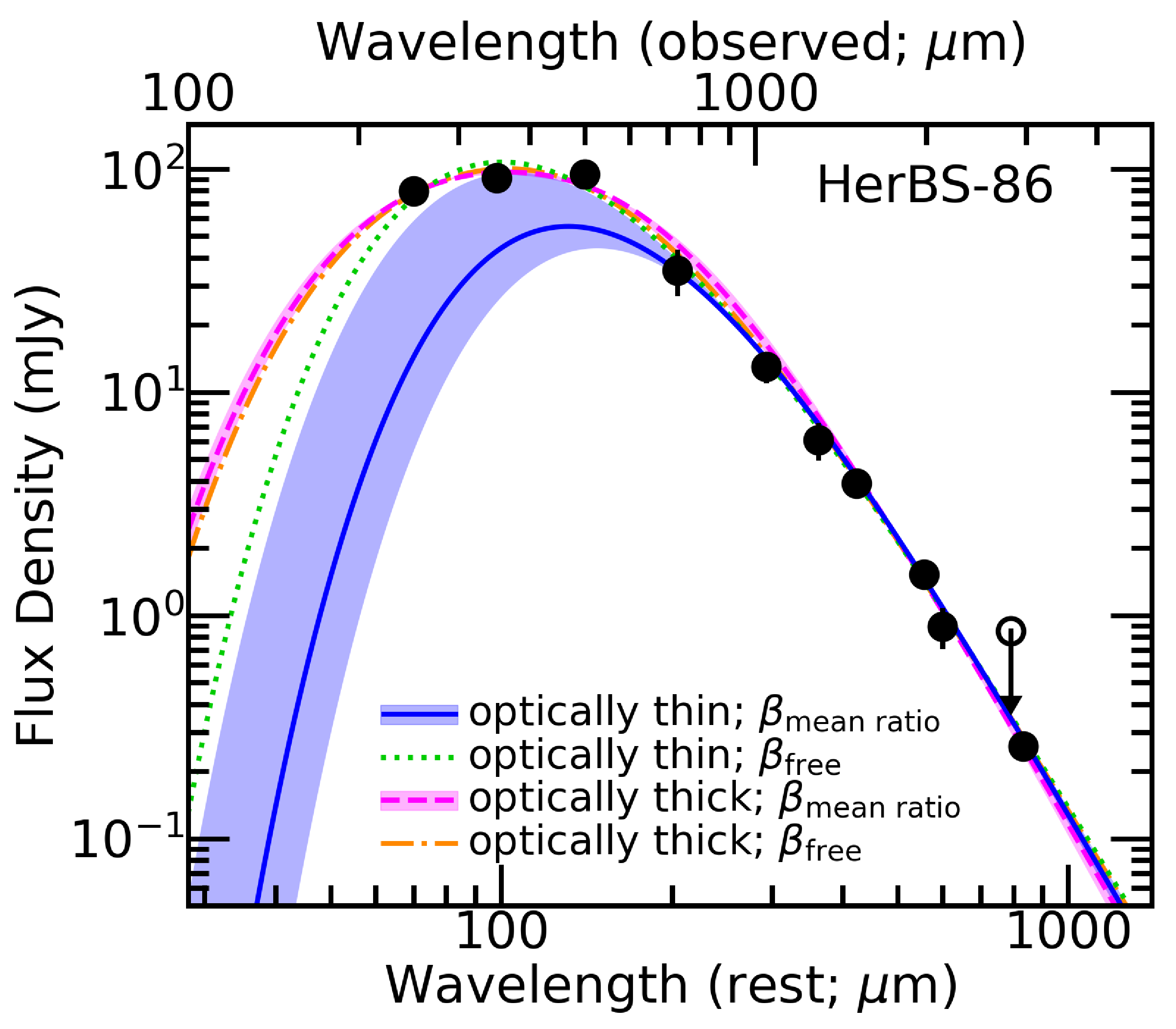} \\ ~\\
\includegraphics[width=5.75cm]{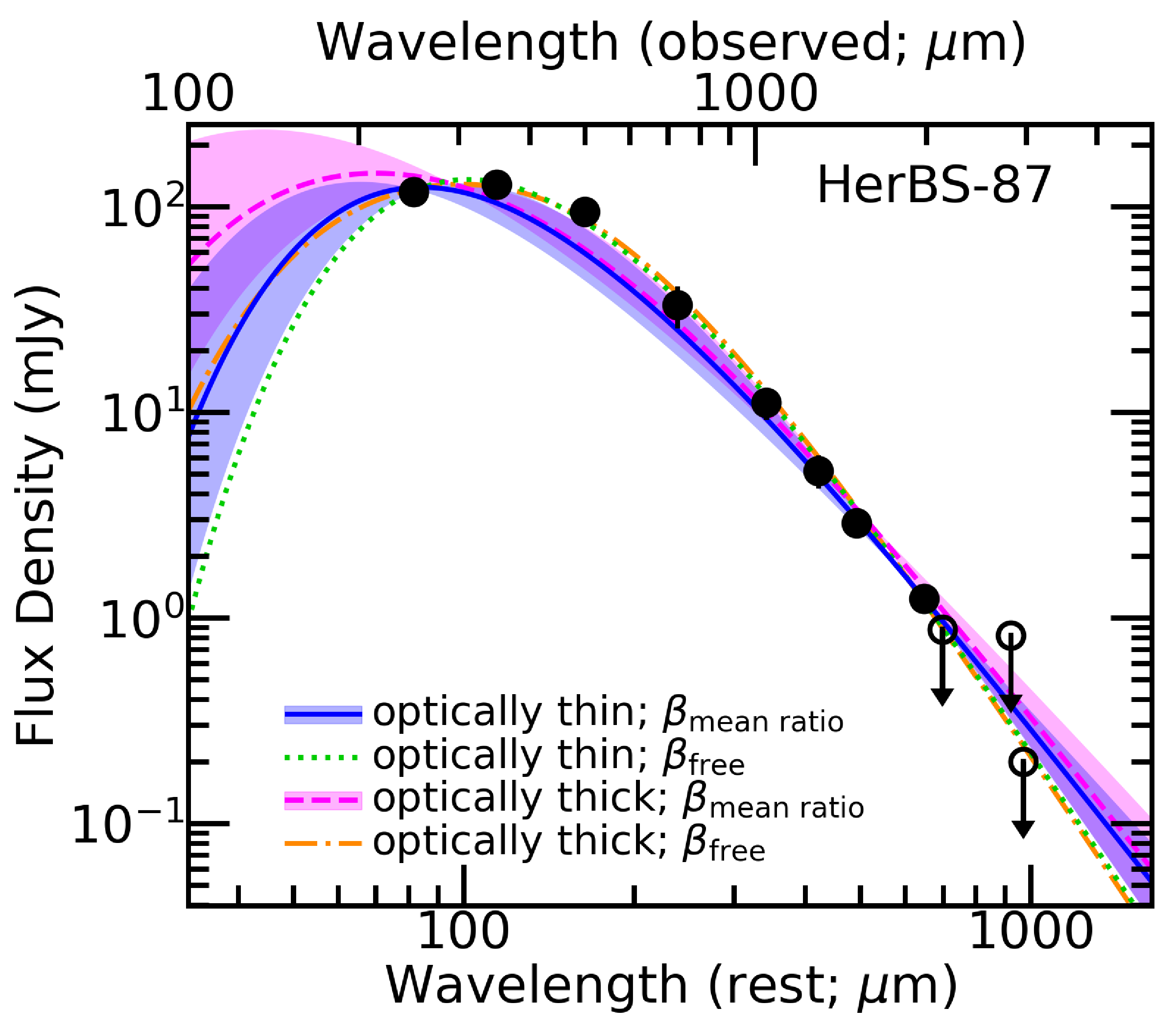} ~
\includegraphics[width=5.75cm]{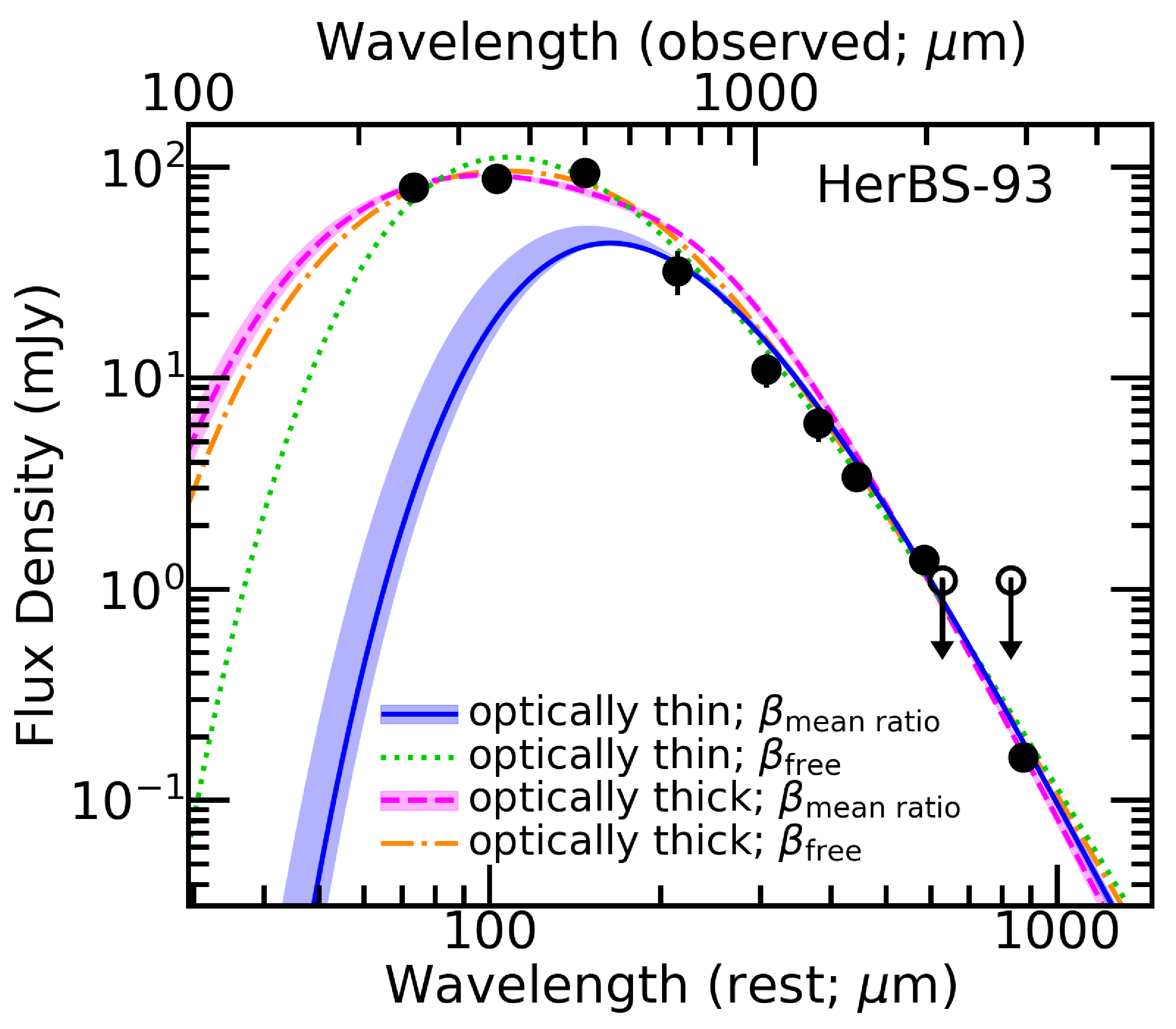} ~
\includegraphics[width=5.75cm]{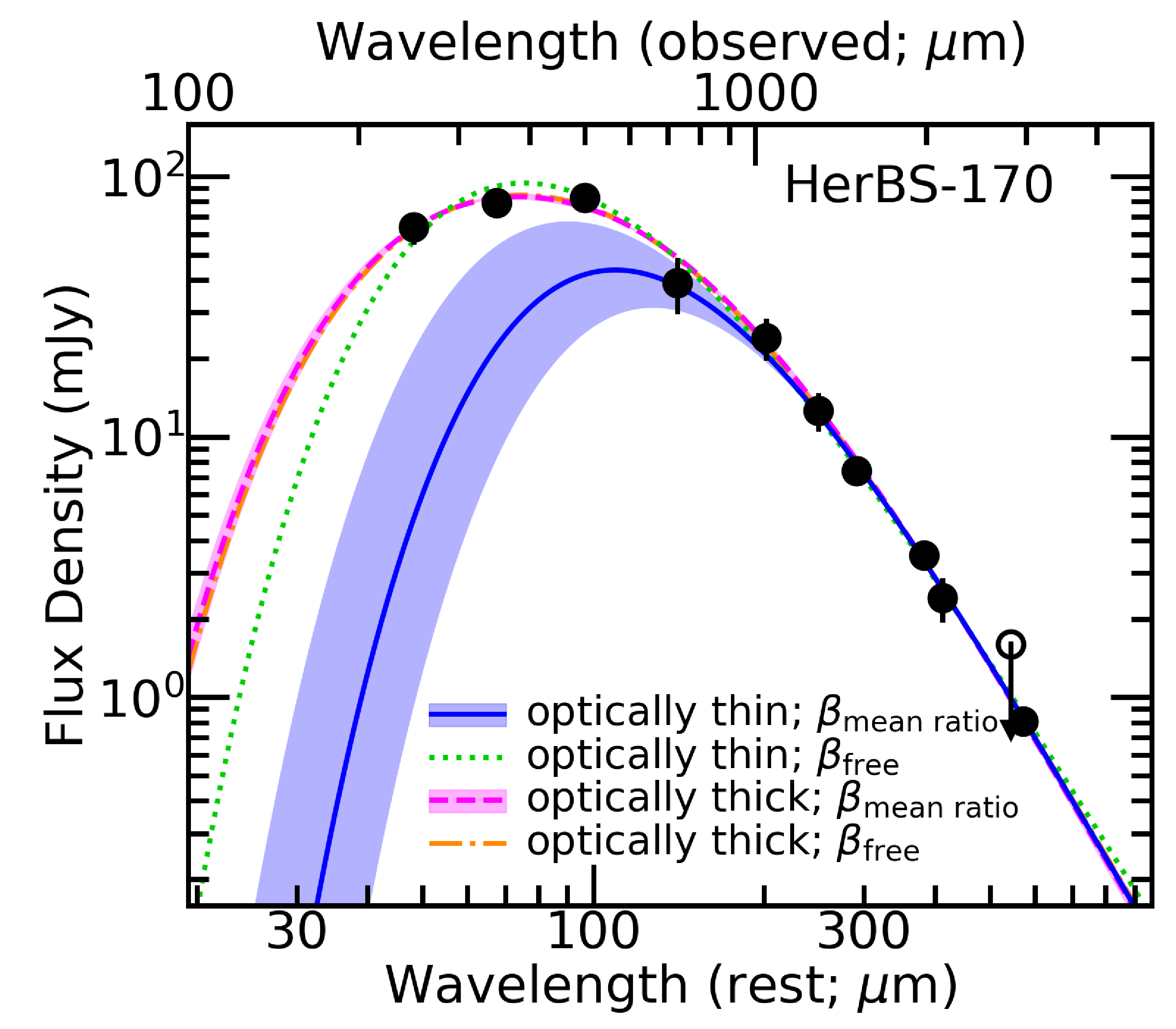} \\ ~\\
\includegraphics[width=5.75cm]{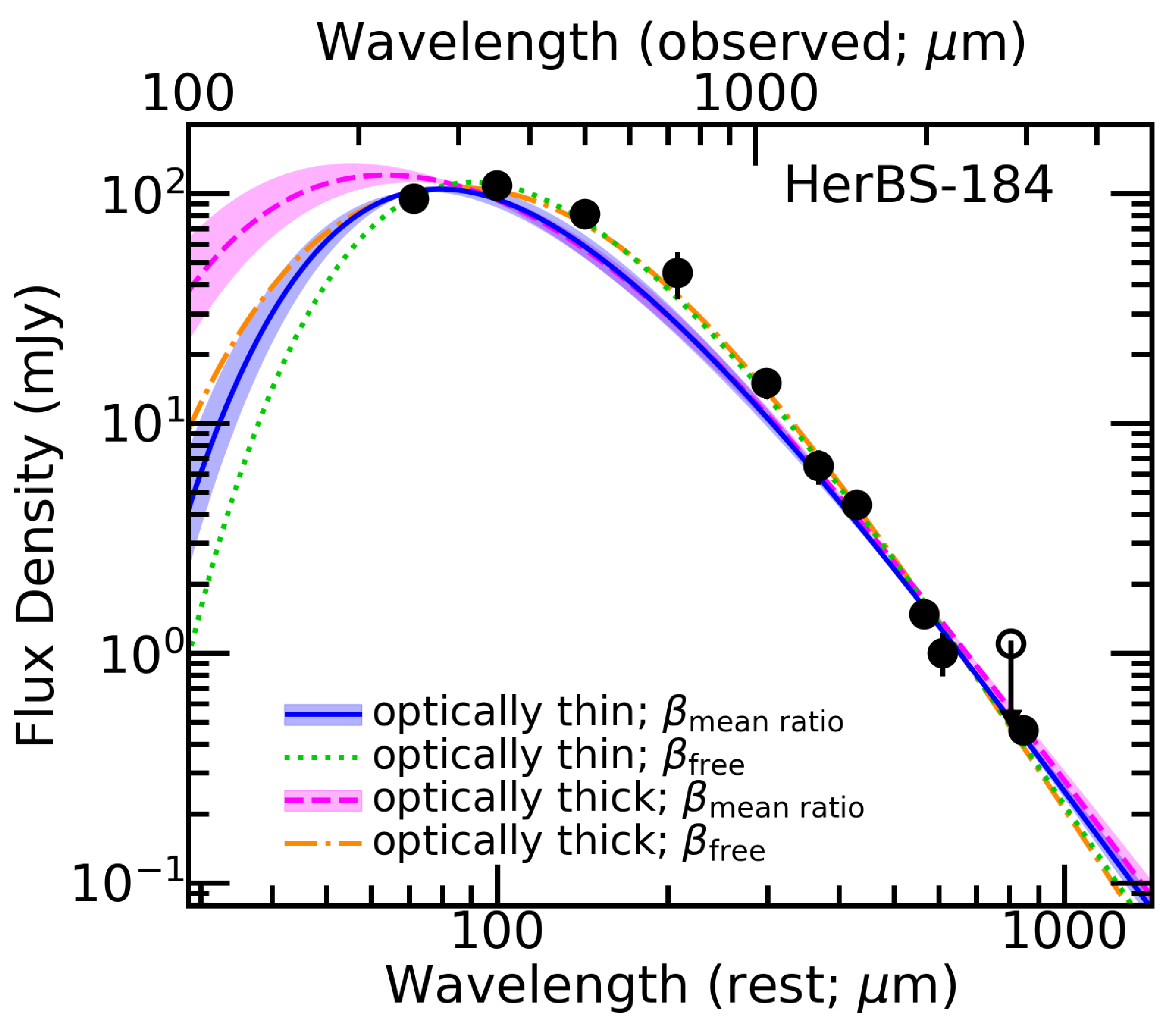} 
\end{center}
\caption{Four different SED fits to the {\it Herschel} and ALMA data for fields containing individual ALMA sources.  The solid blue line represents the optically thin modified blackbodies where $\beta$ fixed to $\beta_\mathrm{mean~ratio}$ and where the best fitting function is consistent with the warmest possible dust emitting at $>1500$~$\mu$m; see Section~\ref{s_discuss_sedapp} for details.  The green dotted line shows an optically thin modified blackbody where $\beta$ is treated as a free parameter.  The magenta dashed line shows an optically thick modified blackbody where $\beta$ fixed to $\beta_\mathrm{mean~ratio}$, and the orange dash dot line shows an optically thick modified blackbody where $\beta$ was allowed to vary in the fit.  The blue and magenta shaded regions show the range of possible best fitting functions based on the uncertainties in $\beta_\mathrm{mean~ratio}$.  Unless error bars are shown, the uncertainties in the data points are smaller than the size of the symbols in the plot.  The $5\sigma$ upper limits are shown as empty symbols with downwards-pointing arrows.}
\label{f_sedfit}
\end{figure*}

The SED fits are shown in Figure~\ref{f_sedfit}.  In the situations where we fixed $\beta$ to $\beta_\mathrm{mean~ratio}$, we show shaded regions in Figure~\ref{f_sedfit} consistent with the uncertainties in $\beta_\mathrm{mean~ratio}$, and we incorporate the uncertainty in $\beta_\mathrm{mean~ratio}$ into the uncertainties in the other quantities.   The derived $T_\mathrm{dust}$, $\beta$, log($M_\mathrm{dust}$), and $\lambda_\mathrm{thick}$ (for the optically thick modified blackbodies) are listed in Table~\ref{t_sedfit_thin} and Table~\ref{t_sedfit_thick}.  For the 3 sources that \citet{bakx2024} identified as gravitationally lensed, we corrected the $M_\mathrm{dust}$ values by their reported magnification terms $\mu$. For the other 4 sources where no evidence of lensing was found, we set $\mu$ to 1 for our calculations, which in effect meant that no magnification correction was applied to the $M_\mathrm{dust}$ values.

\begin{table*}
\centering
\begin{minipage}{139mm}
\caption{Results from fitting optically thin modified blackbodies}
\label{t_sedfit_thin}
\begin{tabular}{@{}lccccccc@{}}
\hline
Object &
  $\mu^a$ &
  \multicolumn{3}{c}{SED fit using $\beta_\mathrm{mean~ratio}$} &
  \multicolumn{3}{c}{SED fit allowing $\beta$ to vary} \\
&
  &
  $T_\mathrm{dust}$ &
  $\beta_\mathrm{mean~ratio}$ &
  log($M_\mathrm{dust}$) &
  $T_\mathrm{dust}$ &
  $\beta_\mathrm{free}$ &
  log($M_\mathrm{dust}$) \\
\hline
HerBS-25 &
  9.2 $\pm$ 1.8 &
  29 $\pm$ 2 &
  2.0 $\pm$ 0.2 &
  9.28 $\pm$ 0.29 &
  31 $\pm$ 2 &
  1.9 $\pm$ 0.1 &
  9.18 $\pm$ 0.27 \\
HerBS-36 &
  4.1 $\pm$ 0.8 &
  28 $\pm$ 2 &
  1.9 $\pm$ 0.1 &
  9.71 $\pm$ 0.14 &
  36 $\pm$ 1 &
  1.6 $\pm$ 0.1 &
  9.39 $\pm$ 0.11 \\
HerBS-86 &
  1.0$^b$ &
  20 $\pm$ 5 &
  2.5 $\pm$ 0.2 &
  10.42 $\pm$ 0.33 &
  28 $\pm$ 2 &
  2.2 $\pm$ 0.1 &
  10.01 $\pm$ 0.11 \\
HerBS-87 &
  8.7 $\pm$ 1.7 &
  38 $\pm$ $_{-9}^{+17}$ &
  1.5 $\pm$ 0.5 &
  8.53 $\pm$ 0.49 &
  28 $\pm$ 1 &
  2.1 $\pm$ 0.1 &
  9.03 $\pm$ 0.25 \\
HerBS-93 &
  1.0$^b$ &
  14 $\pm$ 2 &
  3.2 $\pm$ 0.2 &
  11.09 $\pm$ 0.24 &
  24 $\pm$ 2 &
  2.5 $\pm$ 0.2 &
  10.20 $\pm$ 0.19 \\
HerBS-170 &
  1.0$^b$ &
  27 $\pm$ 6 &
  1.9 $\pm$ 0.1 &
  10.12 $\pm$ 0.27 &
  42 $\pm$ 2 &
  1.5 $\pm$ 0.1 &
  9.64 $\pm$ 0.07 \\
HerBS-184 &
  1.0$^b$ &
  41 $\pm$ 4 &
  1.4 $\pm$ 0.2 &
  9.42 $\pm$ 0.14 &
  34 $\pm$ 3 &
  1.7 $\pm$ 0.1 &
  9.73 $\pm$ 0.12 \\
HerBS-184 (without Band 3) &
  1.0$^b$ &
  22 $\pm$ 8 &
  2.6 $\pm$ 0.5 &
  10.39 $\pm$ 0.59 &
  30 $\pm$ 1 &
  2.1 $\pm$ 0.1 &
  9.91 $\pm$ 0.07 \\
\hline
\end{tabular}
$^a$ These magnification factors are from \citet{bakx2024}.\\
$^b$ When no evidence of magnification by gravitational lensing was found by \citet{bakx2024}, we set the value to 1 for our calculations.
\end{minipage}
\end{table*}

\begin{table*}
\centering
\begin{minipage}{167mm}
\caption{Results from fitting optically thin modified blackbodies}
\label{t_sedfit_thick}
\begin{tabular}{@{}lccccccccc@{}}
\hline
Object &
  $\mu^a$ &
  \multicolumn{4}{c}{SED fit using $\beta_\mathrm{mean~ratio}$} &
  \multicolumn{4}{c}{SED fit allowing $\beta$ to vary} \\
&
  &
  $T_\mathrm{dust}$ &
  $\beta_\mathrm{mean~ratio}$ &
  $\lambda_\mathrm{thick}$ &
  log($M_\mathrm{dust}$) &
  $T_\mathrm{dust}$ &
  $\beta_\mathrm{free}$ &
  $\lambda_\mathrm{thick}$ &
  log($M_\mathrm{dust}$) \\
\hline
HerBS-25 &
  9.2 $\pm$ 1.8 &
  49 $\pm$ 4 &
  2.0 $\pm$ 0.2 &
  200 $\pm$ 30 &
  8.95 $\pm$ 0.26 &
  51 $\pm$ 3 &
  2.1 $\pm$ 0.2 &
  220 $\pm$ 20 &
  9.0 $\pm$ 0.26 \\
HerBS-36 &
  4.1 $\pm$ 0.8 &
  59 $\pm$ 3 &
  1.9 $\pm$ 0.1 &
  240 $\pm$ 20 &
  9.26 $\pm$ 0.1 &
  58 $\pm$ 2 &
  1.8 $\pm$ 0.1 &
  230 $\pm$ 20 &
  9.25 $\pm$ 0.1 \\
HerBS-86 &
  1.0$^b$ &
  50 $\pm$ 6 &
  2.5 $\pm$ 0.2 &
  230 $\pm$ 30 &
  9.77 $\pm$ 0.1 &
  47 $\pm$ 6 &
  2.2 $\pm$ 0.2 &
  200 $\pm$ 40 &
  9.69 $\pm$ 0.08 \\
HerBS-87 &
  8.7 $\pm$ 1.7 &
  72 $\pm$ $_{-20}^{+43}$ &
  1.5 $\pm$ 0.5 &
  220$^c$ &
  8.27 $\pm$ 0.45 &
  51 $\pm$ 3 &
  2.2 $\pm$ 0.1 &
  220$^c$ &
  8.69 $\pm$ 0.24 \\
HerBS-93 &
  1.0$^b$ &
  52 $\pm$ 10 &
  3.2 $\pm$ 0.2 &
  300 $\pm$ 40 &
  10.02 $\pm$ 0.13 &
  48 $\pm$ 9 &
  2.7 $\pm$ 0.3 &
  240 $\pm$ 50 &
  9.82 $\pm$ 0.17 \\
HerBS-170 &
  1.0$^b$ &
  68 $\pm$ 5 &
  1.9 $\pm$ 0.1 &
  200 $\pm$ 30 &
  9.49 $\pm$ 0.04 &
  67 $\pm$ 5 &
  1.8 $\pm$ 0.1 &
  190 $\pm$ 30 &
  9.47 $\pm$ 0.04 \\
HerBS-184 &
  1.0$^b$ &
  79 $\pm$ $_{-13}^{+15}$ &
  1.4 $\pm$ 0.2 &
  220$^c$ &
  9.14 $\pm$ 0.16 &
  59 $\pm$ 6 &
  1.9 $\pm$ 0.1 &
  220$^c$ &
  9.48 $\pm$ 0.11 \\
HerBS-184 (without Band 3) &
  1.0$^b$ &
  57 $\pm$ 7 &
  2.6 $\pm$ 0.5 &
  260 $\pm$ 60 &
  9.75 $\pm$ 0.17 &
  53 $\pm$ 3 &
  2.2 $\pm$ 0.1 &
  220 $\pm$ 20 &
  9.63 $\pm$ 0.04 \\

\hline
\end{tabular}
$^a$ These magnification factors are from \citet{bakx2024}.\\
$^b$ When the magnification was not determined by \citet{bakx2024}, we set the value to 1 for our calculations.\\
$^c$ Because $\lambda_\mathrm{thick}$ was poorly constrained when treated as a free parameter for these two objects, we fixed the value to 220~$\mu$m, which is the approximate average from the SED fits to the other objects.
\end{minipage}
\end{table*}

Just based on the assessment of the plots in Figure~\ref{f_sedfit}, the SED fits generally describe the Rayleigh-Jeans side of the SED accurately. The optically thin modified blackbody with $\beta_\mathrm{mean~ratio}$ illustrates that it is possible to describe most of the emission on the Rayleigh-Jeans side of the SED peak with colder dust emission, in which case the emission on the Wien side would originate from warmer but lower masses of dust (which are not included in the SED fits because they would be poorly constrained).  Blending the emission from these components together could make the SED appear more consistent with optically thin dust with a lower $\beta$.   However, the modified blackbodies using $\beta_\mathrm{mean~ratio}$ did not quite accurately describe the slopes of the SEDs for HerBS-87 and HerBS-184, which we will discuss below.  Also, in general, the two different variant of the optically thick modified blackbody generally yielded very similar curves with very similar parameters.

The problems with some of the modified blackbody fits to the HerBS-87 arise because the $\beta_\mathrm{mean~ratio}$ value of $1.5\pm0.5$ was determined using just two data points (one each from ALMA Band 4 and 5) and has a high uncertainty.  However, based on the upper limits from the other ALMA Band 3 and 4 data\footnote{Using the upper limits as constraints in the SED fits where $\beta$ was fixed to $\beta_\mathrm{mean~ratio}$ actually caused more severe problems with fitting the other data in the SED, so we chose to show the SED fits without including the upper limits.}, the actual $\beta$ may be closer to 2.  The SED fits with a variable $\beta$ to the HerBS-87 data yield values closer to 2 that are more consistent with the upper limits.  Also,  that the lower limits for the shaded regions for the fits with $\beta_\mathrm{mean~ratio}$ (which correspond to $\beta=2$) are also more consistent with the upper limits. 

With HerBS-184 specifically, the SED fits illustrate how the observed frame 101~GHz (rest frame 847~$\mu$m) data point may be flattening near a rest wavelength of $\sim1000$~$\mu$m, which would be more consistent with free-free or synchrotron emission instead of dust emission, and that flattening causes minor problems with fitting the data with just a modified blackbody, causing some inconsistencies in the SED fits, particularly the optically thick modified blackbody using $\beta_\mathrm{mean~ratio}$.  Given this issue, we list in Tables~\ref{t_sedfit_thin} and \ref{t_sedfit_thick} alternate parameters from SED fits to the HerBS-184 data where we excluded the Band 3 data point and, for the modified blackbodies where $\beta$ is fixed, use the $\beta_{\mathrm{Band~5/Band~4}}$ value.

With the optically thin scenarios, allowing $\beta$ to vary when fitting optically thin modified blackbodies generally yields $\beta_\mathrm{free}$ values that are generally lower than $\beta_\mathrm{mean~ratio}$.  For HerBS-87, the $\beta_\mathrm{free}$ value is actually higher than $\beta_\mathrm{mean~ratio}$, but we noted the large uncertainties in $\beta_\mathrm{mean~ratio}$ above.  For HerBS-184, only the fits excluding the Band 3 data yield a $\beta_\mathrm{free}$ value lower than $\beta_\mathrm{mean~ratio}$.  The higher $\beta_\mathrm{mean~ratio}$ values generally correspond to temperatures that are 3-15~K lower.  The lower temperatures also lead to dust masses that are $1.35\times$ to $\sim8\times$ higher, as would be expected from Equation~\ref{e_dustthin}.  However, these higher dust masses would not include the warmer dust emitting at shorter wavelengths.   This also depends on what is used for $\kappa$.  We fixed $\kappa_\nu$ at 200~$\mu$m to the value given by \citet{draine2003}, but if we instead fix the scale of $\kappa_\nu$ at 100~$\mu$m using the value from that paper, the differences in dust masses increase by up to 50\%, while if we fix the scale of $\kappa_\nu$ at 850~$\mu$m, the differences in dust masses are generally $\lesssim 2\times$, and the largest difference in masses is $\sim3\times$.

The SED fits using an optically thick modified blackbody generally yielded similar temperatures and $\lambda_\mathrm{thick}$ results regardless or whether $\beta$ was fixed to $\beta_\mathrm{mean~ratio}$ or allowed to vary.  The only notable cases where $\beta_\mathrm{free}$ was statistically lower than $\beta_\mathrm{mean~ratio}$ were HerBS-86, HerBS-87 (where $\beta_\mathrm{mean~ratio}$ has high uncertainties), and HerBS-93.  Even so, this did not affect the derived temperatures significantly, although it did result in the derived $\lambda_\mathrm{thick}$ shifting to shorter wavelengths.  The derived masses are also also generally the same regardless of whether $\beta$ is allowed to vary; the largest changes were seen for HerBS-93, where the mass corresponding to $\beta_\mathrm{free}$ was $\sim1.6\times$ lower, and HerBS-184 including the Band 3 data, where the mass was $\sim2.2\times$ higher.  In most cases, changing the wavelength at which the scale of $\kappa_\nu$ is fixed does not change the results.  The exception is HerBS-93, where scaling $\kappa_\nu$ at 100~$\mu$m using the \citet{draine2003} value leads to a $\sim2.3\times$ difference between the masses derived using $\beta_\mathrm{mean~ratio}$ or $\beta_\mathrm{free}$, while scaling $\kappa_\nu$ at 850~$\mu$m leads to a difference in dust masses of $\sim5\times$.  Note that the $\lambda_\mathrm{thick}$ derived here, which range from 190 to 300~$\mu$m, are generally higher than the commonly cited value of 100~$\mu$m taken from the work on protoplanetary discs by \citet[][; which may not be applicable to high redshift objects]{draine2006} and are also higher than some of the other values derived for high redshift galaxies \citep[e.g.,][]{spilker2016, simpson2017, dacunha2021, liao2024}, although our results are consistent with other results for high redshift sources \citep[e.g.,][]{riechers2021}.

The comparison between the optically thin and optically thick cases shows that the dust temperatures increase by $\sim2\times$ or more in the optically thick cases.  This results in the dust masses also changing.  When $\beta_\mathrm{mean~ratio}$, the optically thin modified blackbodies yield dust masses that are $\gtrsim2\times$ higher than the optically thick counterparts, but when $\beta$ is allowed to vary, the difference in dust masses is smaller, ranging from $1.4\times$ to $2.4\times$.

Overall, this exercise shows that, in the optically thin case, treating $\beta$ as a free parameter may lead to underestimates of dust masses, but in the optically thick case, the changes are less notable if $\lambda_\mathrm{thick}$ is treated as a free parameter.  However, these models are overly simplistic treatments of the SEDs of these objects.  Unless the dust is extremely optically thick and originates from very compact regions (as is the case for Arp 220; see \citealt{rangwala2011} and \citealt{scoville2017}), then we would expect to see dust with a range of temperatures as well as dust in environments with different opacities.  This could include large masses of optically thin dust with lower temperatures than what is shown in the SED fitting here.  Additionally, one of the reasons why optically thick dust emission is used to model the emission from infrared-bright high-redshift is because, when optically thin dust models are applied, the implied dust masses are too high compared to the sizes of the emitting regions.  This more detailed analysis of the SEDs is beyond the scope of this paper but can potentially be investigated in future works.

\section{Conclusions}

We have derived dust emissivity index $\beta$ values for a set of 21 infrared-bright sources (including gravitationally lensed galaxies) at $1.5<z<4.2$ using the ratios of the ALMA Band 4/Band 3 flux densities (giving (giving $\beta_{\mathrm{4/3~ratio}}$) and Band 5/Band 4 flux densities (giving $\beta_{\mathrm{5/4~ratio}}$).  The weighted average of these yielded values for each galaxy we labelled as $\beta_\mathrm{mean~ratio}$.  The mean $\beta_\mathrm{mean~ratio}$ value is 2.2, and the standard deviation is 0.6.  This technique is largely insensitive to dust temperature variations and therefore avoids issues with degeneracies between $\beta$ and temperature that affect SED fitting.

These $\beta_\mathrm{mean~ratio}$ values are slightly larger than but statistically consistent with the value of $\sim$2 typically used within theoretical models.  However, our $\beta$ values are at the high end of the range of 1.4-2.3 found within the Milky Way, nearby galaxies, and  galaxies at $z>1$.  Most of these other studies have primarily relied on modified blackbody fits to both the peak and Rayleigh-Jeans side of the SED in which both temperature and $\beta$ are free parameters, so the derived $\beta$ values could be affected either by degeneracies with temperature in the SED fitting process or issues with emission from dust at different temperatures being blended near the peak of the SED.  Additional measurements of $\beta$ that are based on just the Rayleigh-Jeans slope of the SED should be applied to other galactic and extragalactic objects to determine whether $\beta$ may actually be greater than 2.  If this is the case or if $\beta$ is found to vary in a systematic way among galaxies, it would have major implications for SED modelling, including dust mass and dust temperature calculations.  We also found no systematic variations in our derived $\beta_\mathrm{mean~ratio}$ values with redshift; this is largely consistent with what has already been found in other studies.

For seven fields in our paper that contained single sources, we fit four different modified blackbodies to the data.  In comparing SED fits using single optically thin modified blackbodies where $\beta$ was fixed to $\beta_\mathrm{mean~ratio}$ or $\beta$ was treated as a free parameter, the fits with the varying $\beta$ generally yielded $\beta$ that were lower than $\beta_\mathrm{mean~ratio}$, temperatures that were higher, and dust masses that were lower.  In the same comparison but with fitting single optically thick modified blackbodies, the temperatures and masses did not change significantly in most cases, although the wavelength at which the dust became optically thick ($\lambda_\mathrm{thick}$) was longer when $\beta$ was fixed to $\beta_\mathrm{mean~ratio}$.  Note that the same results were not obtained for one source where $\beta_\mathrm{mean~ratio}$ was poorly constrained (which led to high uncertainties in dust temperatures) and one source where the Band 3 data may have included emission from sources other than dust (which affected the derived $\beta_\mathrm{mean~ratio}$).  This SED analysis relied on relatively simplistic models of the dust emission; additional analyses with more realistic dust models as well as constraints based on the angular area of the emission would be needed to more accurately characterize the dust.

The best way to expand on the results from this analysis would be to apply the same techniques for deriving $\beta$ to a broader sample of galaxies.  This may require performing additional observations of the Rayleigh-Jeans side of the SEDs for larger samples of objects, including both nearby and high redshift objects.  Our current plans are to expand the analysis to the whole of the BEARS sample using new ALMA observations.  However, it may also be appropriate to revisit data that have already been published that derived $\beta$ by fitting modified blackbodies through the peak and the Rayleigh-Jeans side of the SEDs.  These results could potentially change when when fitting just the Rayleigh-Jeans side of the SEDs or when using the slopes of the Rayleigh-Jeans side of the SED to derive $\beta$.

\section*{Acknowledgements}

We thank the reviewer for their comments on this paper.  GJB acknowledges support from STFC Grants ST/Y004108/1 and ST/T001488/1.  LB and JGN acknowledge the PID2021-125630NB-I00 project funded by MCIN/AEI/10.13039/501100011033/FEDER, UE. LB also acknowledges the CNS2022-135748 project funded by MCIN/AEI/10.13039/501100011033 and by the EU “NextGenerationEU/PRTR”.  MH is supported by Japan Society for the Promotion of Science (JSPS) KEKENHI Grant No. 22KJ1598.  D.R. gratefully acknowledges support from the Collaborative Research Center 1601 (SFB 1601 sub-projects C1, C2, C3, and C6) funded by the Deutsche Forschungsgemeinschaft (DFG) – 500700252. This paper makes use of the following ALMA data: ADS/JAO.ALMA\#2016.2.00133.S, 2018.1.00804.S, 2019.1.01477.S, and 2021.1.01628.S. ALMA is a partnership of ESO (representing its member states), NSF (USA) and NINS (Japan), together with NRC (Canada), MOST and ASIAA (Taiwan), and KASI (Republic of Korea), in cooperation with the Republic of Chile. The Joint ALMA Observatory is operated by ESO, AUI/NRAO and NAOJ.  {\it Herschel} is an ESA space observatory with science instruments provided by European-led Principal Investigator consortia and with important participation from NASA.

\section*{Data Availability}

The {\it Herschel} SPIRE data can be downloaded from \url{https://www.h-atlas.org}\,, while the reduced, calibrated and science-ready ALMA visibility data is available from the ALMA Science Archive at \url{https://almascience.eso.org/alma-data/archive}\,.


\bibliographystyle{mnras}
\bibliography{bendogj} % if your bibtex file is called example.bib

\begin{thebibliography}{}
\makeatletter
\relax
\def\mn@urlcharsother{\let\do\@makeother \do\$\do\&\do\#\do\^\do\_\do\%\do\~}
\def\mn@doi{\begingroup\mn@urlcharsother \@ifnextchar [ {\mn@doi@}
  {\mn@doi@[]}}
\def\mn@doi@[#1]#2{\def\@tempa{#1}\ifx\@tempa\@empty \href
  {http://dx.doi.org/#2} {doi:#2}\else \href {http://dx.doi.org/#2} {#1}\fi
  \endgroup}
\def\mn@eprint#1#2{\mn@eprint@#1:#2::\@nil}
\def\mn@eprint@arXiv#1{\href {http://arxiv.org/abs/#1} {{\tt arXiv:#1}}}
\def\mn@eprint@dblp#1{\href {http://dblp.uni-trier.de/rec/bibtex/#1.xml}
  {dblp:#1}}
\def\mn@eprint@#1:#2:#3:#4\@nil{\def\@tempa {#1}\def\@tempb {#2}\def\@tempc
  {#3}\ifx \@tempc \@empty \let \@tempc \@tempb \let \@tempb \@tempa \fi \ifx
  \@tempb \@empty \def\@tempb {arXiv}\fi \@ifundefined
  {mn@eprint@\@tempb}{\@tempb:\@tempc}{\expandafter \expandafter \csname
  mn@eprint@\@tempb\endcsname \expandafter{\@tempc}}}

\bibitem[\protect\citeauthoryear{{Algera} et~al.,}{{Algera}
  et~al.}{2024}]{algera2024}
{Algera} H. S.~B.,  et~al., 2024, \mn@doi [\mnras] {10.1093/mnras/stae1994},
  \href {https://ui.adsabs.harvard.edu/abs/2024MNRAS.tmp.1955A} {}

\bibitem[\protect\citeauthoryear{{Bakx} et~al.,}{{Bakx}
  et~al.}{2024}]{bakx2024}
{Bakx} T.~J.~L.~C.,  et~al., 2024, \mn@doi [\mnras] {10.1093/mnras/stae2409},
  \href {https://ui.adsabs.harvard.edu/abs/2024MNRAS.535.1533B} {535, 1533}

\bibitem[\protect\citeauthoryear{{Beelen}, {Cox}, {Benford}, {Dowell},
  {Kov{\'a}cs}, {Bertoldi}, {Omont}  \& {Carilli}}{{Beelen}
  et~al.}{2006}]{beelen2006}
{Beelen} A.,  {Cox} P.,  {Benford} D.~J.,  {Dowell} C.~D.,  {Kov{\'a}cs} A.,
  {Bertoldi} F.,  {Omont} A.,   {Carilli} C.~L.,  2006, \mn@doi [\apj]
  {10.1086/500636}, \href
  {https://ui.adsabs.harvard.edu/abs/2006ApJ...642..694B} {642, 694}

\bibitem[\protect\citeauthoryear{{Bendo} et~al.,}{{Bendo}
  et~al.}{2023}]{bendo2023}
{Bendo} G.~J.,  et~al., 2023, \mn@doi [\mnras] {10.1093/mnras/stac3771}, \href
  {https://ui.adsabs.harvard.edu/abs/2023MNRAS.522.2995B} {522, 2995}

\bibitem[\protect\citeauthoryear{{Berta} et~al.,}{{Berta}
  et~al.}{2023}]{berta2023}
{Berta} S.,  et~al., 2023, \mn@doi [\aap] {10.1051/0004-6361/202346803}, \href
  {https://ui.adsabs.harvard.edu/abs/2023A&A...678A..28B} {678, A28}

\bibitem[\protect\citeauthoryear{{Bianchi}}{{Bianchi}}{2013}]{bianchi2013}
{Bianchi} S.,  2013, \mn@doi [\aap] {10.1051/0004-6361/201220866}, \href
  {https://ui.adsabs.harvard.edu/abs/2013A&A...552A..89B} {552, A89}

\bibitem[\protect\citeauthoryear{{Bonavera} et~al.,}{{Bonavera}
  et~al.}{2020}]{bonavera2020}
{Bonavera} L.,  et~al., 2020, \mn@doi [\aap] {10.1051/0004-6361/202038050},
  \href {https://ui.adsabs.harvard.edu/abs/2020A&A...639A.128B} {639, A128}

\bibitem[\protect\citeauthoryear{{Bonavera}, {Cueli}, {Gonz{\'a}lez-Nuevo},
  {Ronconi}, {Migliaccio}, {Lapi}, {Casas}  \& {Crespo}}{{Bonavera}
  et~al.}{2021}]{bonavera2021}
{Bonavera} L.,  {Cueli} M.~M.,  {Gonz{\'a}lez-Nuevo} J.,  {Ronconi} T.,
  {Migliaccio} M.,  {Lapi} A.,  {Casas} J.~M.,   {Crespo} D.,  2021, \mn@doi
  [\aap] {10.1051/0004-6361/202141521}, \href
  {https://ui.adsabs.harvard.edu/abs/2021A&A...656A..99B} {656, A99}

\bibitem[\protect\citeauthoryear{{Boselli} et~al.,}{{Boselli}
  et~al.}{2012}]{boselli2012}
{Boselli} A.,  et~al., 2012, \mn@doi [\aap] {10.1051/0004-6361/201118602},
  \href {https://ui.adsabs.harvard.edu/abs/2012A&A...540A..54B} {540, A54}

\bibitem[\protect\citeauthoryear{{Boudet}, {Mutschke}, {Nayral}, {J{\"a}ger},
  {Bernard}, {Henning}  \& {Meny}}{{Boudet} et~al.}{2005}]{boudet2005}
{Boudet} N.,  {Mutschke} H.,  {Nayral} C.,  {J{\"a}ger} C.,  {Bernard} J.~P.,
  {Henning} T.,   {Meny} C.,  2005, \mn@doi [\apj] {10.1086/432966}, \href
  {https://ui.adsabs.harvard.edu/abs/2005ApJ...633..272B} {633, 272}

\bibitem[\protect\citeauthoryear{{Bouwens} et~al.,}{{Bouwens}
  et~al.}{2020}]{bouwens2020}
{Bouwens} R.,  et~al., 2020, \mn@doi [\apj] {10.3847/1538-4357/abb830}, \href
  {https://ui.adsabs.harvard.edu/abs/2020ApJ...902..112B} {902, 112}

\bibitem[\protect\citeauthoryear{{CASA Team} et~al.,}{{CASA Team}
  et~al.}{2022}]{casateam2022}
{CASA Team} et~al., 2022, \mn@doi [\pasp] {10.1088/1538-3873/ac9642}, \href
  {https://ui.adsabs.harvard.edu/abs/2022PASP..134k4501C} {134, 114501}

\bibitem[\protect\citeauthoryear{{Carlstrom} et~al.,}{{Carlstrom}
  et~al.}{2011}]{carlstrom2011}
{Carlstrom} J.~E.,  et~al., 2011, \mn@doi [\pasp] {10.1086/659879}, \href
  {https://ui.adsabs.harvard.edu/abs/2011PASP..123..568C} {123, 568}

\bibitem[\protect\citeauthoryear{{Cooper}, {Casey}, {Zavala}, {Champagne}, {da
  Cunha}, {Long}, {Spilker}  \& {Staguhn}}{{Cooper} et~al.}{2022}]{cooper2022}
{Cooper} O.~R.,  {Casey} C.~M.,  {Zavala} J.~A.,  {Champagne} J.~B.,  {da
  Cunha} E.,  {Long} A.~S.,  {Spilker} J.~S.,   {Staguhn} J.,  2022, \mn@doi
  [\apj] {10.3847/1538-4357/ac616d}, \href
  {https://ui.adsabs.harvard.edu/abs/2022ApJ...930...32C} {930, 32}

\bibitem[\protect\citeauthoryear{{Cortes} et~al.,}{{Cortes}
  et~al.}{2024}]{cortes2024}
{Cortes} P.~C.,  et~al., 2024, ALMA Cycle 11 Technical Handbook, ALMA Doc.
  11.3.
Joint ALMA Observatory, Vitacura , Santiago, Chile

\bibitem[\protect\citeauthoryear{{Coupeaud} et~al.,}{{Coupeaud}
  et~al.}{2011}]{coupeland2011}
{Coupeaud} A.,  et~al., 2011, \mn@doi [\aap] {10.1051/0004-6361/201116945},
  \href {https://ui.adsabs.harvard.edu/abs/2011A&A...535A.124C} {535, A124}

\bibitem[\protect\citeauthoryear{{Cox} et~al.,}{{Cox} et~al.}{2023}]{cox2023}
{Cox} P.,  et~al., 2023, \mn@doi [\aap] {10.1051/0004-6361/202346801}, \href
  {https://ui.adsabs.harvard.edu/abs/2023A&A...678A..26C} {678, A26}

\bibitem[\protect\citeauthoryear{{Cueli}, {Bonavera}, {Gonz{\'a}lez-Nuevo}  \&
  {Lapi}}{{Cueli} et~al.}{2021}]{cueli2021}
{Cueli} M.~M.,  {Bonavera} L.,  {Gonz{\'a}lez-Nuevo} J.,   {Lapi} A.,  2021,
  \mn@doi [\aap] {10.1051/0004-6361/202039326}, \href
  {https://ui.adsabs.harvard.edu/abs/2021A&A...645A.126C} {645, A126}

\bibitem[\protect\citeauthoryear{{Cueli} et~al.,}{{Cueli}
  et~al.}{2024}]{cueli2024}
{Cueli} M.~M.,  et~al., 2024, \mn@doi [\aap] {10.1051/0004-6361/202449315},
  \href {https://ui.adsabs.harvard.edu/abs/2024A&A...687A.300C} {687, A300}

\bibitem[\protect\citeauthoryear{{\VAN{Cunha}{da}{da}}~Cunha
  et~al.,}{{\VAN{Cunha}{da}{da}}~Cunha et~al.}{2013}]{dacunha2013}
{\VAN{Cunha}{da}{da}}~Cunha E.,  et~al., 2013, \mn@doi [\apj]
  {10.1088/0004-637X/766/1/13}, \href
  {https://ui.adsabs.harvard.edu/abs/2013ApJ...766...13D} {766, 13}

\bibitem[\protect\citeauthoryear{{\VAN{Cunha}{da}{da}}~Cunha
  et~al.,}{{\VAN{Cunha}{da}{da}}~Cunha et~al.}{2021}]{dacunha2021}
{\VAN{Cunha}{da}{da}}~Cunha E.,  et~al., 2021, \mn@doi [\apj]
  {10.3847/1538-4357/ac0ae0}, \href
  {https://ui.adsabs.harvard.edu/abs/2021ApJ...919...30D} {919, 30}

\bibitem[\protect\citeauthoryear{{Demyk} et~al.,}{{Demyk}
  et~al.}{2017a}]{demyk2017a}
{Demyk} K.,  et~al., 2017a, \mn@doi [\aap] {10.1051/0004-6361/201730944}, \href
  {https://ui.adsabs.harvard.edu/abs/2017A&A...606A..50D} {606, A50}

\bibitem[\protect\citeauthoryear{{Demyk} et~al.,}{{Demyk}
  et~al.}{2017b}]{demyk2017b}
{Demyk} K.,  et~al., 2017b, \mn@doi [\aap] {10.1051/0004-6361/201730944}, \href
  {https://ui.adsabs.harvard.edu/abs/2017A&A...606A..50D} {606, A50}

\bibitem[\protect\citeauthoryear{{D{\'e}sert} et~al.,}{{D{\'e}sert}
  et~al.}{2008}]{desert2008}
{D{\'e}sert} F.~X.,  et~al., 2008, \mn@doi [\aap] {10.1051/0004-6361:20078701},
  \href {https://ui.adsabs.harvard.edu/abs/2008A&A...481..411D} {481, 411}

\bibitem[\protect\citeauthoryear{{Draine}}{{Draine}}{2003}]{draine2003}
{Draine} B.~T.,  2003, \mn@doi [\araa]
  {10.1146/annurev.astro.41.011802.094840}, \href
  {https://ui.adsabs.harvard.edu/abs/2003ARA&A..41..241D} {41, 241}

\bibitem[\protect\citeauthoryear{{Draine}}{{Draine}}{2006}]{draine2006}
{Draine} B.~T.,  2006, \mn@doi [\apj] {10.1086/498130}, \href
  {https://ui.adsabs.harvard.edu/abs/2006ApJ...636.1114D} {636, 1114}

\bibitem[\protect\citeauthoryear{{Draine} \& {Lazarian}}{{Draine} \&
  {Lazarian}}{1998}]{draine1998}
{Draine} B.~T.,  {Lazarian} A.,  1998, \mn@doi [\apj] {10.1086/306387}, \href
  {https://ui.adsabs.harvard.edu/abs/1998ApJ...508..157D} {508, 157}

\bibitem[\protect\citeauthoryear{{Duchesne} et~al.,}{{Duchesne}
  et~al.}{2024}]{duchesne2024}
{Duchesne} S.~W.,  et~al., 2024, \mn@doi [\pasa] {10.1017/pasa.2023.60}, \href
  {https://ui.adsabs.harvard.edu/abs/2024PASA...41....3D} {41, e003}

\bibitem[\protect\citeauthoryear{{Dudzevi{\v{c}}i{\={u}}t{\.{e}}}
  et~al.,}{{Dudzevi{\v{c}}i{\={u}}t{\.{e}}} et~al.}{2021}]{dudzeviciute2021}
{Dudzevi{\v{c}}i{\={u}}t{\.{e}}} U.,  et~al., 2021, \mn@doi [\mnras]
  {10.1093/mnras/staa3285}, \href
  {https://ui.adsabs.harvard.edu/abs/2021MNRAS.500..942D} {500, 942}

\bibitem[\protect\citeauthoryear{{Dupac} et~al.,}{{Dupac}
  et~al.}{2003}]{dupac2003}
{Dupac} X.,  et~al., 2003, \mn@doi [\aap] {10.1051/0004-6361:20030575}, \href
  {https://ui.adsabs.harvard.edu/abs/2003A&A...404L..11D} {404, L11}

\bibitem[\protect\citeauthoryear{{Dye} et~al.,}{{Dye} et~al.}{2015}]{dye2015}
{Dye} S.,  et~al., 2015, \mn@doi [\mnras] {10.1093/mnras/stv1442}, \href
  {https://ui.adsabs.harvard.edu/abs/2015MNRAS.452.2258D} {452, 2258}

\bibitem[\protect\citeauthoryear{{Dye} et~al.,}{{Dye} et~al.}{2022}]{dye2022}
{Dye} S.,  et~al., 2022, \mn@doi [\mnras] {10.1093/mnras/stab3569}, \href
  {https://ui.adsabs.harvard.edu/abs/2022MNRAS.510.3734D} {510, 3734}

\bibitem[\protect\citeauthoryear{{Eales}}{{Eales}}{2015}]{eales2015}
{Eales} S.~A.,  2015, \mn@doi [\mnras] {10.1093/mnras/stu2214}, \href
  {https://ui.adsabs.harvard.edu/abs/2015MNRAS.446.3224E} {446, 3224}

\bibitem[\protect\citeauthoryear{{Eales} et~al.,}{{Eales}
  et~al.}{2010}]{eales2010}
{Eales} S.,  et~al., 2010, \mn@doi [\pasp] {10.1086/653086}, \href
  {https://ui.adsabs.harvard.edu/abs/2010PASP..122..499E} {122, 499}

\bibitem[\protect\citeauthoryear{{Galametz} et~al.,}{{Galametz}
  et~al.}{2014}]{galametz2014}
{Galametz} M.,  et~al., 2014, \mn@doi [\mnras] {10.1093/mnras/stu113}, \href
  {https://ui.adsabs.harvard.edu/abs/2014MNRAS.439.2542G} {439, 2542}

\bibitem[\protect\citeauthoryear{{Galliano}}{{Galliano}}{2022}]{galliano2022}
{Galliano} F.,  2022, \mn@doi [Habilitation Thesis]
  {10.48550/arXiv.2202.01868}, \href
  {https://ui.adsabs.harvard.edu/abs/2022HabT.........1G} {p.~1}

\bibitem[\protect\citeauthoryear{{Gonz{\'a}lez-Nuevo}
  et~al.,}{{Gonz{\'a}lez-Nuevo} et~al.}{2017}]{gonzaleznuevo2017}
{Gonz{\'a}lez-Nuevo} J.,  et~al., 2017, \mn@doi [\jcap]
  {10.1088/1475-7516/2017/10/024}, \href
  {https://ui.adsabs.harvard.edu/abs/2017JCAP...10..024G} {2017, 024}

\bibitem[\protect\citeauthoryear{{Gonz{\'a}lez-Nuevo}, {Cueli}, {Bonavera},
  {Lapi}, {Migliaccio}, {Arg{\"u}eso}  \& {Toffolatti}}{{Gonz{\'a}lez-Nuevo}
  et~al.}{2021}]{gonzaleznuevo2021}
{Gonz{\'a}lez-Nuevo} J.,  {Cueli} M.~M.,  {Bonavera} L.,  {Lapi} A.,
  {Migliaccio} M.,  {Arg{\"u}eso} F.,   {Toffolatti} L.,  2021, \mn@doi [\aap]
  {10.1051/0004-6361/202039043}, \href
  {https://ui.adsabs.harvard.edu/abs/2021A&A...646A.152G} {646, A152}

\bibitem[\protect\citeauthoryear{{Gordon} et~al.,}{{Gordon}
  et~al.}{2014}]{gordon2014}
{Gordon} K.~D.,  et~al., 2014, \mn@doi [\apj] {10.1088/0004-637X/797/2/85},
  \href {https://ui.adsabs.harvard.edu/abs/2014ApJ...797...85G} {797, 85}

\bibitem[\protect\citeauthoryear{{Gordon} et~al.,}{{Gordon}
  et~al.}{2021}]{gordon2021}
{Gordon} Y.~A.,  et~al., 2021, \mn@doi [\apjs] {10.3847/1538-4365/ac05c0},
  \href {https://ui.adsabs.harvard.edu/abs/2021ApJS..255...30G} {255, 30}

\bibitem[\protect\citeauthoryear{{Grillo}, {Lombardi}  \& {Bertin}}{{Grillo}
  et~al.}{2008}]{grillo2008}
{Grillo} C.,  {Lombardi} M.,   {Bertin} G.,  2008, \mn@doi [\aap]
  {10.1051/0004-6361:20077534}, \href
  {https://ui.adsabs.harvard.edu/abs/2008A&A...477..397G} {477, 397}

\bibitem[\protect\citeauthoryear{{Hagimoto} et~al.,}{{Hagimoto}
  et~al.}{2023}]{hagimoto2023}
{Hagimoto} M.,  et~al., 2023, \mn@doi [\mnras] {10.1093/mnras/stad784}, \href
  {https://ui.adsabs.harvard.edu/abs/2023MNRAS.521.5508H} {521, 5508}

\bibitem[\protect\citeauthoryear{{H{\"o}gbom}}{{H{\"o}gbom}}{1974}]{hogbom1974}
{H{\"o}gbom} J.~A.,  1974, \aaps, \href
  {https://ui.adsabs.harvard.edu/abs/1974A&AS...15..417H} {15, 417}

\bibitem[\protect\citeauthoryear{{Hunt} et~al.,}{{Hunt}
  et~al.}{2015}]{hunt2015}
{Hunt} L.~K.,  et~al., 2015, \mn@doi [\aap] {10.1051/0004-6361/201424734},
  \href {https://ui.adsabs.harvard.edu/abs/2015A&A...576A..33H} {576, A33}

\bibitem[\protect\citeauthoryear{{Ismail} et~al.,}{{Ismail}
  et~al.}{2023}]{ismail2023}
{Ismail} D.,  et~al., 2023, \mn@doi [\aap] {10.1051/0004-6361/202346804}, \href
  {https://ui.adsabs.harvard.edu/abs/2023A&A...678A..27I} {678, A27}

\bibitem[\protect\citeauthoryear{{James}, {Dunne}, {Eales}  \&
  {Edmunds}}{{James} et~al.}{2002}]{james2002}
{James} A.,  {Dunne} L.,  {Eales} S.,   {Edmunds} M.~G.,  2002, \mn@doi
  [\mnras] {10.1046/j.1365-8711.2002.05660.x}, \href
  {https://ui.adsabs.harvard.edu/abs/2002MNRAS.335..753J} {335, 753}

\bibitem[\protect\citeauthoryear{{Juvela} \& {Ysard}}{{Juvela} \&
  {Ysard}}{2012}]{juvela2012}
{Juvela} M.,  {Ysard} N.,  2012, \mn@doi [\aap] {10.1051/0004-6361/201118596},
  \href {https://ui.adsabs.harvard.edu/abs/2012A&A...541A..33J} {541, A33}

\bibitem[\protect\citeauthoryear{{Juvela} et~al.,}{{Juvela}
  et~al.}{2011}]{juvela2011}
{Juvela} M.,  et~al., 2011, \mn@doi [\aap] {10.1051/0004-6361/201015916}, \href
  {https://ui.adsabs.harvard.edu/abs/2011A&A...527A.111J} {527, A111}

\bibitem[\protect\citeauthoryear{{Juvela}, {Montillaud}, {Ysard}  \&
  {Lunttila}}{{Juvela} et~al.}{2013}]{juvela2013}
{Juvela} M.,  {Montillaud} J.,  {Ysard} N.,   {Lunttila} T.,  2013, \mn@doi
  [\aap] {10.1051/0004-6361/201220910}, \href
  {https://ui.adsabs.harvard.edu/abs/2013A&A...556A..63J} {556, A63}

\bibitem[\protect\citeauthoryear{{Kelly}, {Shetty}, {Stutz}, {Kauffmann},
  {Goodman}  \& {Launhardt}}{{Kelly} et~al.}{2012}]{kelly2012}
{Kelly} B.~C.,  {Shetty} R.,  {Stutz} A.~M.,  {Kauffmann} J.,  {Goodman} A.~A.,
    {Launhardt} R.,  2012, \mn@doi [\apj] {10.1088/0004-637X/752/1/55}, \href
  {https://ui.adsabs.harvard.edu/abs/2012ApJ...752...55K} {752, 55}

\bibitem[\protect\citeauthoryear{{Kirkpatrick} et~al.,}{{Kirkpatrick}
  et~al.}{2014}]{kirkpatrick2014}
{Kirkpatrick} A.,  et~al., 2014, \mn@doi [\apj] {10.1088/0004-637X/789/2/130},
  \href {https://ui.adsabs.harvard.edu/abs/2014ApJ...789..130K} {789, 130}

\bibitem[\protect\citeauthoryear{{Lamperti} et~al.,}{{Lamperti}
  et~al.}{2019}]{lamperti2019}
{Lamperti} I.,  et~al., 2019, \mn@doi [\mnras] {10.1093/mnras/stz2311}, \href
  {https://ui.adsabs.harvard.edu/abs/2019MNRAS.489.4389L} {489, 4389}

\bibitem[\protect\citeauthoryear{{Li}}{{Li}}{2004}]{li2004}
{Li} A.,  2004, in {Block} D.~L.,  {Puerari} I.,  {Freeman} K.~C.,  {Groess}
  R.,   {Block} E.~K.,  eds,  Astrophysics and Space Science Library Vol. 319,
  Penetrating Bars Through Masks of Cosmic Dust. p.~535,
  \mn@doi{10.1007/978-1-4020-2862-5_47}

\bibitem[\protect\citeauthoryear{{Liao} et~al.,}{{Liao}
  et~al.}{2024}]{liao2024}
{Liao} C.-L.,  et~al., 2024, \mn@doi [\apj] {10.3847/1538-4357/ad148c}, \href
  {https://ui.adsabs.harvard.edu/abs/2024ApJ...961..226L} {961, 226}

\bibitem[\protect\citeauthoryear{{Magnelli} et~al.,}{{Magnelli}
  et~al.}{2012}]{magnelli2012}
{Magnelli} B.,  et~al., 2012, \mn@doi [\aap] {10.1051/0004-6361/201118312},
  \href {https://ui.adsabs.harvard.edu/abs/2012A&A...539A.155M} {539, A155}

\bibitem[\protect\citeauthoryear{{McKay}, {Barger}, {Cowie}, {Bauer}  \&
  {Rosenthal}}{{McKay} et~al.}{2023}]{mckay2023}
{McKay} S.~J.,  {Barger} A.~J.,  {Cowie} L.~L.,  {Bauer} F.~E.,   {Rosenthal}
  M.~J.~N.,  2023, \mn@doi [\apj] {10.3847/1538-4357/acd1e5}, \href
  {https://ui.adsabs.harvard.edu/abs/2023ApJ...951...48M} {951, 48}

\bibitem[\protect\citeauthoryear{{McMullin}, {Waters}, {Schiebel}, {Young}  \&
  {Golap}}{{McMullin} et~al.}{2007}]{mcmullin2007}
{McMullin} J.~P.,  {Waters} B.,  {Schiebel} D.,  {Young} W.,   {Golap} K.,
  2007, in {Shaw} R.~A.,  {Hill} F.,   {Bell} D.~J.,  eds,  Astronomical
  Society of the Pacific Conference Series Vol. 376, Astronomical Data Analysis
  Software and Systems XVI. p.~127

\bibitem[\protect\citeauthoryear{{Miller} et~al.,}{{Miller}
  et~al.}{2018}]{miller2018}
{Miller} T.~B.,  et~al., 2018, \mn@doi [\nat] {10.1038/s41586-018-0025-2},
  \href {https://ui.adsabs.harvard.edu/abs/2018Natur.556..469M} {556, 469}

\bibitem[\protect\citeauthoryear{{Negrello} et~al.,}{{Negrello}
  et~al.}{2017}]{negrello2017}
{Negrello} M.,  et~al., 2017, \mn@doi [\mnras] {10.1093/mnras/stw2911}, \href
  {https://ui.adsabs.harvard.edu/abs/2017MNRAS.465.3558N} {465, 3558}

\bibitem[\protect\citeauthoryear{{Oteo} et~al.,}{{Oteo}
  et~al.}{2018}]{oteo2018}
{Oteo} I.,  et~al., 2018, \mn@doi [\apj] {10.3847/1538-4357/aaa1f1}, \href
  {https://ui.adsabs.harvard.edu/abs/2018ApJ...856...72O} {856, 72}

\bibitem[\protect\citeauthoryear{{Paradis} et~al.,}{{Paradis}
  et~al.}{2010}]{paradis2010}
{Paradis} D.,  et~al., 2010, \mn@doi [\aap] {10.1051/0004-6361/201015301},
  \href {https://ui.adsabs.harvard.edu/abs/2010A&A...520L...8P} {520, L8}

\bibitem[\protect\citeauthoryear{{Pilbratt} et~al.,}{{Pilbratt}
  et~al.}{2010}]{pilbratt2010}
{Pilbratt} G.~L.,  et~al., 2010, \mn@doi [\aap] {10.1051/0004-6361/201014759},
  \href {https://ui.adsabs.harvard.edu/abs/2010A&A...518L...1P} {518, L1}

\bibitem[\protect\citeauthoryear{{Planck Collaboration XIV}}{{Planck
  Collaboration XIV}}{2014}]{planck2014}
{Planck Collaboration XIV} 2014, \mn@doi [\aap] {10.1051/0004-6361/201322367},
  \href {https://ui.adsabs.harvard.edu/abs/2014A&A...564A..45P} {564, A45}

\bibitem[\protect\citeauthoryear{{Planck Collaboration XXII}}{{Planck
  Collaboration XXII}}{2015}]{planck2015}
{Planck Collaboration XXII} 2015, \mn@doi [\aap] {10.1051/0004-6361/201424088},
  \href {https://ui.adsabs.harvard.edu/abs/2015A&A...576A.107P} {576, A107}

\bibitem[\protect\citeauthoryear{{Planck Collaboration XXIII}}{{Planck
  Collaboration XXIII}}{2011}]{planckxxiii2011}
{Planck Collaboration XXIII} 2011, \mn@doi [\aap]
  {10.1051/0004-6361/201116472}, \href
  {https://ui.adsabs.harvard.edu/abs/2011A&A...536A..23P} {536, A23}

\bibitem[\protect\citeauthoryear{{Rangwala} et~al.,}{{Rangwala}
  et~al.}{2011}]{rangwala2011}
{Rangwala} N.,  et~al., 2011, \mn@doi [\apj] {10.1088/0004-637X/743/1/94},
  \href {https://ui.adsabs.harvard.edu/abs/2011ApJ...743...94R} {743, 94}

\bibitem[\protect\citeauthoryear{{Reuter} et~al.,}{{Reuter}
  et~al.}{2020}]{reuter2020}
{Reuter} C.,  et~al., 2020, \mn@doi [\apj] {10.3847/1538-4357/abb599}, \href
  {https://ui.adsabs.harvard.edu/abs/2020ApJ...902...78R} {902, 78}

\bibitem[\protect\citeauthoryear{{Riechers}, {Cooray}, {P{\'e}rez-Fournon}  \&
  {Neri}}{{Riechers} et~al.}{2021}]{riechers2021}
{Riechers} D.~A.,  {Cooray} A.,  {P{\'e}rez-Fournon} I.,   {Neri} R.,  2021,
  \mn@doi [\apj] {10.3847/1538-4357/abf6d7}, \href
  {https://ui.adsabs.harvard.edu/abs/2021ApJ...913..141R} {913, 141}

\bibitem[\protect\citeauthoryear{{Schreiber}, {Elbaz}, {Pannella}, {Ciesla},
  {Wang}  \& {Franco}}{{Schreiber} et~al.}{2018}]{schreiber2018}
{Schreiber} C.,  {Elbaz} D.,  {Pannella} M.,  {Ciesla} L.,  {Wang} T.,
  {Franco} M.,  2018, \mn@doi [\aap] {10.1051/0004-6361/201731506}, \href
  {https://ui.adsabs.harvard.edu/abs/2018A&A...609A..30S} {609, A30}

\bibitem[\protect\citeauthoryear{{Scoville} et~al.,}{{Scoville}
  et~al.}{2017}]{scoville2017}
{Scoville} N.,  et~al., 2017, \mn@doi [\apj] {10.3847/1538-4357/836/1/66},
  \href {https://ui.adsabs.harvard.edu/abs/2017ApJ...836...66S} {836, 66}

\bibitem[\protect\citeauthoryear{{Shetty}, {Kauffmann}, {Schnee}  \&
  {Goodman}}{{Shetty} et~al.}{2009a}]{shetty2009a}
{Shetty} R.,  {Kauffmann} J.,  {Schnee} S.,   {Goodman} A.~A.,  2009a, \mn@doi
  [\apj] {10.1088/0004-637X/696/1/676}, \href
  {https://ui.adsabs.harvard.edu/abs/2009ApJ...696..676S} {696, 676}

\bibitem[\protect\citeauthoryear{{Shetty}, {Kauffmann}, {Schnee}, {Goodman}  \&
  {Ercolano}}{{Shetty} et~al.}{2009b}]{shetty2009b}
{Shetty} R.,  {Kauffmann} J.,  {Schnee} S.,  {Goodman} A.~A.,   {Ercolano} B.,
  2009b, \mn@doi [\apj] {10.1088/0004-637X/696/2/2234}, \href
  {https://ui.adsabs.harvard.edu/abs/2009ApJ...696.2234S} {696, 2234}

\bibitem[\protect\citeauthoryear{{Simpson} et~al.,}{{Simpson}
  et~al.}{2017}]{simpson2017}
{Simpson} J.~M.,  et~al., 2017, \mn@doi [\apj] {10.3847/1538-4357/aa65d0},
  \href {https://ui.adsabs.harvard.edu/abs/2017ApJ...839...58S} {839, 58}

\bibitem[\protect\citeauthoryear{{Smith} et~al.,}{{Smith}
  et~al.}{2012}]{smith2012}
{Smith} M.~W.~L.,  et~al., 2012, \mn@doi [\apj] {10.1088/0004-637X/756/1/40},
  \href {https://ui.adsabs.harvard.edu/abs/2012ApJ...756...40S} {756, 40}

\bibitem[\protect\citeauthoryear{{Spilker} et~al.,}{{Spilker}
  et~al.}{2016}]{spilker2016}
{Spilker} J.~S.,  et~al., 2016, \mn@doi [\apj] {10.3847/0004-637X/826/2/112},
  \href {https://ui.adsabs.harvard.edu/abs/2016ApJ...826..112S} {826, 112}

\bibitem[\protect\citeauthoryear{{Swinbank} et~al.,}{{Swinbank}
  et~al.}{2010}]{swinbank2010}
{Swinbank} A.~M.,  et~al., 2010, \mn@doi [\nat] {10.1038/nature08880}, \href
  {https://ui.adsabs.harvard.edu/abs/2010Natur.464..733S} {464, 733}

\bibitem[\protect\citeauthoryear{{Tabatabaei} et~al.,}{{Tabatabaei}
  et~al.}{2014}]{tabatabaei2014}
{Tabatabaei} F.~S.,  et~al., 2014, \mn@doi [\aap]
  {10.1051/0004-6361/201321441}, \href
  {https://ui.adsabs.harvard.edu/abs/2014A&A...561A..95T} {561, A95}

\bibitem[\protect\citeauthoryear{{Treu}}{{Treu}}{2010}]{treu2010}
{Treu} T.,  2010, \mn@doi [\araa] {10.1146/annurev-astro-081309-130924}, \href
  {https://ui.adsabs.harvard.edu/abs/2010ARA&A..48...87T} {48, 87}

\bibitem[\protect\citeauthoryear{{Tripodi} et~al.,}{{Tripodi}
  et~al.}{2024}]{tripodi2024}
{Tripodi} R.,  et~al., 2024, \mn@doi [arXiv e-prints]
  {10.48550/arXiv.2401.04211}, \href
  {https://ui.adsabs.harvard.edu/abs/2024arXiv240104211T} {p. arXiv:2401.04211}

\bibitem[\protect\citeauthoryear{{Tsukui}, {Wisnioski}, {Krumholz}  \&
  {Battisti}}{{Tsukui} et~al.}{2023}]{tsukui2023}
{Tsukui} T.,  {Wisnioski} E.,  {Krumholz} M.~R.,   {Battisti} A.,  2023,
  \mn@doi [\mnras] {10.1093/mnras/stad1464}, \href
  {https://ui.adsabs.harvard.edu/abs/2023MNRAS.523.4654T} {523, 4654}

\bibitem[\protect\citeauthoryear{{Urquhart} et~al.,}{{Urquhart}
  et~al.}{2022}]{urquhart2022}
{Urquhart} S.~A.,  et~al., 2022, \mn@doi [\mnras] {10.1093/mnras/stac150},
  \href {https://ui.adsabs.harvard.edu/abs/2022MNRAS.511.3017U} {511, 3017}

\bibitem[\protect\citeauthoryear{{Valiante} et~al.,}{{Valiante}
  et~al.}{2016}]{valiante2016}
{Valiante} E.,  et~al., 2016, \mn@doi [\mnras] {10.1093/mnras/stw1806}, \href
  {https://ui.adsabs.harvard.edu/abs/2016MNRAS.462.3146V} {462, 3146}

\bibitem[\protect\citeauthoryear{{Vieira} et~al.,}{{Vieira}
  et~al.}{2013}]{viera2013}
{Vieira} J.~D.,  et~al., 2013, \mn@doi [\nat] {10.1038/nature12001}, \href
  {https://ui.adsabs.harvard.edu/abs/2013Natur.495..344V} {495, 344}

\bibitem[\protect\citeauthoryear{{Ward}, {Eales}, {Ivison}  \&
  {Arumugam}}{{Ward} et~al.}{2024}]{ward2024}
{Ward} B.~A.,  {Eales} S.~A.,  {Ivison} R.~J.,   {Arumugam} V.,  2024, \mn@doi
  [\mnras] {10.1093/mnras/stae405}, \href
  {https://ui.adsabs.harvard.edu/abs/2024MNRAS.530.4887W} {530, 4887}

\bibitem[\protect\citeauthoryear{{Wei{\ss}} et~al.,}{{Wei{\ss}}
  et~al.}{2013}]{weiss2013}
{Wei{\ss}} A.,  et~al., 2013, \mn@doi [\apj] {10.1088/0004-637X/767/1/88},
  \href {https://ui.adsabs.harvard.edu/abs/2013ApJ...767...88W} {767, 88}

\bibitem[\protect\citeauthoryear{{Whitworth} et~al.,}{{Whitworth}
  et~al.}{2019}]{whitworth2019}
{Whitworth} A.~P.,  et~al., 2019, \mn@doi [\mnras] {10.1093/mnras/stz2166},
  \href {https://ui.adsabs.harvard.edu/abs/2019MNRAS.489.5436W} {489, 5436}

\bibitem[\protect\citeauthoryear{{Witstok}, {Jones}, {Maiolino}, {Smit}  \&
  {Schneider}}{{Witstok} et~al.}{2023}]{witstok2023}
{Witstok} J.,  {Jones} G.~C.,  {Maiolino} R.,  {Smit} R.,   {Schneider} R.,
  2023, \mn@doi [\mnras] {10.1093/mnras/stad1470}, \href
  {https://ui.adsabs.harvard.edu/abs/2023MNRAS.523.3119W} {523, 3119}

\bibitem[\protect\citeauthoryear{{Yang} \& {Phillips}}{{Yang} \&
  {Phillips}}{2007}]{yang2007}
{Yang} M.,  {Phillips} T.,  2007, \mn@doi [\apj] {10.1086/514810}, \href
  {https://ui.adsabs.harvard.edu/abs/2007ApJ...662..284Y} {662, 284}

\makeatother
\end{thebibliography}

%\begin{thebibliography}{99}
%\end{thebibliography}

%%%%%%%%%%%%%%%%%%%%%%%%%%%%%%%%%%%%%%%%%%%%%%%%%%

%%%%%%%%%%%%%%%%% APPENDICES %%%%%%%%%%%%%%%%%%%%%

\appendix

\section{ALMA Band 5 images}
\label{a_map}

\begin{figure*}
\begin{center}
\includegraphics[width=5.5cm]{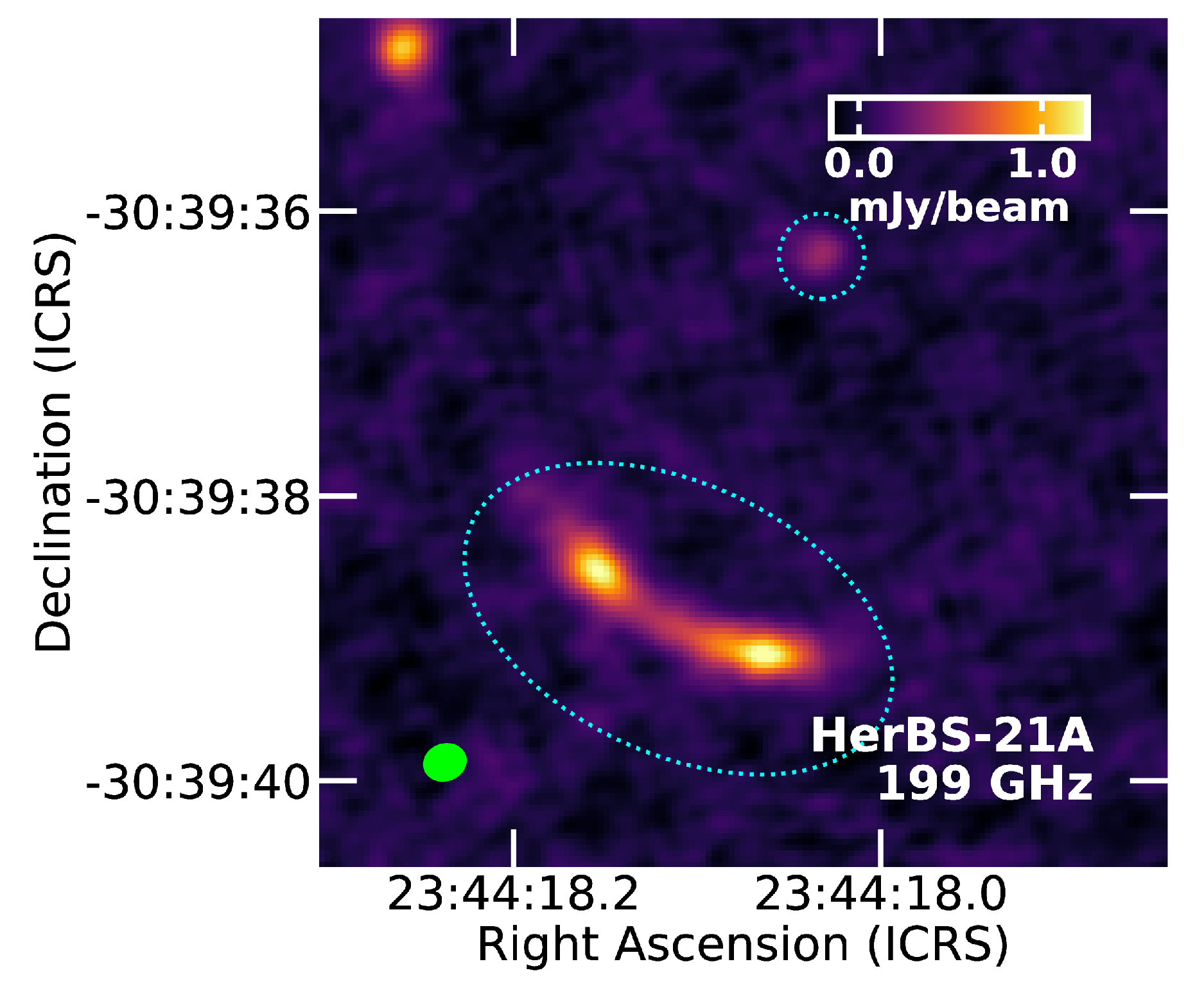} ~
\includegraphics[width=5.5cm]{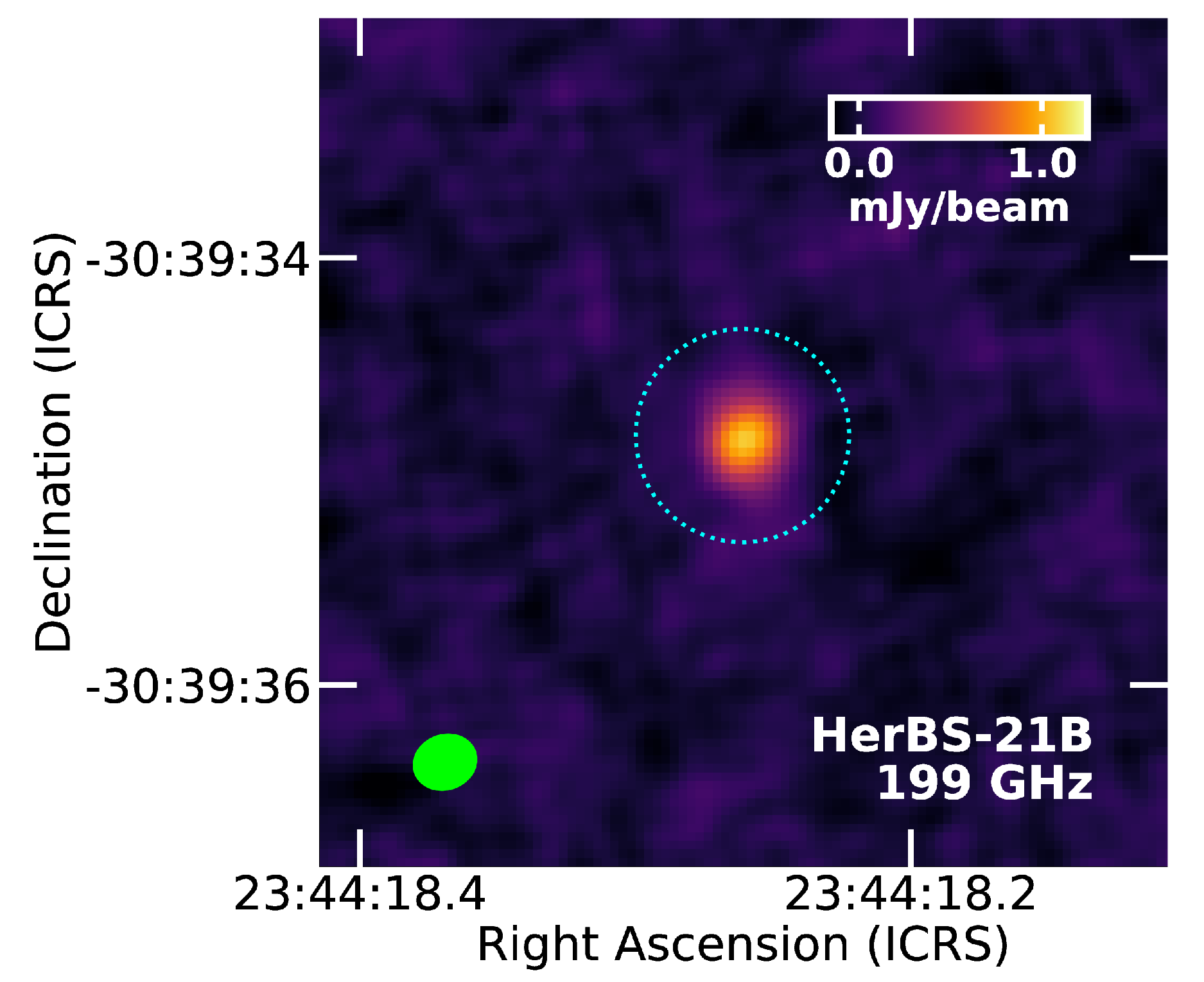} ~
\includegraphics[width=5.5cm]{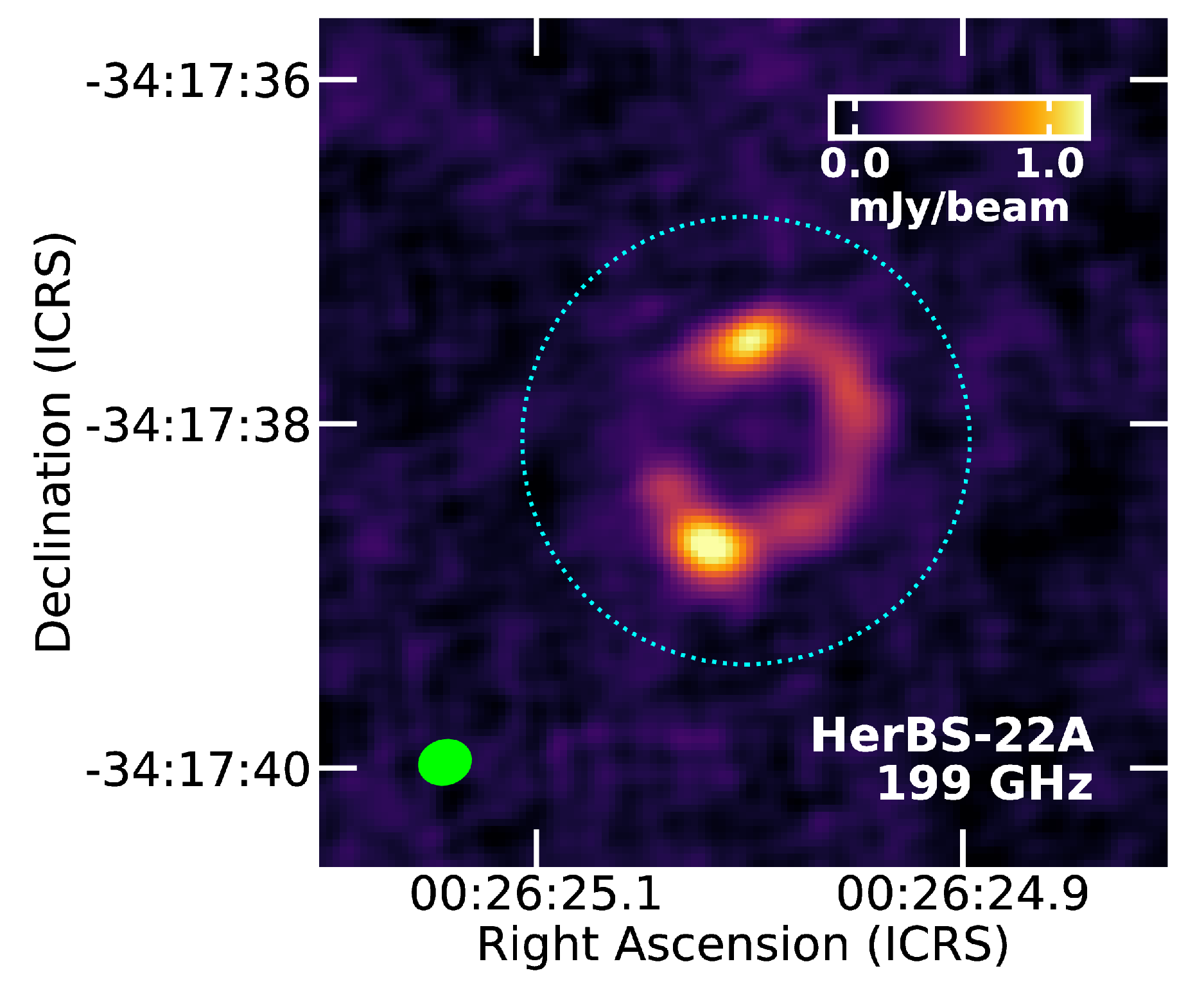} \\ ~ \\
\includegraphics[width=5.5cm]{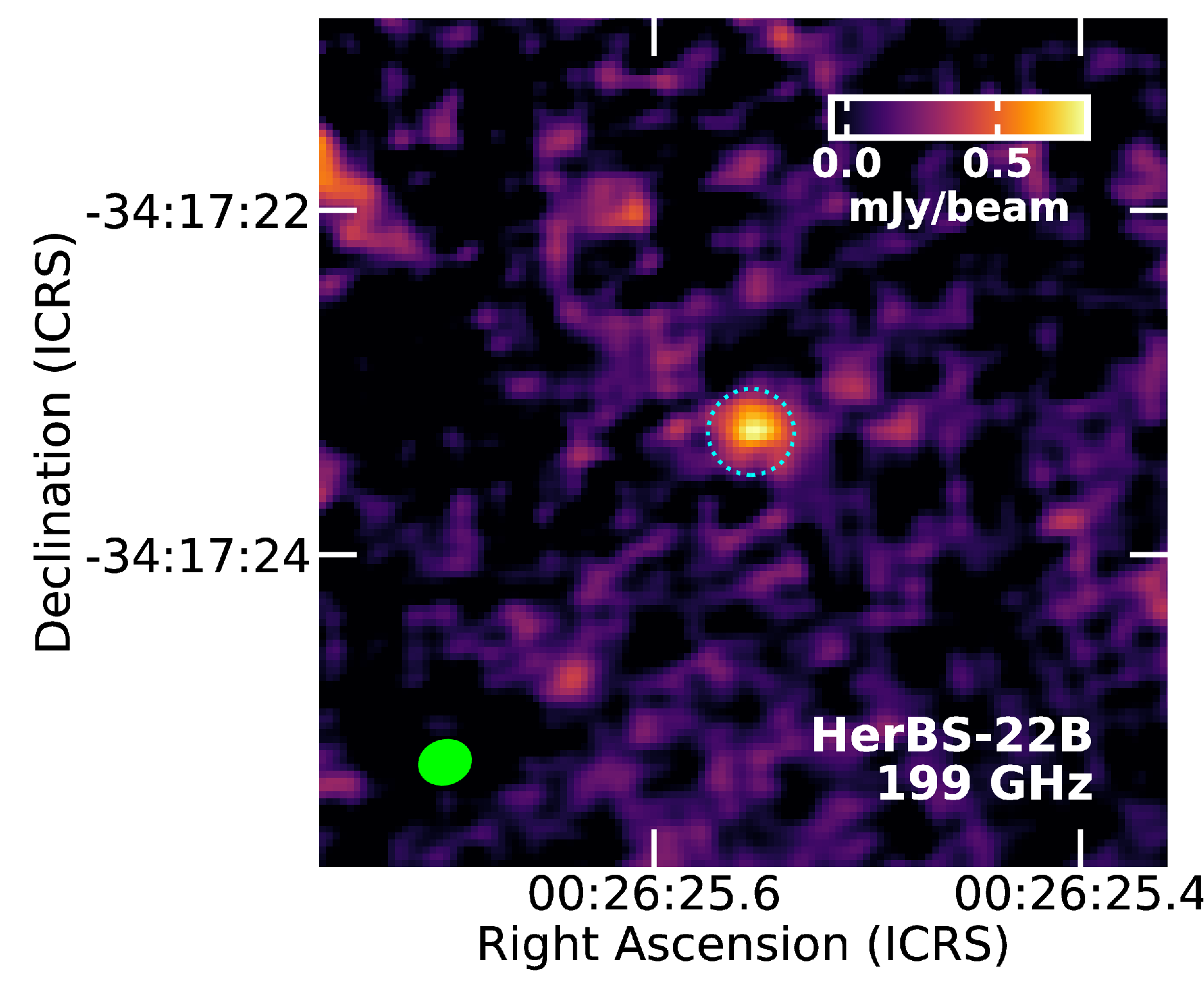} ~
\includegraphics[width=5.5cm]{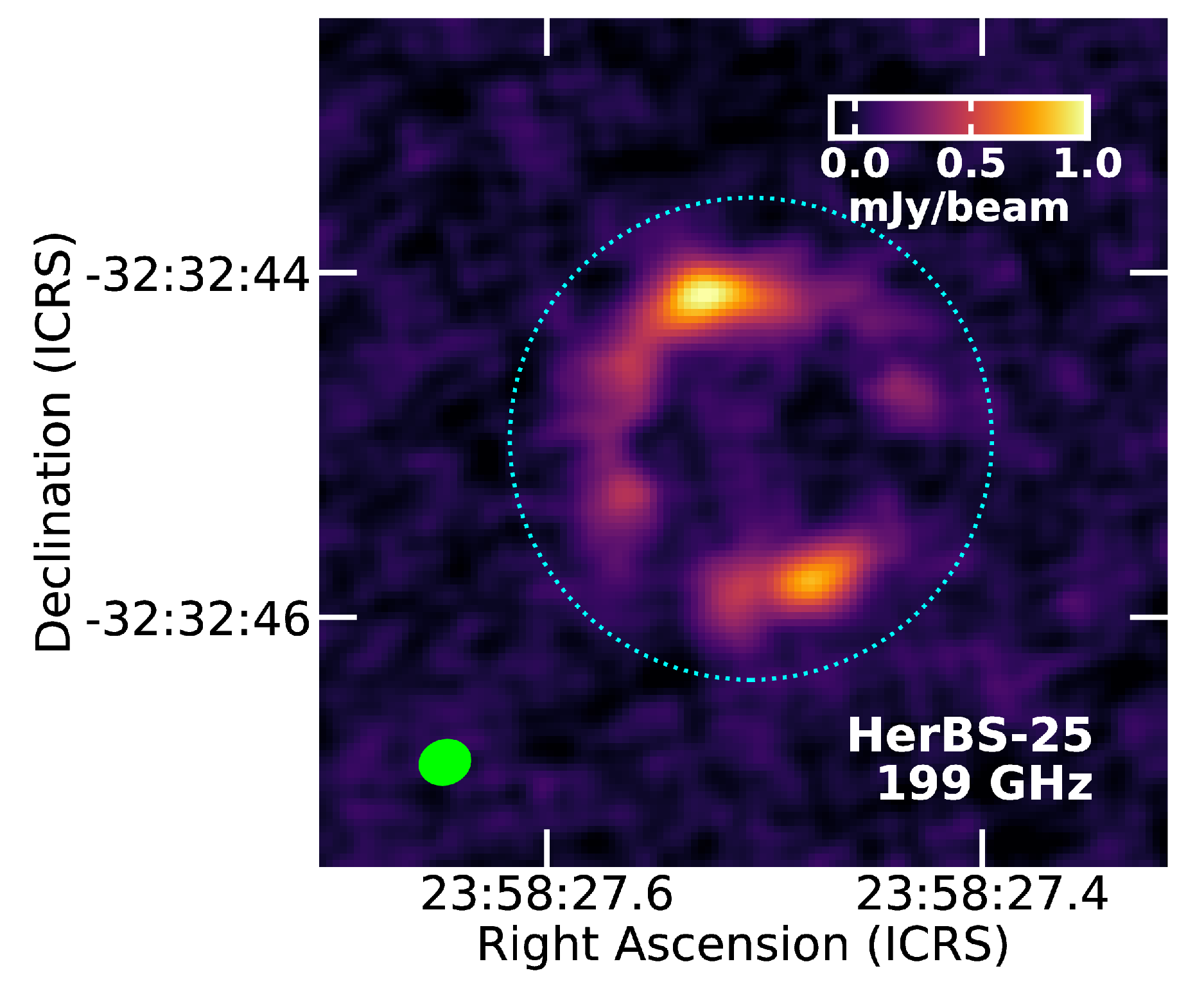} ~
\includegraphics[width=5.5cm]{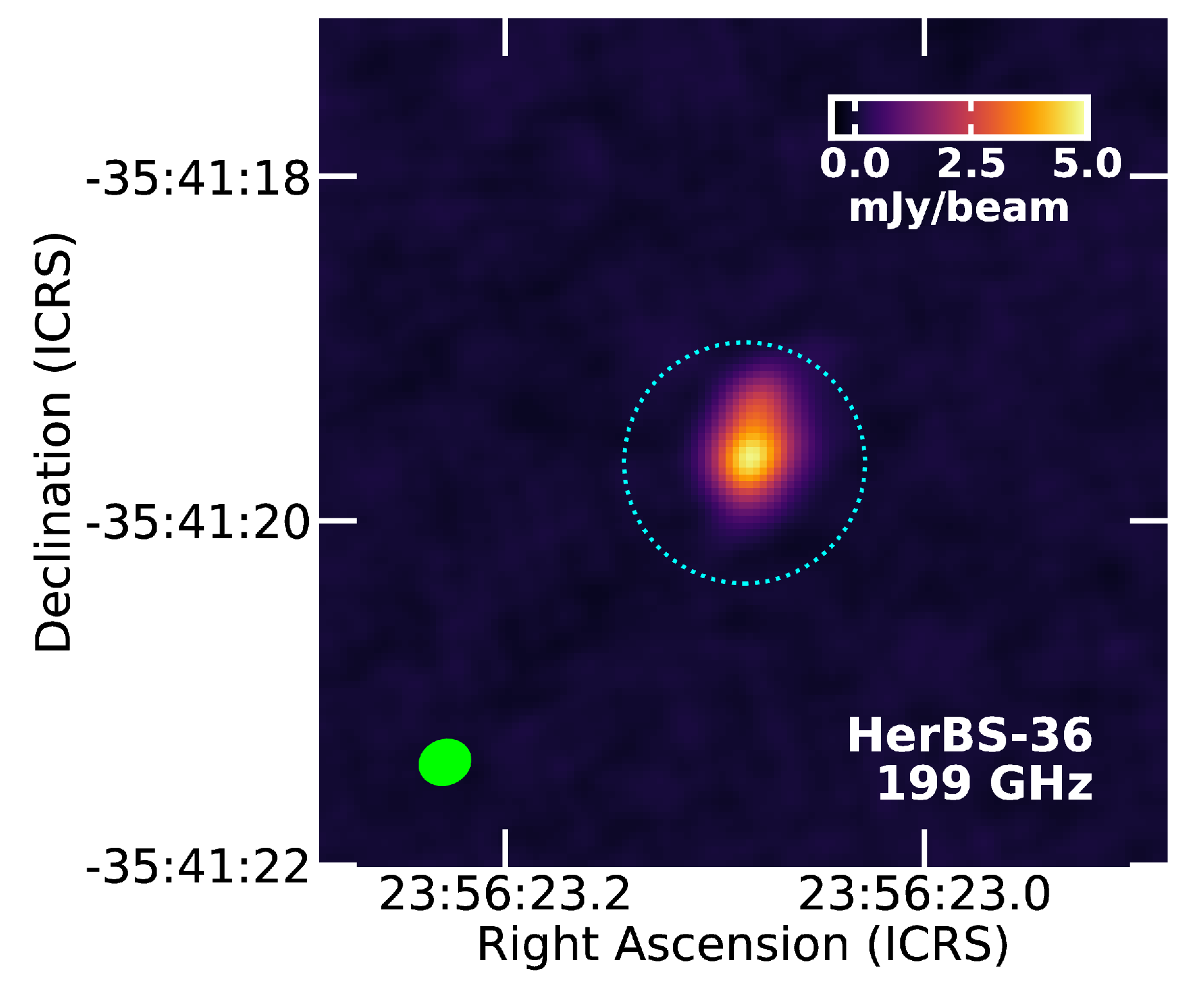} \\ ~ \\
\includegraphics[width=5.5cm]{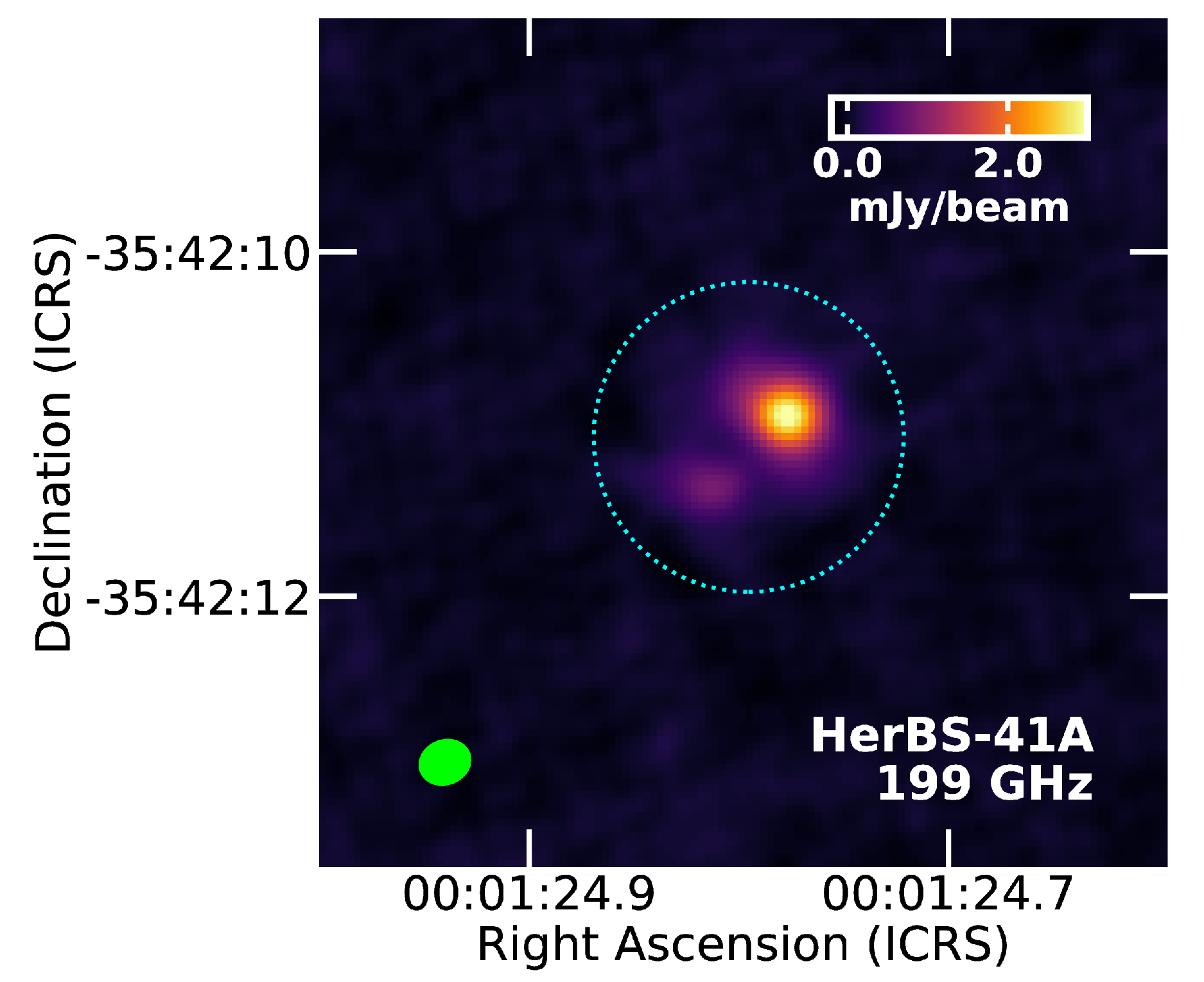} ~
\includegraphics[width=5.5cm]{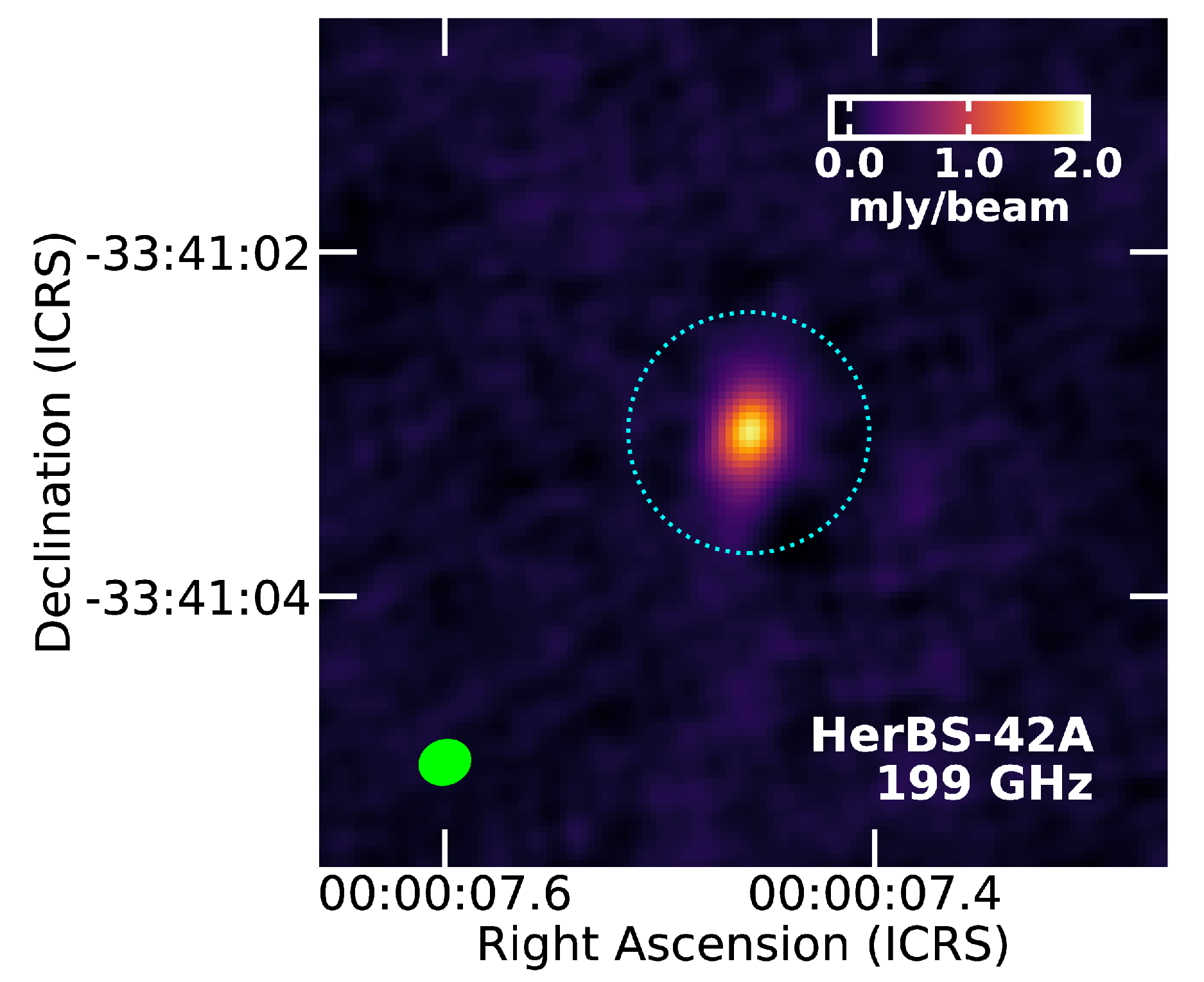} ~
\includegraphics[width=5.5cm]{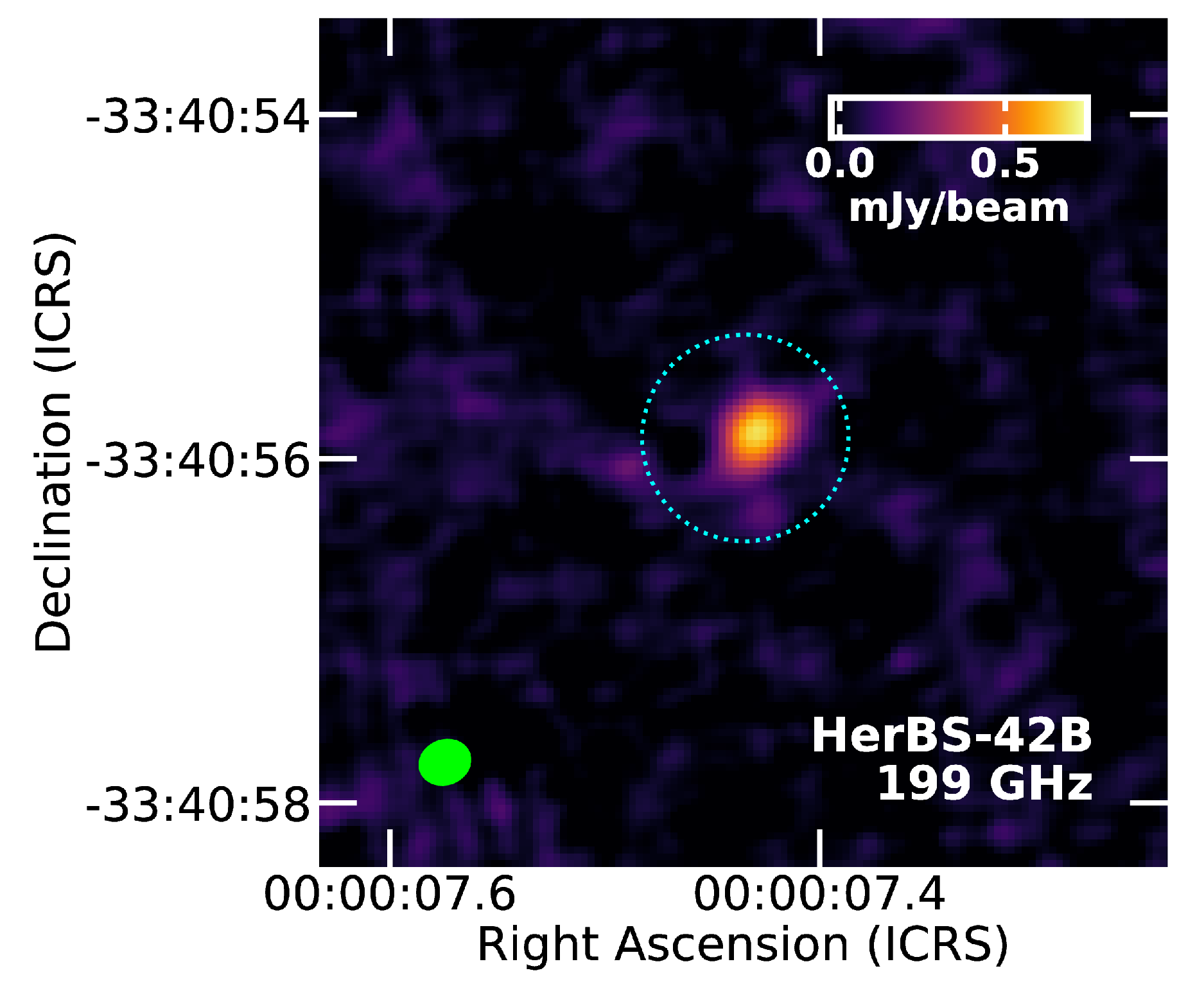} \\ ~ \\
\includegraphics[width=5.5cm]{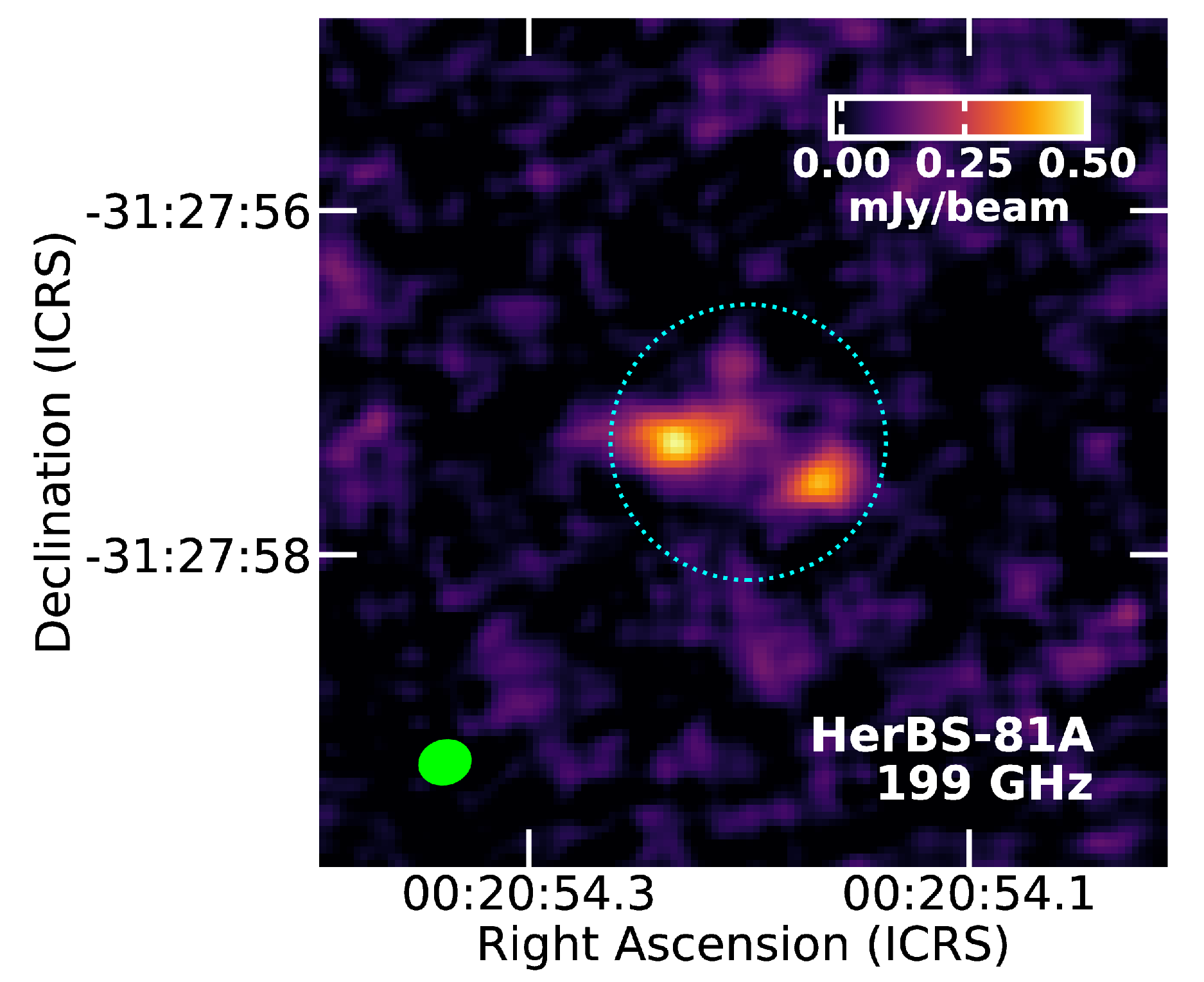} ~
\includegraphics[width=5.5cm]{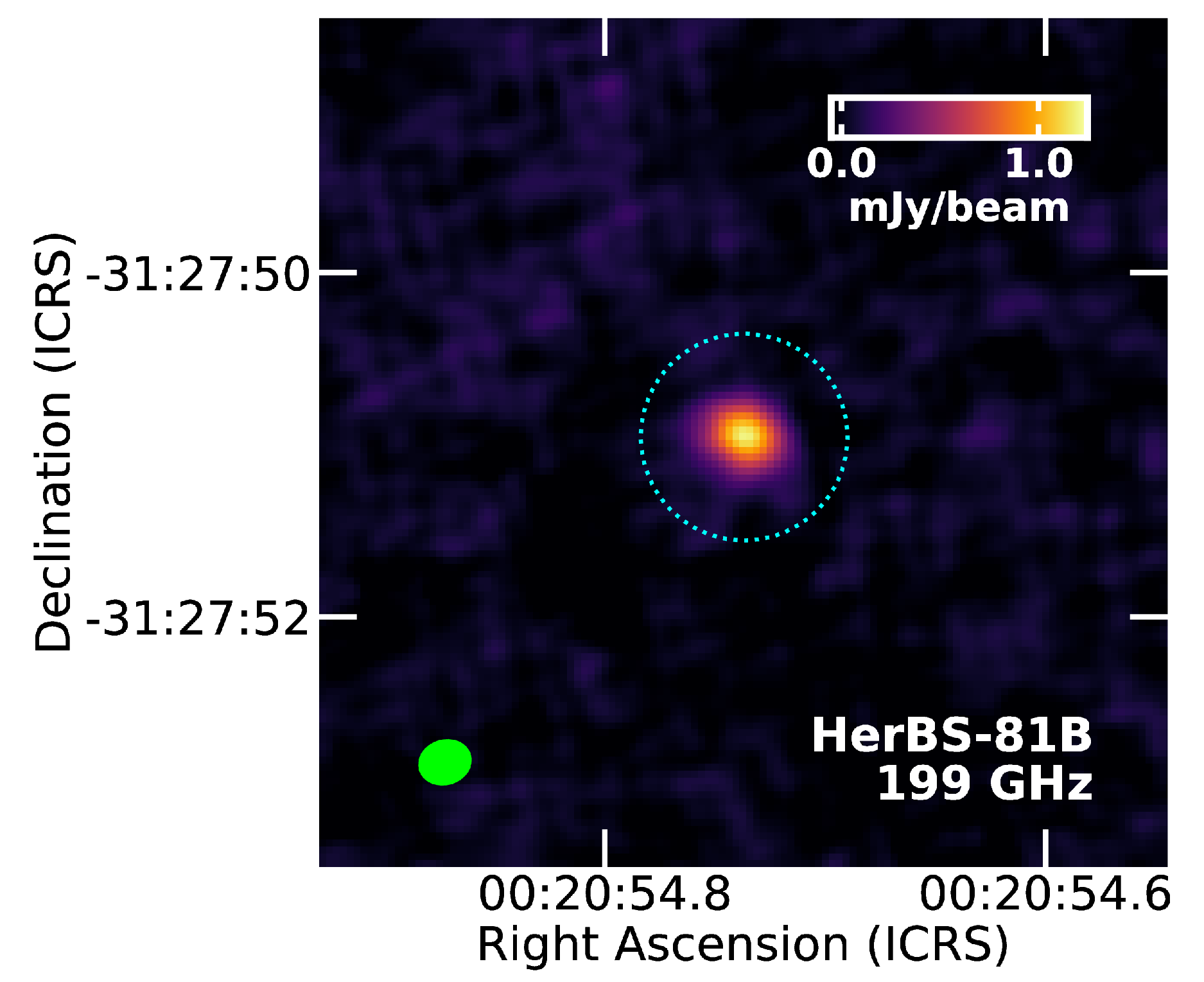} ~
\includegraphics[width=5.5cm]{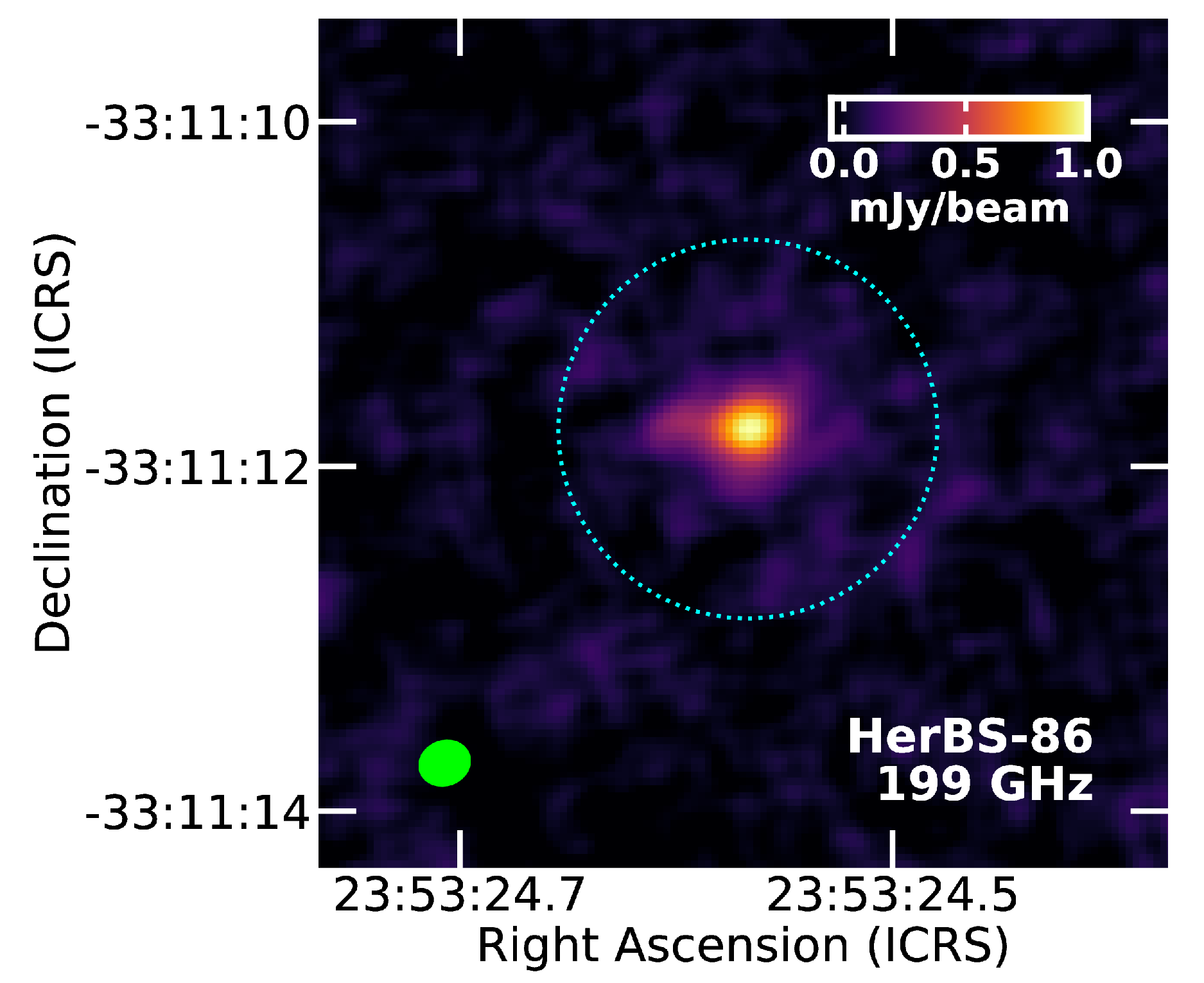} 
\end{center}
\caption{ALMA Band 5 (199 GHz) continuum images of the sources detected in the sample.  All images are $5\times5$~arcsec except for HerBS-21A, where the image size was increased to $6\times6$~arcsec to fit both images more easily within the panel.  All images use linear colour scales.  The measurement apertures used for the Band 5 data are shown as dotted cyan circles or ellipses; note that apertures of different sizes may have been used for the other bands.  The FWHM of the beams are shown as green ellipses in the bottom left corner of each panel.}
\label{f_map}
\end{figure*}

\addtocounter{figure}{-1}
\begin{figure*}
\begin{center}
\includegraphics[width=5.5cm]{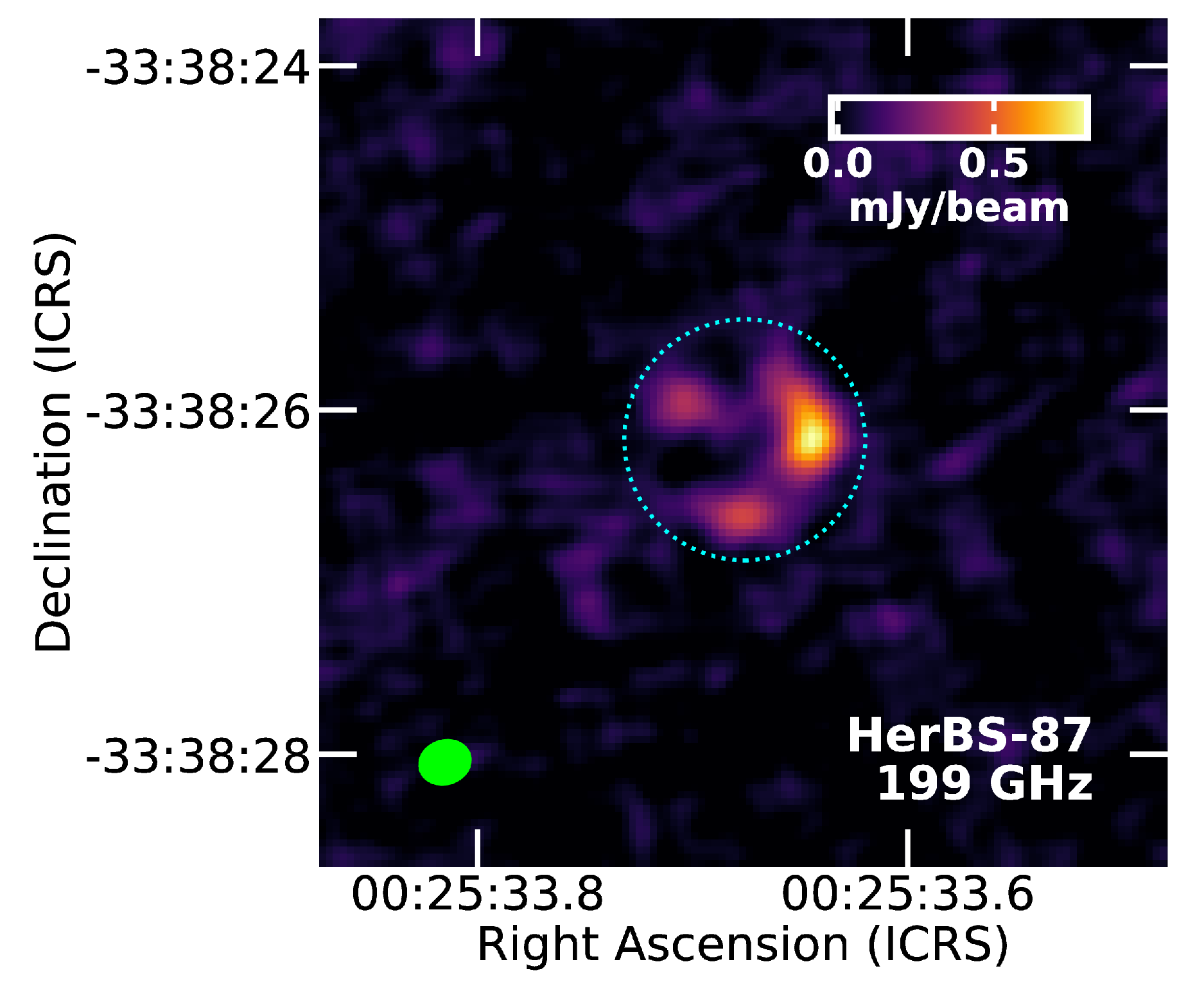} ~
\includegraphics[width=5.5cm]{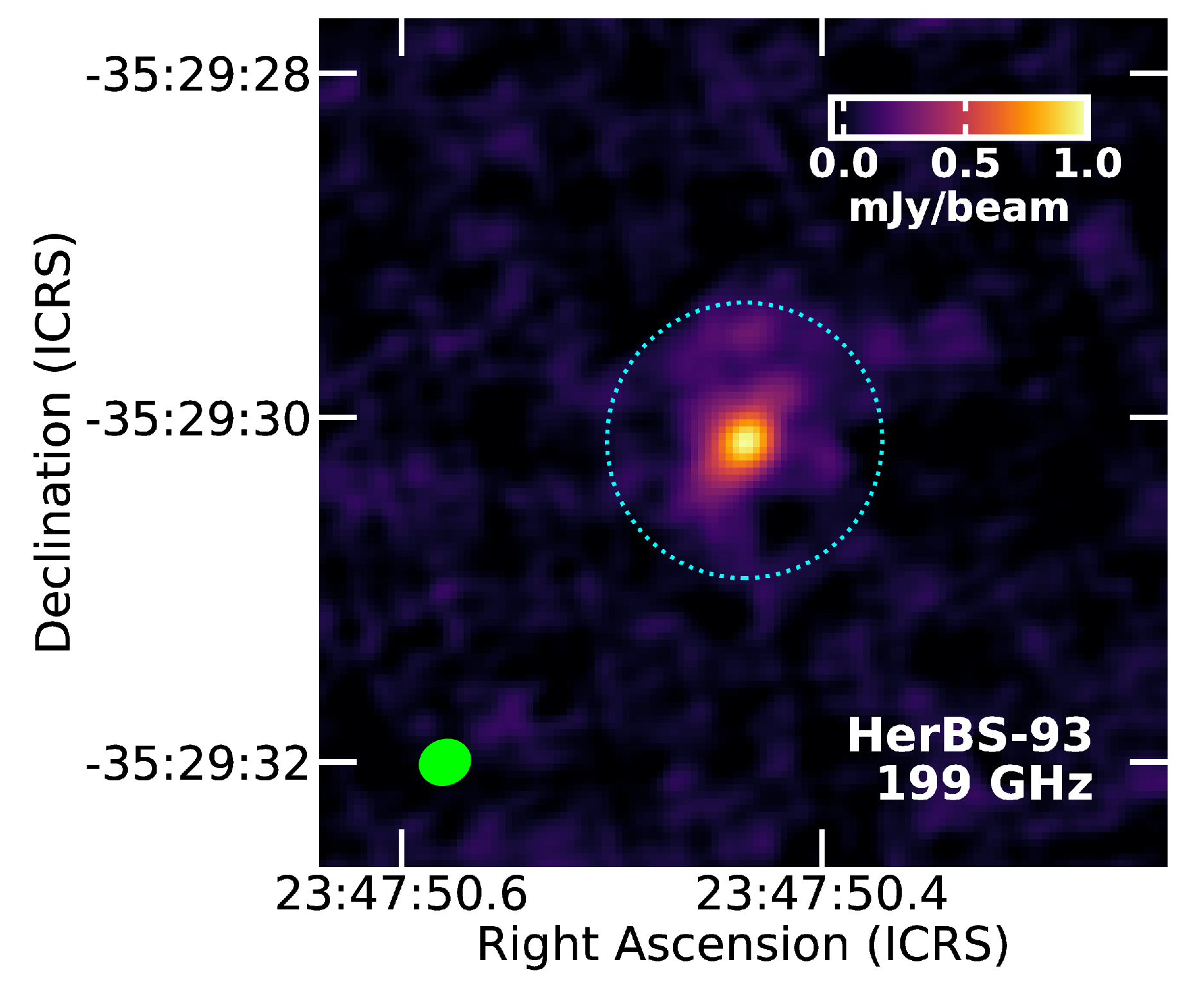} ~
\includegraphics[width=5.5cm]{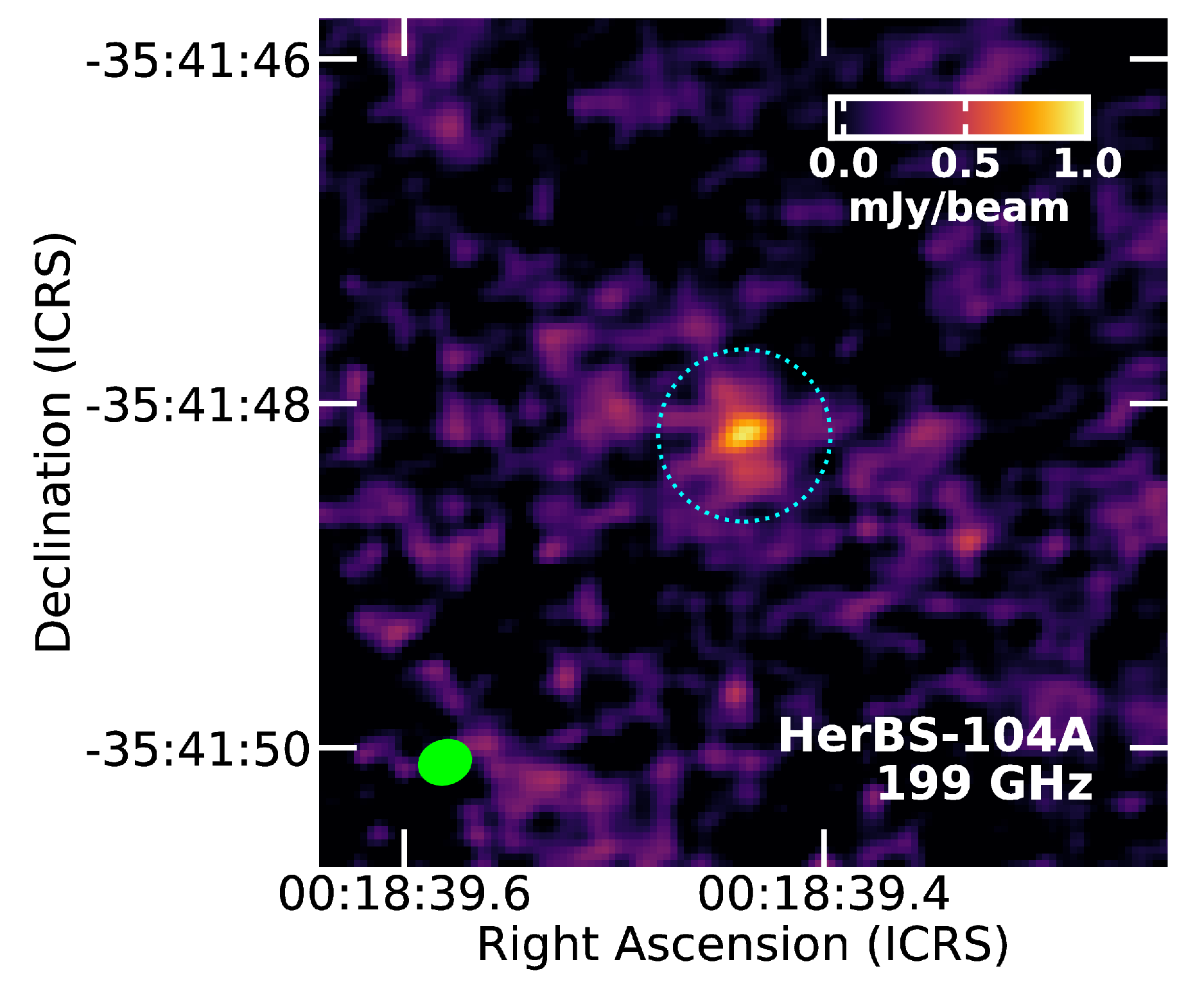} \\ ~ \\
\includegraphics[width=5.5cm]{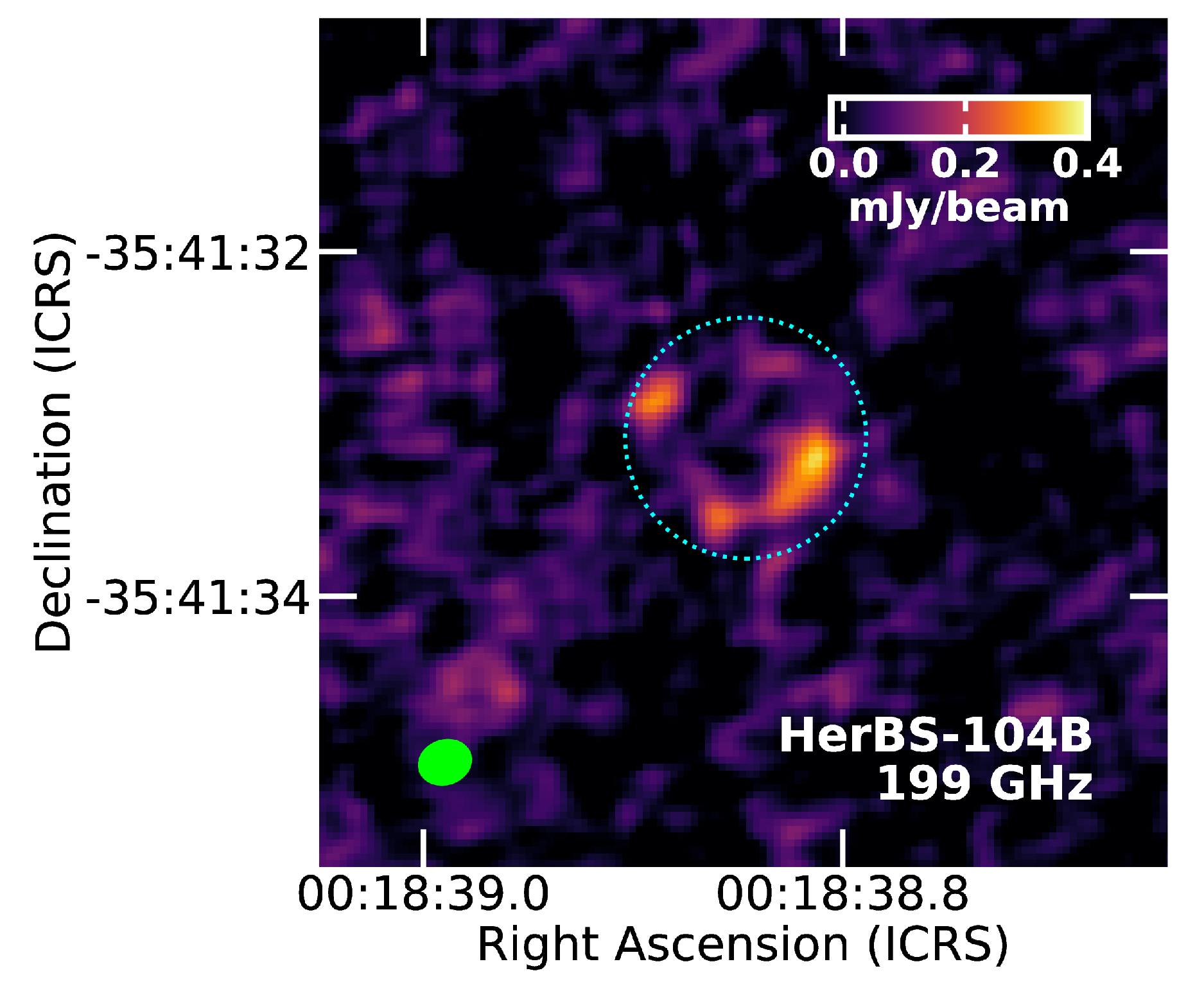} ~
\includegraphics[width=5.5cm]{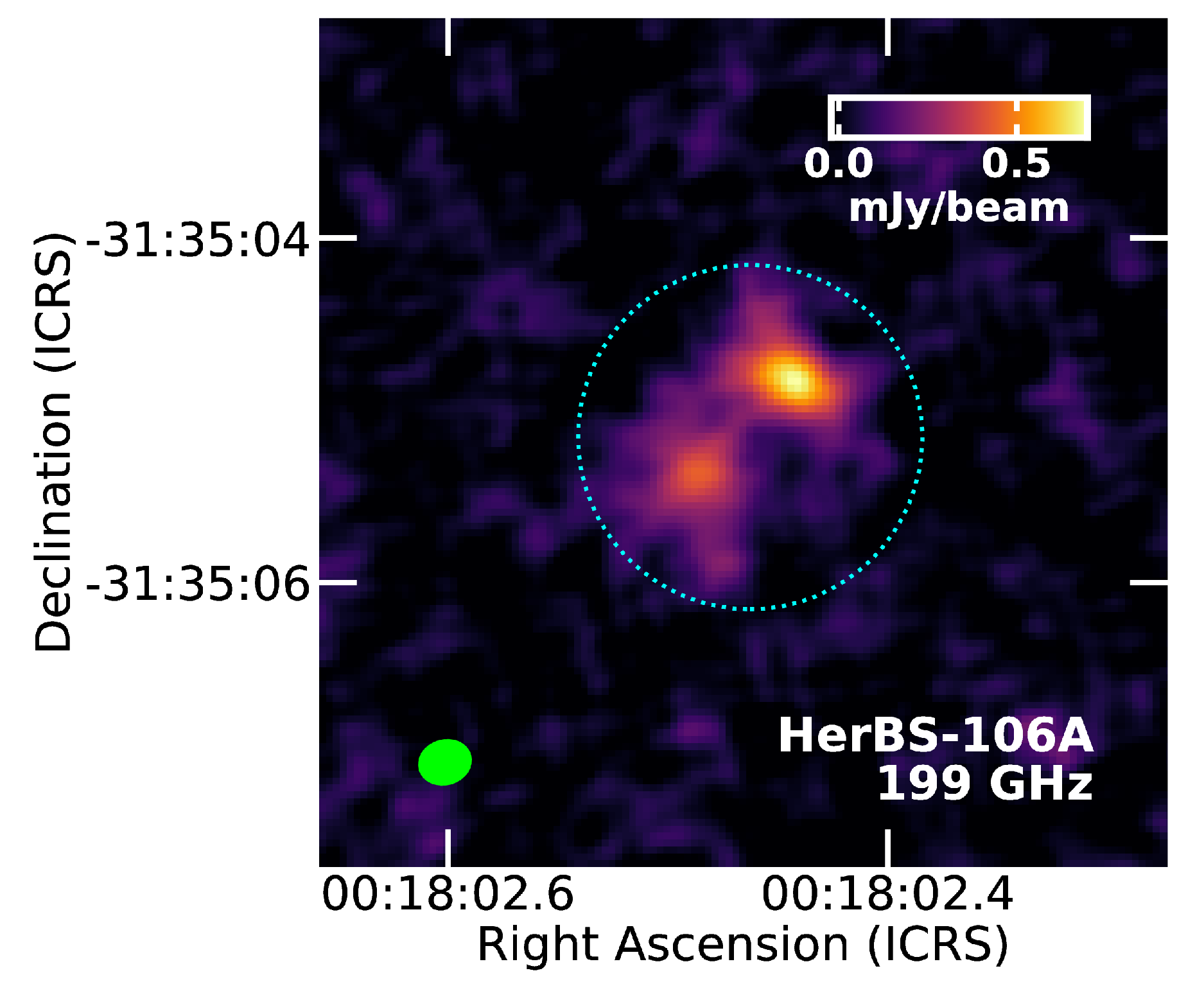} ~
\includegraphics[width=5.5cm]{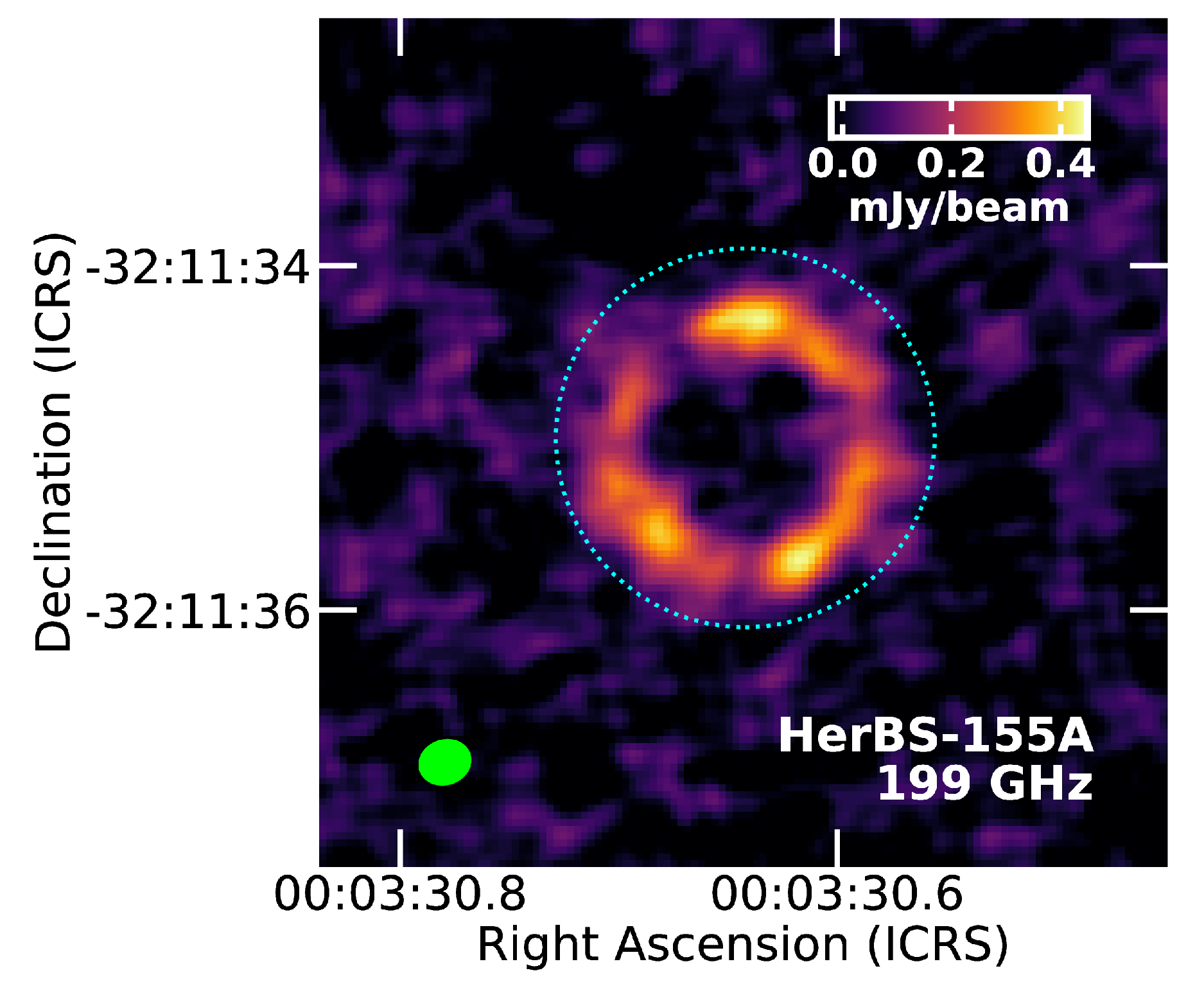} \\ ~ \\
\includegraphics[width=5.5cm]{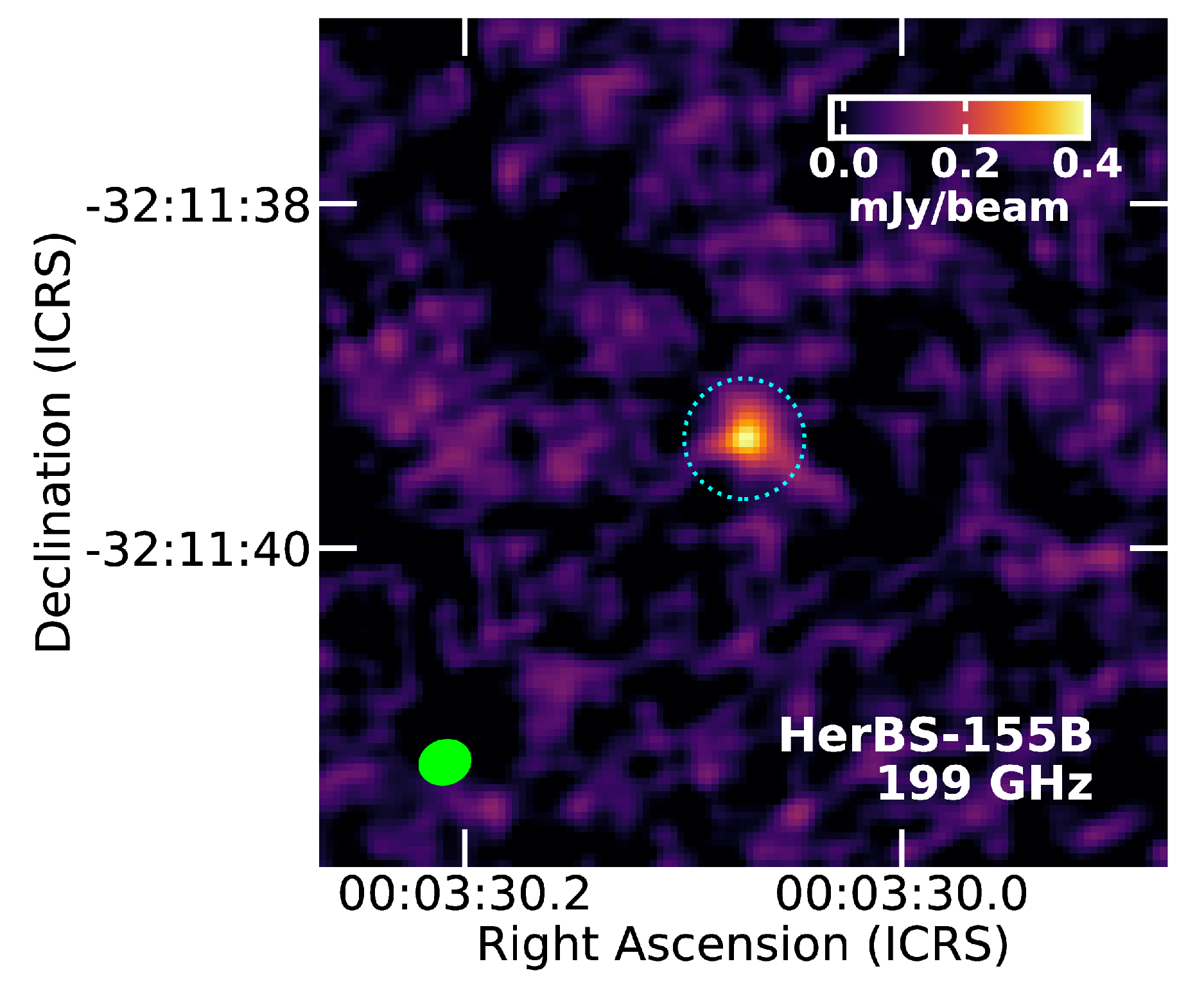} ~
\includegraphics[width=5.5cm]{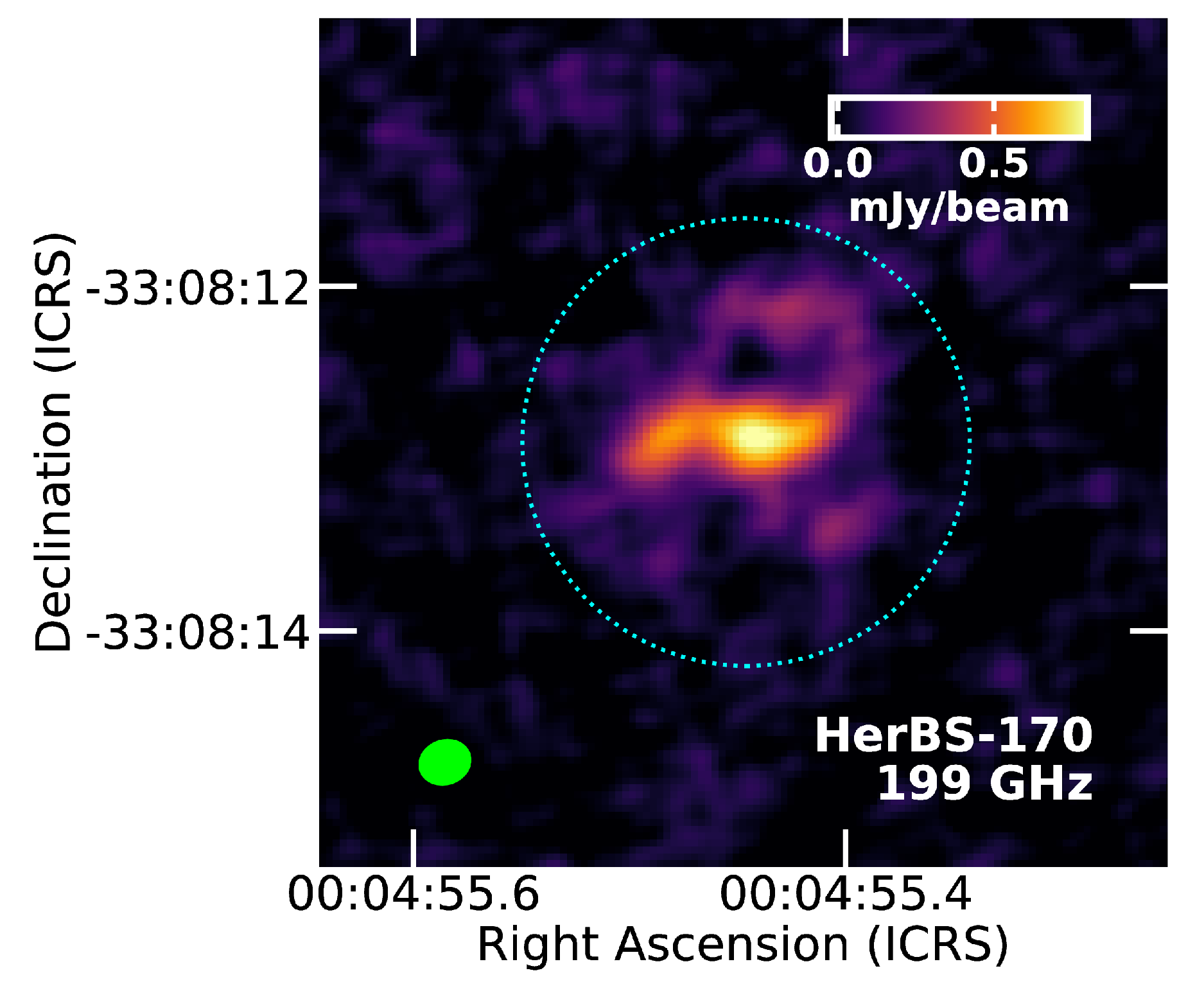} ~
\includegraphics[width=5.5cm]{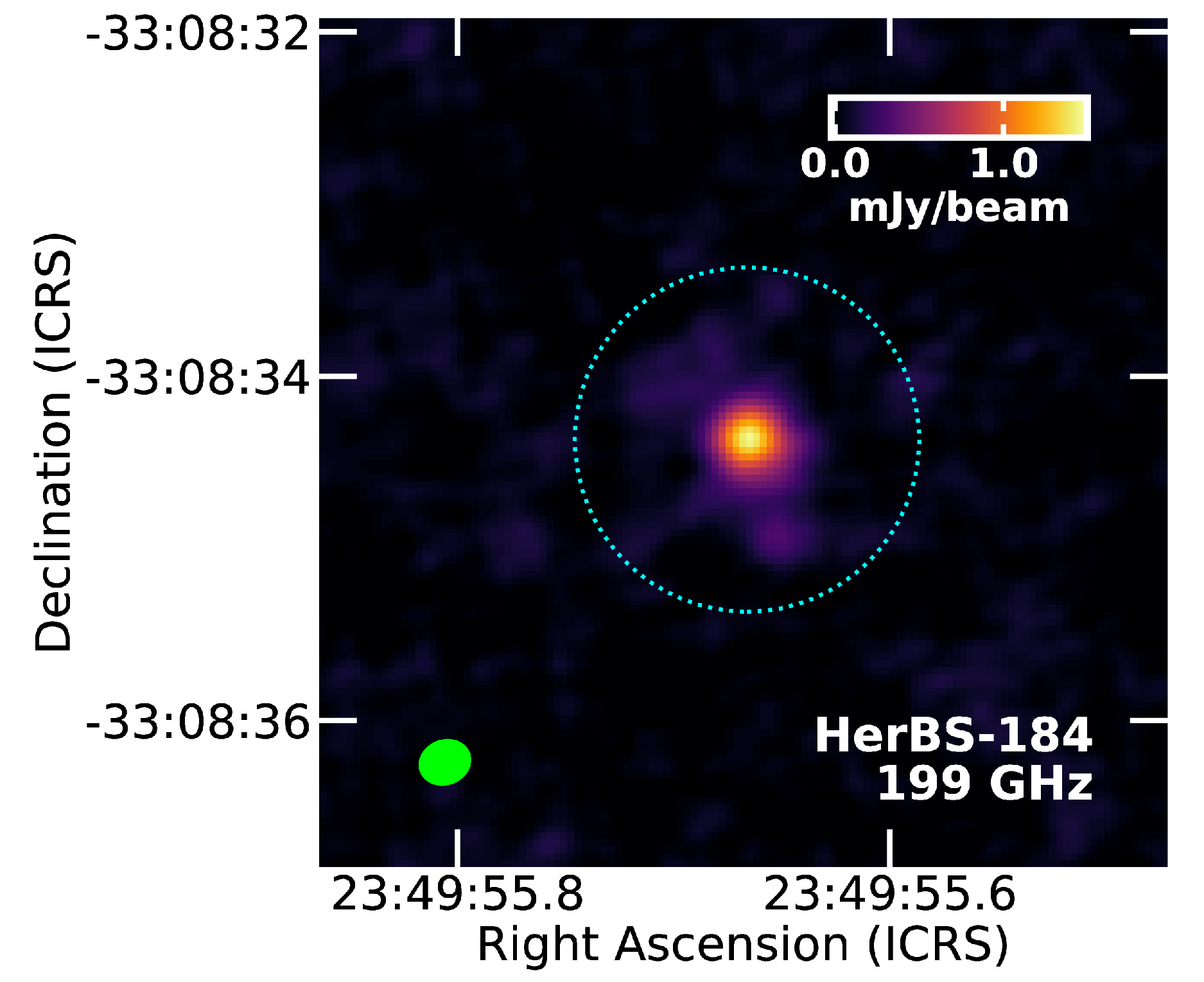} 
\hspace{11cm} ~ ~ 
\end{center}
\caption{Continued.}
\end{figure*}

\newpage
\section{SEDs for the entire sample}
\label{a_sed}

\begin{figure*}
\begin{center}
\includegraphics[width=17cm]{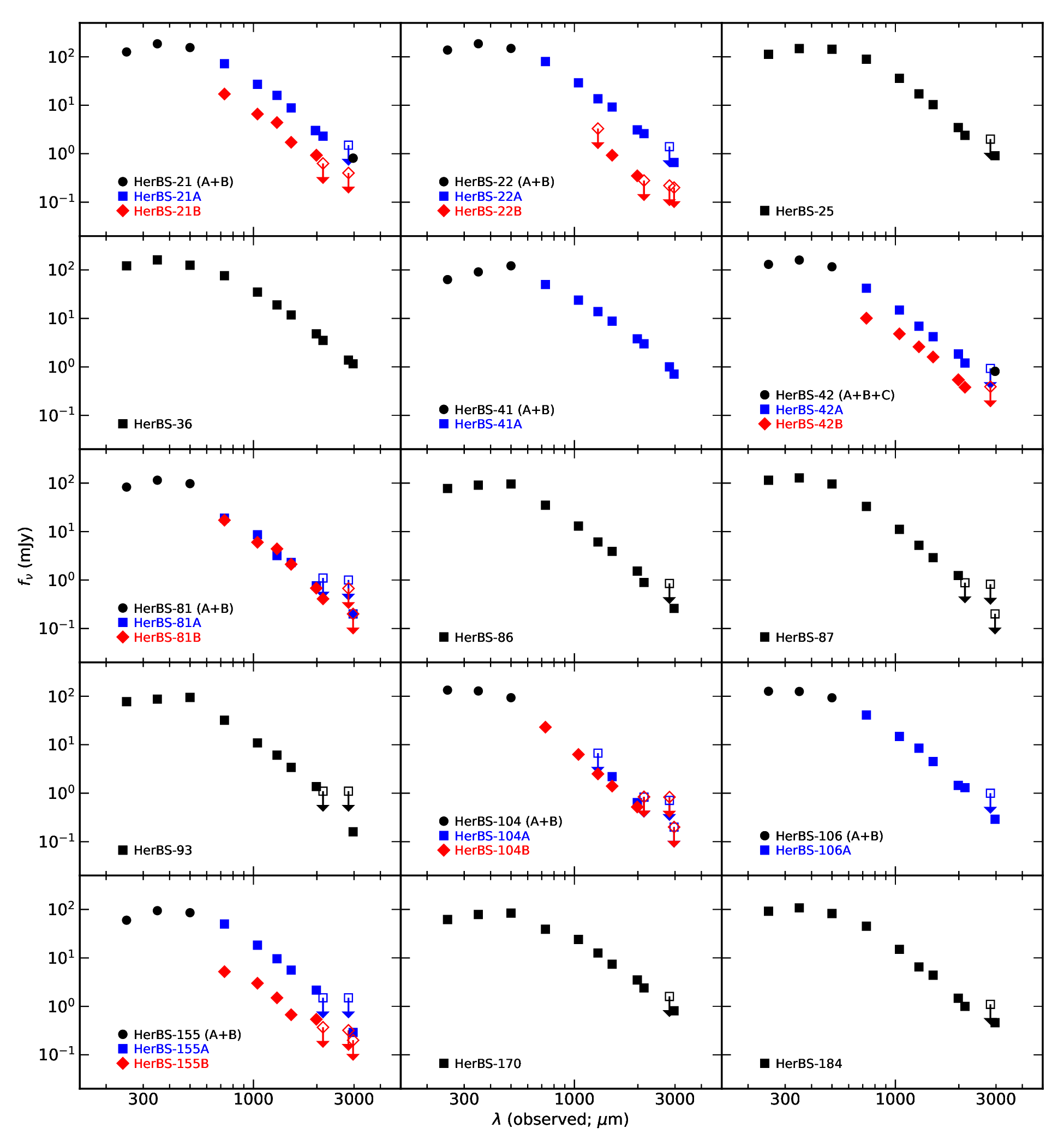}
\end{center}
\caption{SEDs of the objects in the sample sorted by field based on the ALMA data from this paper, the ALMA data from \citet{bendo2023}, and the {\it Herschel} data from H-ATLAS \citep[as reported by][]{valiante2016}.  When a field contains one ALMA source, we assume the {\it Herschel} emission is associated with that source.  When more than one ALMA source was detected, we use separate symbols for each source and then also use black circles for the {\it Herschel} data (and for the 101~GHz 7~m data for the HerBS-42 field) to indicate that the emission could originate from multiple sources.  The $5\sigma$ upper limits are shown as empty symbols with downward arrows.}
\label{f_sed}
\end{figure*}

\bsp
\label{lastpage}
\end{document}